\documentclass[english,aps,article,twocolumn,superscriptaddress]{revtex4-1}
\usepackage[T1]{fontenc}
\usepackage[latin9]{inputenc}
\usepackage{amsmath}
\usepackage{amssymb}
\usepackage{graphicx}
\usepackage{esint}

\makeatletter
\renewcommand\vec{\mathbf}
\renewcommand{\Im}{\mathrm{Im}}
\renewcommand{\Re}{\mathrm{Re}}
\newcommand{\Dt}{D_\mathrm{t}}

\makeatother

\usepackage{babel}
\begin{document}

\title{Symmetry, stability, and computation of degenerate lasing modes}

\author{David~Liu}

\email{daveliu@mit.edu}

\affiliation{Department of Mathematics, Massachusetts Institute of Technology,
Cambridge, Massachusetts 02139, USA}

\affiliation{Department of Physics, Massachusetts Institute of Technology, Cambridge,
Massachusetts 02139, USA}

\author{Bo~Zhen}

\affiliation{Department of Physics, Massachusetts Institute of Technology, Cambridge,
Massachusetts 02139, USA}

\author{Li~Ge}

\affiliation{Department of Engineering Science and Physics, College of Staten
Island, and The Graduate Center, CUNY, Staten Island, New York 10314,
USA}

\author{Felipe~Hernandez}

\affiliation{Department of Mathematics, Massachusetts Institute of Technology,
Cambridge, Massachusetts 02139, USA}

\author{Adi~Pick}

\affiliation{Department of Physics, Harvard University, Cambridge, Massachusetts
02138, USA}

\author{Stephan~Burkhardt}

\affiliation{Institute for Theoretical Physics, Vienna University of Technology
(TU Wien), A-1040 Vienna, Austria}

\author{Matthias~Liertzer}

\affiliation{Institute for Theoretical Physics, Vienna University of Technology
(TU Wien), A-1040 Vienna, Austria}

\author{Stefan~Rotter}

\affiliation{Institute for Theoretical Physics, Vienna University of Technology
(TU Wien), A-1040 Vienna, Austria}

\author{Steven~G.~Johnson}

\email{stevenj@math.mit.edu}

\affiliation{Department of Mathematics, Massachusetts Institute of Technology,
Cambridge, Massachusetts 02139, USA}

\affiliation{Department of Physics, Massachusetts Institute of Technology, Cambridge,
Massachusetts 02139, USA}
\begin{abstract}
We present a general method to obtain the stable lasing solutions
for the steady-state \emph{ab-initio} lasing theory (SALT) for the
case of a degenerate symmetric laser in two dimensions (2d). We find
that under most regimes (with one pathological exception), the stable
solutions are clockwise and counterclockwise circulating modes, generalizing
previously known results of ring lasers to all 2d rotational symmetry
groups. Our method uses a combination of semi-analytical solutions
close to lasing threshold and numerical solvers to track the lasing
modes far above threshold. Near threshold, we find closed-form expressions
for both circulating modes and other types of lasing solutions as
well as for their linearized Maxwell\textendash Bloch eigenvalues,
providing a simple way to determine their stability without having
to do a full nonlinear numerical calculation. Above threshold, we
show that a key feature of the circulating mode is its ``chiral''
intensity pattern, which arises from spontaneous symmetry-breaking
of mirror symmetry, and whose symmetry group requires that the degeneracy
persists even when nonlinear effects become important. Finally, we
introduce a numerical technique to solve the degenerate SALT equations
far above threshold even when spatial discretization artificially
breaks the degeneracy.
\end{abstract}
\maketitle

\section{Introduction\label{sec:Introduction}}

\begin{figure}
\centerline{\includegraphics[scale=1.1]{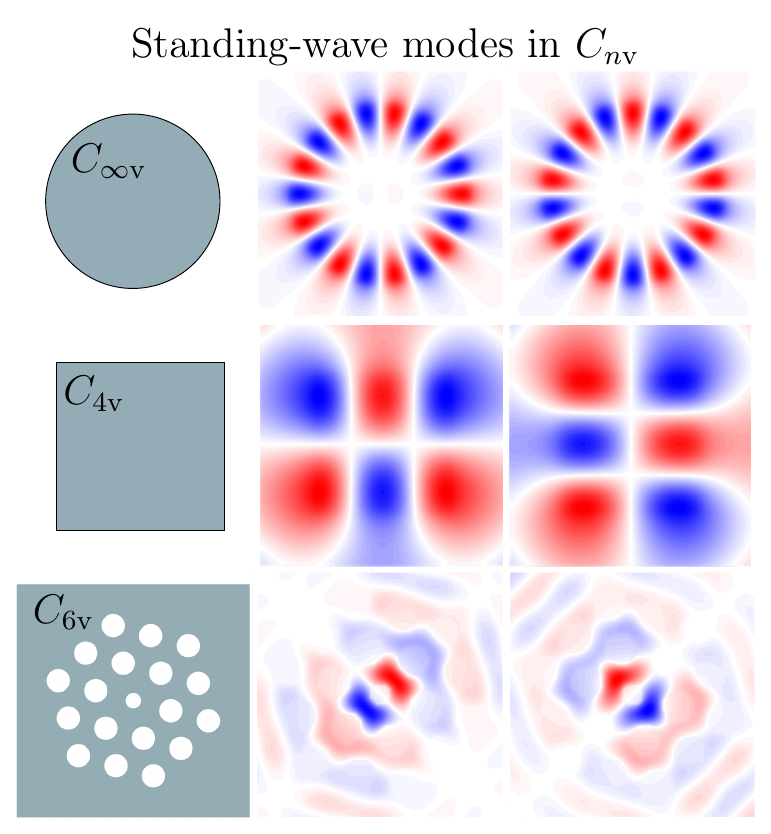}}

\caption{(Color online) Degenerate pairs of standing-wave modes in a laser.
For this uniform dielectric disk (top), which has $C_{\infty\mathrm{v}}$
symmetry, the two eigenfunctions (of which only the real part is shown)
are proportional to $\cos(\ell\phi)$ and $\sin(\ell\phi)$ (here,
$\ell=9$). For a homogeneous dielectric square ($\varepsilon=5$
inside a square of length 1), the eigenfunctions are $\frac{\pi}{2}$
rotations of one another. Here, the order of the irreducible representation
(irrep) is $\ell=1$, which is the only possibility for $C_{4\mathrm{v}}$.
The $\frac{\pi}{2}$ rotation is an exact symmetry of the geometry,
so there is an exact degeneracy even for the numerical grid. For $C_{6\mathrm{v}}$,
the symmetry group of the regular hexagon, as in this example of TE
modes (with the transverse magnetic field $H_{z}$ shown) in a 2d
slab with air holes (described in further detail in Sec.~\ref{sub:Examples}),
the two eigenfunctions have no immediately obvious symmetry operation
that transforms between them, but are in fact still degenerate (here
$\ell=1$, and there also exists an $\ell=2$ irrep with its own degenerate
pair). In all three cases (and in fact for all $C_{n\mathrm{v}}$),
the standing-wave modes have mirror planes that are $\frac{\pi}{2}$
rotations from one another, and the two standing-wave modes have opposite
parities across these mirror planes. \label{fig:overview}}
\end{figure}

Many lasers are formed from high-symmetry microcavity geometries that
have degenerate resonant modes, most famously ring and disc resonators
in which the clockwise and counterclockwise circulating modes are
degenerate (having the same complex resonant frequency). In a linear
system, any superposition of these solutions also satisfies Maxwell's
equations, but above-threshold lasers have nonlinear gain that allows
only certain superpositions; it is well known that the only stable
lasing solutions of a ring are the circulating solutions $\mathbf{E}\sim e^{im\phi}$,
as opposed to the standing-wave modes $\mathbf{E}\sim\sin m\phi$,
$\cos m\phi$ \cite{ref-cao_review,modeswitching2014,ref-matsko,ref-pnas_whispering}.
However, more recent microcavities often have other symmetry groups
supporting degeneracies \cite{inui_group}, such as the 6-fold symmetry
that commonly occurs in photonic-crystal resonators \cite{englund},
as seen in Fig.~\ref{fig:overview}, or more generally the $C_{n\mathrm{v}}$
symmetry group ($n$-fold rotations and $n$ mirror planes) for $n>2$~\cite{inui_group},
and much less is known about the lasing solutions in such cases. Figure~\ref{fig:overview}
gives examples of degenerate lasing modes in $C_{n\mathrm{v}}$ geometries.
Previous work \cite{ref-rotter_degeneracy} showed how the steady-state
degenerate solutions of SALT (steady-state ab-initio lasing theory
\cite{tureci_self-consistent_2006,tureci_strong_2008,tureci_ab_2009,ge_quantitative_2008,ge_steady-state_2010})
could be found from an educated guess of a superposition of the threshold
degenerate modes, and how their stability could be computed \emph{numerically}.
In this work, we show rigorously using degenerate perturbation theory
on the SALT equations that the circulating modes used in Ref.~\cite{ref-rotter_degeneracy},
along with standing-wave modes that are linear combinations of the
clockwise and counterclockwise circulating modes, are the \emph{only
}solutions to SALT in the $C_{n\mathrm{v}}$ degenerate case. We complement
those results with semi-analytical \emph{closed-form} expressions
for the stability eigenvalues of the Maxwell--Bloch equations linearized
about these lasing solutions (Sec.~\ref{sec:Threshold-perturbation-theory}).
We find that the only stable solutions right above threshold (with
one isolated exception that is unattainable under normal circumstances)
are typically the circulating ones. An important observation of our
paper is that $C_{n\mathrm{v}}$ symmetries experience a spontaneous
symmetry breaking due to nonlinearity above threshold, and analysis
of the resulting ``chiral'' symmetry~\cite{dichroism} is key to
stability of the lasing mode. These analytical solutions then give
us a starting point for a \emph{numerical} method to compute the degenerate
solutions far above threshold, extending our earlier work on computational
methods for non-degenerate SALT~\cite{direct_salt}. Our numerical
method, in turn, relies on a new semi-analytical technique (Sec.~\ref{sub:Forcing-the-degeneracy})
to address problems created by numerical symmetry breaking (e.g.,
by a low-symmetry computational grid) that would otherwise spoil the
nonlinear SALT solutions.

In Ref.~\cite{ref-rotter_degeneracy}, a full linear-stability analysis
(Sec.~\ref{sub:Stability-analysis summary}) was applied numerically
to the Maxwell--Bloch equations of lasing in order to check whether
the steady state was stable, and stability of the solution was also
analyzed when the degeneracy was broken by a perturbation. This generalized
many earlier works on ring-laser solutions and perturbations thereof~\cite{ref-risken_instability,ref-bidirectional,ref-MBring,ref-breaking_cylindrical_symmetry,ref-tamm}.
It reproduced the stability of the circulating solution near threshold,
and found that far above the lasing threshold (where nonlinearities
are strong) the circulating solution may become unstable (replaced
by an oscillating limit-cycle solution). Conversely, it was found
there that slightly breaking the symmetry caused the (now nearly degenerate)
solution to become unstable (e.g., oscillating between clockwise and
counterclockwise modes) in the vicinity of the threshold, but that
a stable solution re-appears further above threshold by means of cooperative
frequency locking~\cite{ref-lugiato-josa,ref-tamm}. The present
paper complements those results in two ways. First, near threshold,
we are able to both \emph{solve for} the steady-state lasing modes
(Sec.~\ref{sec:Threshold-perturbation-theory}) and evaluate their
stability (Sec.~\ref{sub:Stability-analysis summary}) \emph{analytically},
by perturbation theory in the basis of the degenerate linear solutions
at lasing threshold, and we generalize the notion of a circulating
laser mode to other symmetry groups and establish its stability near
threshold. Even the degeneracy itself is somewhat unusual above threshold,
because the nonlinear gain spontaneously breaks some of the symmetry
in noncircular ($C_{n\mathrm{v}}$ with $n\neq\infty$) geometries,
leaving one with a ``chiral'' degeneracy as discussed in Sec.~\ref{sub:Stability-analysis summary},
Ref.~\cite{dichroism}, and Appendix~\ref{sec: appendix allowed lasing modes}.
Second, we develop a numerical solution technique for \emph{far above}
threshold in Sec.~\ref{sub:Forcing-the-degeneracy}, generalizing
earlier SALT methods, where a numerical nonlinear solver is the only
option; in this regime, our focus is on finding a degenerate lasing
mode in SALT (if one exists), and we defer to the results of Ref.~\cite{ref-rotter_degeneracy}
for checking its stability after a solution far above threshold is
found.

Above threshold, the SALT equations \cite{tureci_self-consistent_2006,ge_steady-state_2010,tureci_ab_2009}
provide an elegant formulation of the problem of steady-state lasing
modes: they analytically eliminate the time dependence from the Maxwell\textendash Bloch
equations to obtain a nonlinear Maxwell-like eigenproblem $\nabla\times\nabla\times\mathbf{E}_{\mu}=\omega_{\mu}^{2}\varepsilon_{\mathrm{SALT}}\mathbf{E}_{\mu}$
for the lasing electric fields $\mathbf{E}_{\mu}$ and frequencies
$\omega_{\mu}$, in which the permittivity $\varepsilon_{\mathrm{SALT}}$
depends nonlinearly on both the field and frequency (here, the speed
of light $c$ has been set to unity). This equation can be efficiently
solved numerically by adapting standard techniques from computational
electromagnetism \cite{direct_salt}. As described below and also
in previous work~\cite{ref-rotter_degeneracy,modeswitching2014},
the SALT framework applies very naturally to lasing of degenerate
microcavities, assuming a stable degenerate steady state exists, but
it turns out that there are two complications. First, in order to
apply a numerical nonlinear solver to a large system of nonlinear
equations like SALT, one needs to have a good ``starting guess''
for the solution. In the non-degenerate case, the starting guess is
supplied by the threshold solution, but for a degenerate threshold
there are infinitely many superpositions. Picking the wrong starting
guess, e.g., the $\sin(m\phi)$ mode in a ring, would lead SALT to
converge to an unstable solutions, but our near-threshold perturbutation
theory supplies us with a correct guess (which turns out to be the
$C_{n\mathrm{v}}$ analog of the circulating solution in the ring). 

Second, there is a tricky complication that arises purely from \emph{numerical}
effects when a practical computational method is applied to spatially
discretize the SALT equations. In principle, what one would like to
find from a degenerate SALT solver is a lasing mode (e.g., the clockwise
circulating mode of a ring) with a passive pole (a pole of the Green's
function linearized around the lasing solution) that coincides with
the lasing frequency (there will be two possible lasing modes, e.g.,
clockwise and counterclockwise, but only one solution can exist at
a time with nonzero amplitude due to the nonlinearity; which one is
found will depend on the starting ``guess'' of the SALT solver).
However, when one discretizes a microcavity geometry for a numerical
solver, e.g., in a finite-difference or finite-element method, often
the discretization itself breaks the symmetry and hence breaks the
degeneracy slightly, causing the passive pole to separate from the
lasing frequency. In a linear eigenproblem, this is at worst a minor
annoyance, because from the symmetry group one can easily identify
resonance modes that ``should'' be degenerate \cite{inui_group,tinkham}.
In the nonlinear problem, however, the splitting can prevent the desired
solution (e.g., the circulating mode) from existing in the SALT equations,
because the solver can no longer pick arbitrary superpositions of
the formerly degenerate modes, as described in Sec.~\ref{sub:Forcing-the-degeneracy}.
(If the discretization breaks the degeneracy, but the pump strength
is high enough, a single-mode circulating solution \emph{may} still
come back into existance, due to strong nonlinear self-interaction
effects \cite{ref-rotter_degeneracy}. This effect can provide a fast
and easy way to initially evaluate the field profile of a discretized
geometry that is not exactly symmetric, and hence does not have an
exact degeneracy. However, the conditions under which this effect
can happen are not completely understood, as we explain in Sec.~\ref{sec:discretization_breaking},
and we wish to deal with the discretization symmetry breaking in a
more systematic and provably reliable manner.) To fix the problem
of broken degeneracy from discretization, we found a simple way to
uniquely restore the degeneracy in a way that both guarantees convergence
to the correct solution (as the discretization is refined), that generalizes
to an arbitrary number of lasing modes (in Sec.~\ref{sub:multimode discretization method}),
and that is, at worst, a few times more computationally expensive
than our non-degenerate solver.

\section{Background}

\subsection{Review of SALT}

The equations of SALT are derived from the Maxwell--Bloch equations
\cite{ref-haken_laser_theory,ref-sauermann,haken_light:_1985,ref-lamb}
(with the rotating-wave approximation):

\begin{align}
-\varepsilon\ddot{\mathbf{E}}^{+} & =\nabla\times\nabla\times\mathbf{E}^{+}+\ddot{\mathbf{P}}^{+}\nonumber \\
i\dot{\mathbf{P}}^{+} & =(\omega_{a}-i\gamma_{\perp})\mathbf{P}^{+}+\gamma_{\perp}\mathbf{E}^{+}D\label{eq:maxwell--bloch}\\
\dot{D} & =\gamma_{\parallel}(D_{0}-D)+\mathrm{Im}(\mathbf{E}^{-}\cdot\mathbf{P}^{+}),\nonumber 
\end{align}
where $\mathbf{E}^{+}(\mathbf{x},t)$ is the ``positive-frequency''
component of the electric field {[}with $\mathbf{E}^{-}=\mathbf{E}^{+\star}$
and the physical field given by $2\mathrm{Re}(\mathbf{E}^{+})${]},
$\varepsilon(\mathbf{x})$ is the ``cold-cavity'' permittivity (not
including the gain transition), $\mathbf{P}(\mathbf{x},t)$ is the
polarization describing a transition (of frequency $\omega_{a}$ and
linewidth $\gamma_{\perp}$) between two atomic energy levels, $D(\mathbf{x},t)$
is the population inversion between those two levels (with relaxation
rate $\Gamma_{\Vert}$), and $D_{0}(\mathbf{x})$ is the strength
of a pumping process driving the inversion. Additionally, for convenience,
one chooses units such that the following factors are set to unity:
the dipole moment matrix element of the two level system, Planck's
constant $\hbar$, and the speed of light $c$. Using the stationary
inversion approximation $D(\mathbf{x},t)\approx D(\mathbf{x})$ \cite{fu_multifrequency_1991,tureci_self-consistent_2006}
along with an ansatz of a finite number of lasing modes
\begin{equation}
\mathbf{E}^{+}(\mathbf{x},t)=\sum_{\nu}\mathbf{E}_{\nu}(\mathbf{x})e^{-i\omega_{\nu}t},\label{eq:positive frequency component}
\end{equation}
where $\omega_{\nu}$ are the real mode frequencies, the second equation
in Eq. (\ref{eq:maxwell--bloch}) is solved to eliminate $\mathbf{P}^{+}$
as an unknown, and the third equation becomes
\begin{equation}
\dot{D}=\gamma_{\parallel}(D_{0}-D)+D\,\Im\left[\sum_{\mu\nu}\Gamma(\omega_{\nu})\mathbf{E}_{\mu}^{\star}\cdot\mathbf{E}_{\nu}e^{i(\omega_{\mu}-\omega_{\nu})t}\right],\label{eq:Ddot MB equation}
\end{equation}
where $\Gamma(\omega_{\mu})\equiv\gamma_{\perp}/(\omega_{\mu}-\omega_{a}+i\gamma_{\perp})$.
In order for the stationary inversion approximation $\dot{D}=0$ to
be valid, the oscillating terms on the right-hand side of Eq.~(\ref{eq:Ddot MB equation})
must average to zero on a timescale much faster than the relaxation
timescale $1/\gamma_{\parallel}$. In order to do so, the beating
frequencies $\omega_{\mu}-\omega_{\nu}$ must be either exactly zero
or much faster than the relaxation rate $\gamma_{\parallel}$ \cite{fu_multifrequency_1991,tureci_self-consistent_2006,ge_quantitative_2008};
that is, two modes is either exactly degenerate or situated very far
apart from each other in frequency space, with the latter case resulting
in the time-dependent beating component of the inversion having a
negligible amplitude compared to the stationary component \cite{haken_light:_1985}.
When these conditions are met, Eq. (\ref{eq:maxwell--bloch}) reduces
to the SALT equation \cite{ge_steady-state_2010,tureci_self-consistent_2006,tureci_ab_2009}

\begin{equation}
\nabla\times\nabla\times\mathbf{E}_{\mu}=\omega_{\mu}^{2}\left[\varepsilon+\Gamma(\omega_{\mu})D\right]\mathbf{E}_{\mu},\label{eq:salt general}
\end{equation}
for the unknowns $\mathbf{E}_{\mu}$ and $\omega_{\nu}$, where $D(\mathbf{x})$
is the steady-state population inversion, which depends nonlinearly
on the electric fields and lasing frequencies of all lasing modes:
\begin{equation}
D(\mathbf{x})=\frac{D_{0}(\mathbf{x})}{1+\gamma_{\parallel}^{-1}{\displaystyle \sum_{\nu}}\left|\Gamma(\omega_{\nu})\mathbf{E}_{\nu}\right|^{2}}.\label{eq:D(x) inversion}
\end{equation}
The intensity term in the denominator of Eq.~(\ref{eq:D(x) inversion})
is known as the ``spatial hole-burning'' \cite{tureci_self-consistent_2006,fu_multifrequency_1991,haken_light:_1985}
term; it represents the saturation of the gain medium due to the total
time-averaged intensity of all the lasing modes. Once Eq.~(\ref{eq:salt general})
is solved for all the lasing modes $\mathbf{E}_{\mu}$ and frequencies
$\omega_{\mu}$, one typically checks that the ``passive'' poles,
i.e. the eigenvalues $\tilde{\omega}_{\mu}$ of the linearized SALT
equation
\begin{equation}
\nabla\times\nabla\times\tilde{\mathbf{E}}_{\mu}=\tilde{\omega}_{\mu}\left[\varepsilon+\Gamma(\tilde{\omega}_{\mu})D\right]\tilde{\mathbf{E}}_{\mu},\label{eq:passive pole salt}
\end{equation}
are not above the real axis. As long as $\ensuremath{|\omega_{\nu}-\tilde{\omega}_{\mu}|\gg\gamma_{\parallel}}$
(where $\omega_{\nu}$ are the lasing frequencies), this is a good
indicator that the SALT solution is stable. However, a rigorous evaluation
of the stability of the SALT solution requires a linear stability
analysis based on the MB equations \cite{ref-rotter_degeneracy}.
(In Sec.~\ref{sub:Stability-analysis summary}, we give analytical
results for this stability analysis for the near-threshold degenerate
case.)

\subsection{Effects of exact degeneracies}

So far, most cases in which SALT has been applied have dealt with
either single lasing modes or multimode regimes in which frequencies
are far apart. When two lasing frequencies are close but not exactly
degenerate, there is non-negligible beating and SALT is invalid. However,
when two lasing modes are \emph{exactly} degenerate, we find that
SALT is still perfectly valid, because there is an exact steady-state
solution of the MB equations (for a single lasing mode), provided
that interference between the two degenerate modes is taken into account.
Of course, it is possible that a degeneracy in the linear regime may
split in the presence of the laser nonlinearity above threshold. However,
if a degeneracy persists (and we have observed that it is guaranteed
to do so for $C_{n\mathrm{v}}$ symmetry-induced degeneracies, because
of the ``chiral'' symmetry of the lasing mode as discussed in Appendix.~\ref{sec: degeneracy in Cn}),
our method will find it. The literature on degenerate lasing modes
has almost invariably dealt with whispering-gallery modes in microdisks
and ring resonators \cite{ref-cao_review,modeswitching2014,ref-matsko,ref-pnas_whispering}.
Many of these earlier works discussed the stability of traveling-wave
modes in ring resonators under perturbations that break the symmetry
\cite{ref-risken_instability,ref-bidirectional,ref-MBring,ref-breaking_cylindrical_symmetry,ref-tamm}.
A very limited number of other works on degenerate lasing modes in
other geometries exist \cite{ref-braun}, which were mostly experimental
and focused on the linear cavity rather than the nonlinear lasing
regime. However, the microdisk is just one of many examples of a setting
where one can find degenerate resonant modes that can lase: there
are a great variety of other symmetric geometries where degeneracies
can occur \cite{inui_group,tinkham,sakoda}. So far, the  problem
of above-threshold degenerate modes in lasers has not been studied
systematically for the general $C_{n\mathrm{v}}$ case. 

The presence of degenerate eigenvalues is typically a direct consequence
of symmetry. For systems with $C_{n\mathrm{v}}$ symmetry for $n>2$
($n$-fold rotational symmetry with $n$ mirror planes, the symmetry
of the regular $n$-gon), the existence of 2d irreducible representations
(irreps) of the symmetry group corresponds to 2-fold degeneracies.
Below, we therefore refer to 2-fold degenerate modes (at lasing threshold)
as corresponding to a 2d irrep, and we exploit some known properties
of these irreps in deriving selection rules \cite{inui_group} for
overlap integrals. For systems with $C_{n}$ symmetry ($n$-fold rotational
symmetry without mirror symmetry, e.g., a ``chiral'' spiral structure
with $n$ arms), the combination of group theory and electromagnetic
reciprocity again supports 2-fold degenerate solutions \cite{dichroism}
(see also Appendix~\ref{sec: degeneracy in Cn} for a review). Even
with $C_{n\mathrm{v}}$ symmetry, we explain below that the nonlinear
hole-burning term for lasers above threshold typically breaks the
mirror symmetry, so the reciprocity argument for $C_{n}$ symmetry
is crucial to maintaining the degeneracy of the lasing mode and a
passive pole. Figure~\ref{fig:overview} shows three examples of
symmetric geometries, along with examples of degenerate eigenfunctions.

Ordinarily, SALT assumes that all distinct modes have distinct frequencies,
i.e. $\omega_{\mu}\neq\omega_{\nu}$ when $\mu\neq\nu$, which gives
the stationary-inversion expression Eq.~(\ref{eq:D(x) inversion})
when higher-frequency $\omega_{\mu}-\omega_{\nu}$ ($\nu\neq\mu$)
terms are dropped. However, when there are degeneracies, the MB equations
will have terms of the form $\mathbf{E}_{\mu}\cdot\mathbf{E}_{\nu}^{\star}$
where $\mu\neq\nu$, since $\omega_{\mu}=\omega_{\nu}$ and one can
no longer drop the $e^{i(\omega_{\mu}-\omega_{\nu})t}$ term. The
correct expression for the stationary inversion will then be
\begin{equation}
D=\frac{D_{0}}{1+\gamma_{\parallel}^{-1}\sum^{\prime}\Gamma_{\mu}\mathbf{E}_{\mu}\cdot\Gamma_{\nu}^{\star}\mathbf{E}_{\nu}^{\star}},\label{eq:Dcorrect}
\end{equation}
where $\Gamma_{\mu}\equiv\Gamma(\omega_{\mu})$ and $\sum^{\prime}$
indicates a summation over all $\mu$ and $\nu$ for which $\omega_{\mu}=\omega_{\nu}$,
not just for $\mu=\nu$. To illustrate the difference between the
two, we examine a case in which there are three lasing modes, two
of which are degenerate with each other ($\omega_{1}=\omega_{2}\neq\omega_{3}$).
Equation~(\ref{eq:D(x) inversion}) will have
\begin{equation}
\left|\Gamma_{1}\mathbf{E}_{1}\right|^{2}+\left|\Gamma_{2}\mathbf{E}_{2}\right|^{2}+\left|\Gamma_{3}\mathbf{E}_{3}\right|^{2}
\end{equation}
in the denominator, while Eq.~(\ref{eq:Dcorrect}) will have
\begin{equation}
\left|\Gamma_{1}(\mathbf{E}_{1}+\mathbf{E}_{2})\right|^{2}+\left|\Gamma_{3}\mathbf{E}_{3}\right|^{2}.\label{eq:E123}
\end{equation}
From Eq.~(\ref{eq:E123}) we see that the degenerate pair acts as
a single mode that is a superposition of $\mathbf{E}_{1}$ and $\mathbf{E}_{2}$.
This means that the solution to the lasing degenerate problem can
be portrayed in two equivalent pictures. First, we can think of the
linear combination $\mathbf{E}=\mathbf{E}_{1}+\mathbf{E}_{2}$ as
a single mode that satisfies the equation
\begin{align}
-\nabla\times\nabla\times\mathbf{E} & =\omega_{1}^{2}\left(\varepsilon+D\Gamma_{1}\right)\mathbf{E}\nonumber \\
D & \equiv\frac{D_{0}}{1+\gamma_{\parallel}^{-1}\left(\left|\Gamma_{1}\mathbf{E}\right|^{2}+\left|\Gamma_{3}\mathbf{E}_{3}\right|^{2}\right)}\label{eq:1mode_salt}
\end{align}
{[}where the external pump $D_{0}(\mathbf{x})$ may be spatially dependent,
as noted before{]}. Second, we can think of the two modes as \emph{separately
}satisfying the two equations
\begin{align}
-\nabla\times\nabla\times\mathbf{E}_{1,2} & =\omega_{1}^{2}\left(\varepsilon+D\Gamma_{1}\right)\mathbf{E}_{1,2}\nonumber \\
D & \equiv\frac{D_{0}}{1+\gamma_{\parallel}^{-1}\left(\left|\Gamma_{1}(\mathbf{E}_{1}+\mathbf{E}_{2})\right|^{2}+\left|\Gamma_{3}\mathbf{E}_{3}\right|^{2}\right)}.\label{eq:2mode_salt}
\end{align}
The existence of a solution to Eq.~(\ref{eq:1mode_salt}) is a necessary
but not sufficient condition for the existence of a solution to Eq.~(\ref{eq:2mode_salt}).
The reason is that Eq.~(\ref{eq:2mode_salt}) enforces a double eigenvalue
of the linearized eigenproblem (i.e. a double pole of the Green's
function) on the real-$\omega$ axis, whereas Eq.~(\ref{eq:1mode_salt})
only enforces a single eigenvalue.

Prior to lasing, suppose that we have a 2-fold degenerate solution,
corresponding to a double pole in the Green function. As the gain
increases, and even when the system passes threshold and becomes nonlinear,
poles can shift (and degeneracies may split) but poles do not appear
or disappear discontinuously, so we should always expect there to
be two poles (in the linearized Green's function around the SALT solution)
arising from the original degenerate pair. Given this fact, if we
solve the single-mode SALT equations as in Eq.~(\ref{eq:1mode_salt}),
there is the danger that the other pole is unstable. As we show in
Appendix.~\ref{sec: appendix stability calculations}, close to lasing
threshold the zeroth order stability analysis (in the pump strength
increment) simply depends on the SALT eigenproblem: if a SALT pole
lies above the real-$\omega$ axis, then a lasing solution is necessarily
unstable, whereas SALT poles below the real axis cannot induce instability.
If a pole lies \emph{on} the real axis, higher-order calculations
are required to check stability as described in Sec.~\ref{sub:Stability-analysis summary}.

On the other hand, if we find a solution of the two-mode SALT equations
as in Eq.~(\ref{eq:2mode_salt}), then by construction we have placed
both poles together on the real-$\omega$ axis and the other passive
pole by itself is not a source of instability (and the overall stability
of the Maxwell\textendash Bloch equations can be checked as in Ref.
\cite{ref-rotter_degeneracy}). However, Eq.~(\ref{eq:2mode_salt})
has a drawback: the hole-burning term now depends on the relative
phase of $\mathbf{E}_{1}$ and $\mathbf{E}_{2}$. In the original
SALT equations, even for multimode problems, the phase was irrelevant
and was chosen arbitrarily in order to obtain a solvable system of
equations. If we remove the arbitrary phase choice, our equations
(derived in Ref. \cite{direct_salt}) become underdetermined. However,
if we solve the single-mode equation (Eq.~(\ref{eq:1mode_salt}))
but simultaneously constrain the other pole (the linearly independent
degenerate partner) to be degenerate with the lasing pole, then we
will effectively have solved Eq.~(\ref{eq:2mode_salt}), and in the
following sections we will explain how to implement this constraint.

\section{Threshold perturbation theory\label{sec:Threshold-perturbation-theory}}

In this section, we analyze the SALT (Eq.~(\ref{eq:salt general}))
and Maxwell--Bloch equations (Eq. (\ref{eq:maxwell--bloch})) just
above the lasing threshold in order to obtain insight into the nature
of the degenerate solutions, as well as to determine the correct initial
guess for the above-threshold regime (e.g., when solving for lasing
modes using the method of Ref. \cite{direct_salt}). For a regime
with a single steady-state lasing mode $\mathbf{E}^{+}(\mathbf{x},t)=\mathbf{E}(\mathbf{x})e^{-i\omega t}$
with frequency $\omega$, one obtains a stationary inversion \cite{tureci_self-consistent_2006,tureci_ab_2009,ge_steady-state_2010}
$D(\mathbf{x},t)=D(\mathbf{x})$ and the single-mode SALT nonlinear
eigenproblem
\begin{equation}
\nabla\times\nabla\times\mathbf{E}=\omega^{2}\left[\varepsilon+\frac{D_{0}\Gamma(\omega)}{1+\gamma_{\parallel}^{-1}\left|\Gamma(\omega)\mathbf{E}\right|^{2}}\right]\mathbf{E}.\label{eq:single-mode-salt}
\end{equation}
The first lasing threshold occurs when $D_{0}$ is increased to a
value $D_{\mathrm{t}}$ where a complex eigenvalue $\omega$ of this
SALT equation with infinitesimal $\mathbf{E}$ hits the real-$\omega$
axis ($\Im\,\omega=0$) \cite{tureci_self-consistent_2006,ge_steady-state_2010}.

Now, we will consider the \emph{near-threshold }problem $D_{0}=D_{\mathrm{t}}(1+d)$
for $0\le d\ll1$, for the case where the threshold mode ($d=0$)
is doubly degenerate, and expand the solutions to lowest order in
$d$. First (Sec.~\ref{sub:Perturbative-SALT-solutions}), we will
solve the SALT equations perturbatively in $d$, in order to find
the steady-state lasing solutions near threshold, regardless of whether
they are stable. Then (Sec.~\ref{sub:Stability-analysis summary}),
we will plug those solutions into the full Maxwell--Bloch equations,
again expanding to lowest-order in $d$, in order to evaluate the
dynamical stability of the SALT modes. This yields a small $4\times4$
eigenproblem, whose eigenvalues determine the stability, and whose
matrix elements are integrals of the threshold solutions. In the case
of $C_{n\mathrm{v}}$ symmetry, we know enough about the modes in
order to simplify many of these calculations analytically, to conclude:
(i) the only SALT solutions are either standing-wave or circulating
solutions (defined below); (ii) the standing-wave modes are unstable
for all $C_{n\mathrm{v}}$ cases (except for a small group of isolated,
realistically unattainable examples when $n$ is a multiple of four),
and otherwise the stability can be determined by evaluating a simple
integral of the threshold modes.

\subsection{Perturbative lasing solutions near threshold\label{sub:Perturbative-SALT-solutions}}

\begin{figure}[!h]
\centerline{\includegraphics[scale=0.44]{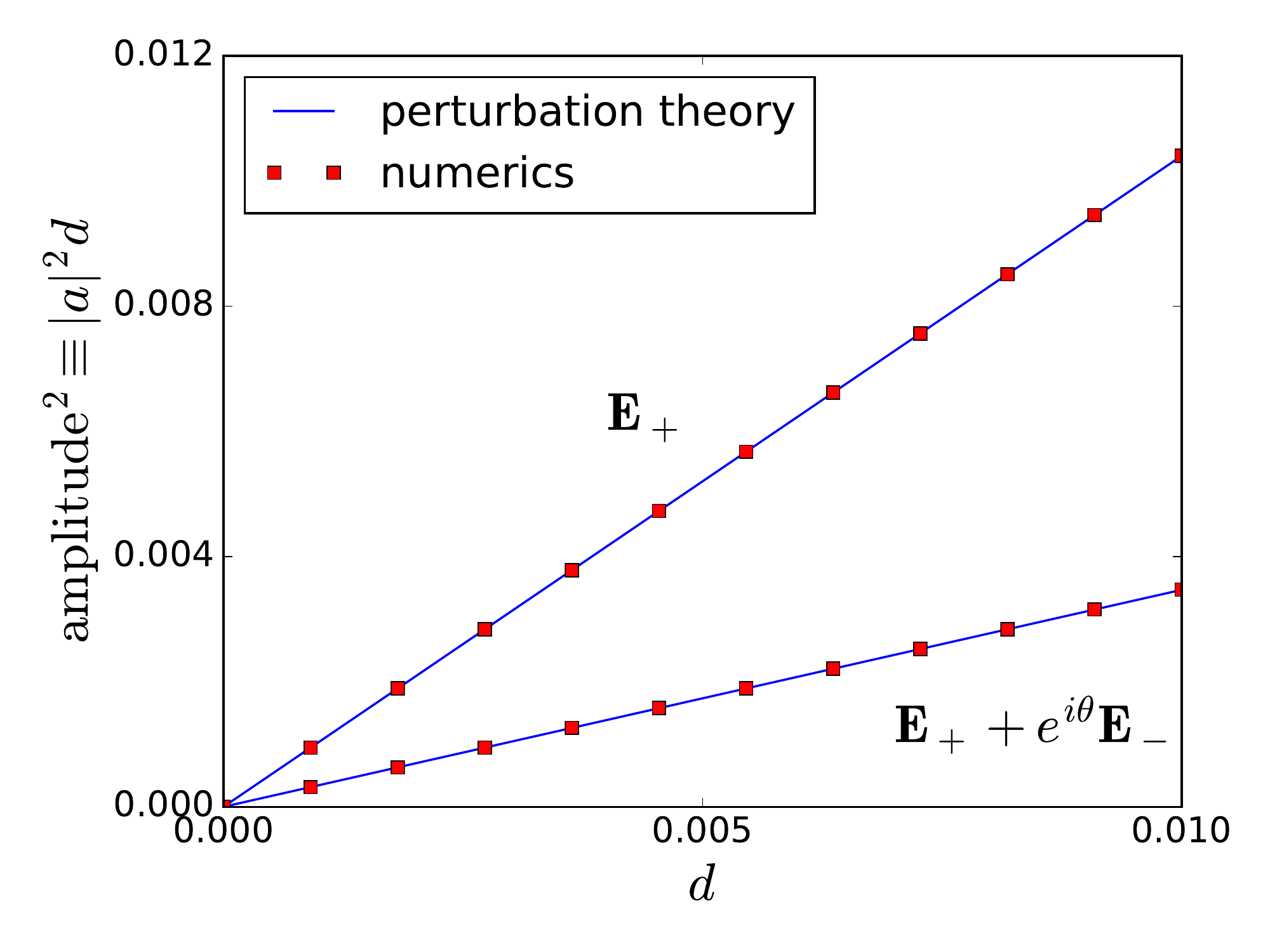}}

\centerline{\includegraphics[scale=0.43]{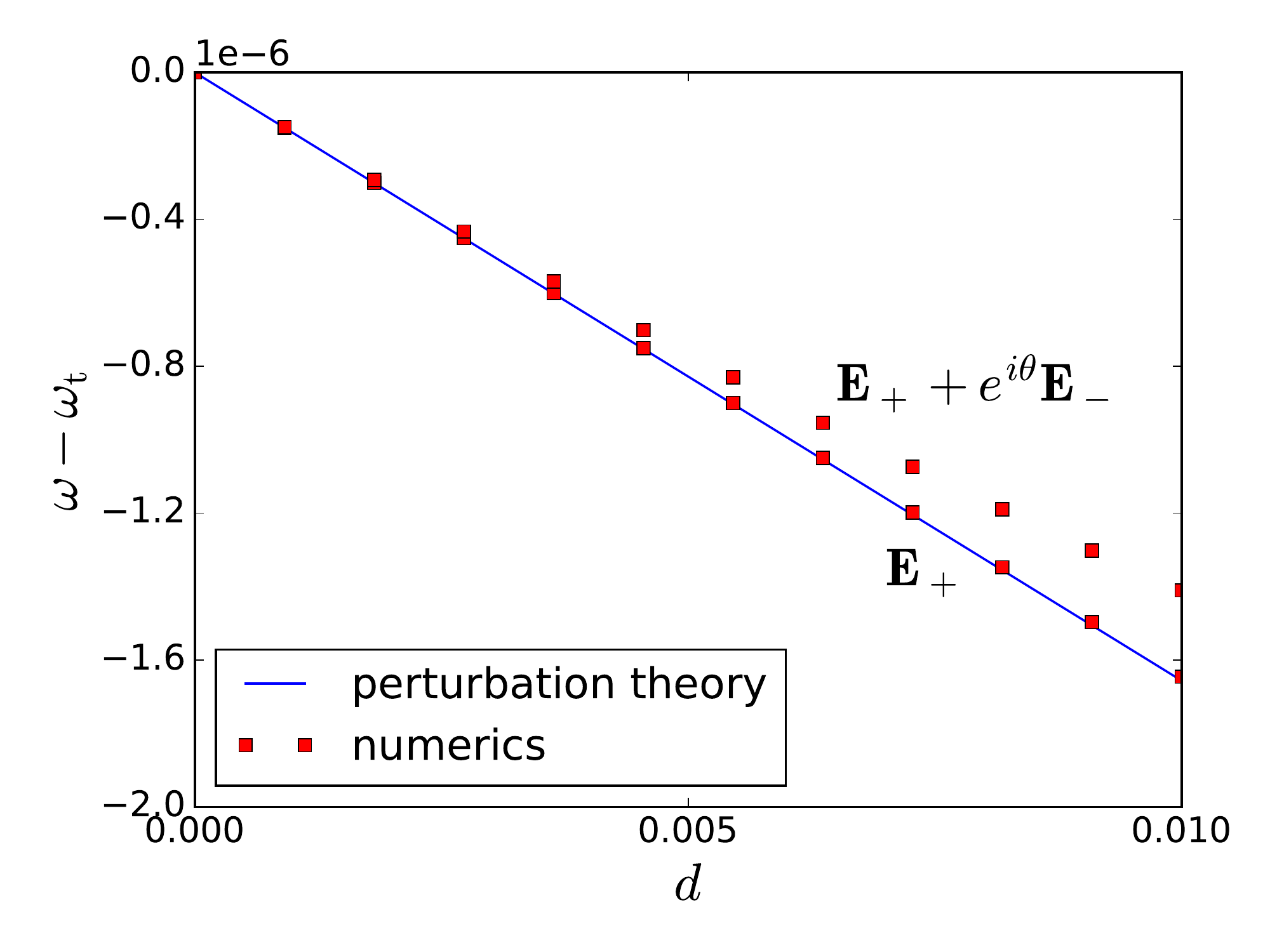}}\caption{(Color online) Lasing amplitudes (top) and frequency shifts (bottom)
for 1d laser (periodic geometry with $0<x<1$) with uniform dielectric
$\varepsilon(x)=1+0.3i$ and gain profile $D_{0}(x)=D_{\mathrm{t}}(1+d)[1+0.2\cos(2\pi nx)]$.
Here, we have chosen $n=5$, so the gain has $C_{5\mathrm{v}}$ symmetry,
and the discretization had $N=150$ grid points. Data points were
obtained by solving Eq.~(\ref{eq:single-mode-salt}) (SALT) numerically
using Newton's method \cite{direct_salt}, while theoretical lines
were provided by Eq.~(\ref{eq:circulating lasing}) and Eq.~(\ref{eq:n !=00003D 4 lasing}).
For the standing mode, numerical results are independent of relative
phase $\theta$, as predicted by Eq.~(\ref{eq:n !=00003D 4 lasing}).
Agreement between numerics and theory is excellent for mode amplitudes
(up to at least $d\approx100$ {[}not shown in the figure{]}, and
possibly much higher) but only good at $d<0.005$ for frequency shifts.
For the numerical data in the amplitude plot, only the magnitude of
the $\mathbf{E}_{+}$ component (obtained by taking the normalized
inner product of the lasing mode $\mathbf{E}$ with $\mathbf{E}_{+}$)
of both the circulating and standing-wave modes are shown. For the
$\mathbf{E}_{-}$ component, the circulating mode had magnitude zero
and the standing mode had the same magnitude as the $\mathbf{E}_{+}$
component. \label{fig:n=00003D5 numerical results}}
\end{figure}

We begin with the situation of a degenerate threshold, where two modes
$\mathbf{E}_{1}$ and $\mathbf{E}_{2}$ (such as any of the pairs
in Fig.~\ref{fig:overview}) hit threshold at the same pump strength
$D_{\mathrm{t}}$ and same frequency $\omega_{\mathrm{t}}$. Since
the frequencies are the same, we can consider any linear superposition
of the two modes as a single mode. With infinitesimal amplitude, Eq.~(\ref{eq:single-mode-salt})
is
\begin{equation}
\nabla\times\nabla\times\mathbf{E}_{1,2}=\omega_{\mathrm{t}}^{2}\left(\varepsilon+D_{\mathrm{t}}\Gamma_{\mathrm{t}}\right)\mathbf{E}_{1,2},\label{eq:threshold salt}
\end{equation}
where $\Gamma_{\mathrm{t}}=\Gamma(\omega_{\mathrm{t}})$. Now we perturb
the pump strength to bring the mode slightly above threshold, with
$D_{0}=D_{\mathrm{t}}(1+d)$ and $0<d\ll1$. We then expect the lasing
mode slightly above threshold to be of the form
\begin{align}
\mathbf{E} & =\Gamma_{\mathrm{t}}^{-1}\sqrt{\gamma_{\parallel}d}\left(a_{1}\mathbf{E}_{1}+a_{2}\mathbf{E}_{2}\right)+d^{3/2}\delta\mathbf{E},\nonumber \\
\omega & =\omega_{\mathrm{t}}+\omega_{1}d+O(d^{2})\label{eq:lasing_ansatz}
\end{align}
where the complex coefficients $a_{1,2}$ and the real eigenvalue
shift $\omega_{1}$ are to be determined. The linear relation between
$d$ and intensity $|\mathbf{E}|^{2}$ has previously been shown for
lasing modes above threshold in SPA-SALT approximation \cite{ge_steady-state_2010}.
Inserting Eq.~(\ref{eq:lasing_ansatz}) into Eq.~(\ref{eq:single-mode-salt}),
expanding to lowest order in $d$, and taking the inner product of
both sides with $\mathbf{E}_{1}$ and $\mathbf{E}_{2}$ (as performed
in detail in Appendix~\ref{sec: appendix allowed lasing modes}),
we obtain the pair of nonlinear equations for $a_{1,2}$ and $\omega_{1}$:
\begin{multline}
0=\int d^{3}x\,\mathbf{E}_{1,2}\cdot\left(a_{1}\mathbf{E}_{1}+a_{2}\mathbf{E}_{2}\right)\times\\
\left[\omega_{1}\frac{\partial}{\partial\omega_{\mathrm{t}}}\omega_{\mathrm{t}}^{2}(\varepsilon+D_{\mathrm{t}}\Gamma_{\mathrm{t}})+\omega_{\mathrm{t}}^{2}D_{\mathrm{t}}\Gamma_{\mathrm{t}}\left(1-\left|a_{1}\mathbf{E}_{1}+a_{2}\mathbf{E}_{2}\right|^{2}\right)\right]\label{eq:a eigenproblem}
\end{multline}
To proceed, we must choose a basis $\mathbf{E}_{1,2}$ to work with
(the end result, Eq.~(\ref{eq:lasing_ansatz}) turns out to be independent
of the choice, as expected). One possible choice is the even and odd
(with respect to the mirror planes of the $C_{n\mathrm{v}}$ geometry)
standing-wave modes, which we denote as $\mathbf{E}_{\mathrm{even}}$
and $\mathbf{E}_{\mathrm{odd}}$ (as in Fig.~\ref{fig:overview}).
However, it turns out that another choice makes the analytical solution
of Eq.~(\ref{eq:a eigenproblem}) significantly easier to obtain,
due to various convenient symmetry properties. In particular, we construct
a basis of \emph{clockwise }and \emph{counterclockwise }``circulating''
modes (analogous to $e^{\pm i\ell\phi}$ modes in a ring)
\begin{equation}
\mathbf{E}_{\pm}=\sum_{k=1}^{n}\exp\left(\pm\frac{2\pi i\ell k}{n}\right)R_{k/n}\mathbf{E}_{\mathrm{even}},\label{eq:chiral definition}
\end{equation}
where $\ell$ is given by the 2d irrep that $\mathbf{E}_{\mathrm{even}}$
belongs to {[}$\ell$ ranges from 1 to $\mathrm{floor}(\frac{n-1}{2})${]},
and $R_{k/n}$ is a counterclockwise rotation of the vector field
$\mathbf{E}_{\mathrm{even}}(\mathbf{x})$ in the plane of the $C_{n\mathrm{v}}$
symmetry by $2\pi k/n$ (if $\ell$ is chosen to be the wrong integer,
then $\mathbf{E}_{\pm}$ vanishes because Eq.~(\ref{eq:chiral definition})
is a projection operator \cite{inui_group,tinkham}). With this definition,
$\mathbf{E}_{+}$ and $\mathbf{E}_{-}$ are mirror flips of one another,
and they span the same space as $\mathbf{E}_{\mathrm{even}}$ and
$\mathbf{E}_{\mathrm{odd}}$. One important property of the $\mathbf{E}_{\pm}$
is that they transform according to the chiral $1d$ irreps of the
$C_{n}$ symmetry group, i.e. $R_{1/n}\mathbf{E}_{\pm}=\exp\left(\mp\frac{2\pi i\ell k}{n}\right)\mathbf{E}_{\pm}$.
This fact will turn out to greatly simplify some upcoming calculations.
Choosing $\mathbf{E}_{1}=\mathbf{E}_{+}$ and $\mathbf{E}_{2}=\mathbf{E}_{-}$
and exploiting the symmetry properties of this basis, we see (as shown
in detail in Appendix~\ref{sec: appendix allowed lasing modes})
that Eq.~(\ref{eq:a eigenproblem}) reduces to
\begin{multline}
0=a_{\mp}\left(\omega_{1}H+G_{D}\right)-\\
a_{\mp}\left|a_{\pm}\right|^{2}\left(I+J\right)-a_{\mp}\left|a_{\mp}\right|^{2}I-a_{\pm}^{2}a_{\mp}^{\star}K,\label{eq:reduced a plus minus}
\end{multline}
where the coefficients $G_{D}$, $H$, $I$, $J$, and $K$ are simple
overlap integrals of the threshold modes, given in closed form in
Eq.~(\ref{eq:G and H definitions}) and Eq.~(\ref{eq:I J K definitions}).
Equation.~(\ref{eq:reduced a plus minus}) can be solved in closed
form (as performed in Appendix~\ref{sec: appendix allowed lasing modes}),
and yields only a few solutions. First, there are the purely circulating
modes, given by
\begin{align}
\mathbf{E} & =\Gamma_{\mathrm{t}}^{-1}\sqrt{\frac{\left(\omega_{1}H+G_{D}\right)\gamma_{\parallel}d}{I}}\mathbf{E}_{\pm}\nonumber \\
\omega_{1} & =-\frac{\mathrm{Im}\left(G_{D}/I\right)}{\mathrm{Im}\left(H/I\right)}.\label{eq:circulating lasing}
\end{align}
The other solutions to Eq.~(\ref{eq:reduced a plus minus}) are \emph{standing-wave
}modes, as to be expected, and it turns out their form depends (as
explained in Appendix~\ref{sec: appendix allowed lasing modes})
crucially on whether the irrep of the degenerate pair satisfies $n=4\ell$,
where again $\ell$ is the order of the 2d irrep. For $C_{n\mathrm{v}}$
with $n\neq4\ell$, the other modes are
\begin{align}
\mathbf{E} & =\Gamma_{\mathrm{t}}^{-1}\sqrt{\frac{(\omega_{1}H+G_{D})\gamma_{\parallel}d}{2I+J}}\left(\mathbf{E}_{+}+e^{i\theta}\mathbf{E}_{-}\right)\nonumber \\
\omega_{1} & =-\frac{\mathrm{Im}\left(G_{D}/[2I+J]\right)}{\mathrm{Im}\left(H/[2I+J]\right)}\label{eq:n !=00003D 4 lasing}
\end{align}
where $\theta$ is an arbitrary phase angle. (During numerical solution
of Eq.~(\ref{eq:single-mode-salt}), we have found that for some
$n$, there seem to be constraints on $\theta$, namely having to
be a multiple of $\frac{\pi}{n}$. These constraints probably come
from equations that are higher-order in $d$ than our perturbation
theory. However, they are inconsequential because it turns out that
these standing-wave modes are always unstable, regardless of $\theta$.)
On the other hand, for the case of $n=4\ell$, there are \emph{two
}sets of standing-wave modes. The first is
\begin{align}
\mathbf{E} & =\Gamma_{\mathrm{t}}^{-1}\sqrt{\frac{(\omega_{1}H+G_{D})\gamma_{\parallel}d}{2I+J+K}}\left(\mathbf{E}_{+}\pm\mathbf{E}_{-}\right)\nonumber \\
\omega_{1} & =-\frac{\mathrm{Im}\left(G_{D}/[2I+J+K]\right)}{\mathrm{Im}\left(H/[2I+J+K]\right)}\label{eq:E+ + E- lasing}
\end{align}
and the other is
\begin{align}
\mathbf{E} & =\Gamma_{\mathrm{t}}^{-1}\sqrt{\frac{(\omega_{1}H+G_{D})\gamma_{\parallel}d}{2I+J-K}}\left(\mathbf{E}_{+}\pm i\mathbf{E}_{-}\right)\nonumber \\
\omega_{1} & =-\frac{\mathrm{Im}\left(G_{D}/[2I+J-K]\right)}{\mathrm{Im}\left(H/[2I+J-K]\right)}.\label{eq:E+ + i E- lasing}
\end{align}
These four sets of solutions turn out to constitute all the solutions
of Eq.~(\ref{eq:reduced a plus minus}) for the general $C_{n\mathrm{v}}$
case, and this completes the solution of Eq.~(\ref{eq:single-mode-salt})
slightly above threshold. Figure.~\ref{fig:n=00003D5 numerical results}
shows a comparison of numerical results for a 1d $C_{5\mathrm{v}}$
laser with the predictions of perturbation theory. 

In order to find out which of these solutions is the one that actually
lases in a real $C_{n\mathrm{v}}$ system, we must test stability
of each of these solutions. It turns out that this test can also be
done mostly analytically, using perturbation theory on the linearized
Maxwell--Bloch equations, as we will present in the next section.

\subsection{Perturbative stability analysis\label{sub:Stability-analysis summary}}

While these three forms of lasing modes all solve SALT near threshold,
and hence are ``fixed-point'' equilibria of the Maxwell--Bloch equations,
one intuitively expects that only the circulating mode will be stable.
The reason is that standing-wave modes for the $C_{n\mathrm{v}}$
group have zero amplitude along certain lines, most obviously along
$x=0$ for the $\mathbf{E}_{\mathrm{odd}}$ mode, and those zero-amplitude
regions are making no use of the gain. This allows the opposite-symmetry
standing-wave mode to grow exponentially into these nulls, and this
is the reason why the sine and cosine modes are unstable in a ring.
(This fact is only true for \emph{exact }degeneracies; for real systems,
which almost always break the degeneracy, there may be small regions
near threshold where the standing-wave mode is stable \cite{ref-sorel}.)
To quantify this intuition, we perform linear-stability analysis,
along the same lines as the numerical procedure in Ref.~\cite{ref-rotter_degeneracy}.
We linearize the Maxwell--Bloch equations for small perturbations
around the SALT modes, by inserting
\begin{align}
\mathbf{E}^{+}(\mathbf{x},t) & =\left[\mathbf{E}(\mathbf{x})+\delta\mathbf{E}(\mathbf{x},t)\right]e^{-i\omega t}\nonumber \\
\mathbf{P}^{+}(\mathbf{x},t) & =\left[\mathbf{P}(\mathbf{x})+\delta\mathbf{P}(\mathbf{x},t)\right]e^{-i\omega t}\label{eq:ansatz-1}\\
D(\mathbf{x},t) & =D(\mathbf{x})+\delta D(\mathbf{x},t)\nonumber 
\end{align}
into the Maxwell--Bloch equations (where $\mathbf{E}$ is any of the
SALT solutions obtained in Sec.~\ref{sub:Perturbative-SALT-solutions},
$D(\mathbf{x})$ is the stationary inversion given in (\ref{eq:single-mode-salt}),
$\omega$ is the lasing frequency given in Eq.~(\ref{eq:lasing_ansatz}),
and $\mathbf{P}(\mathbf{x})=\Gamma(\omega)\mathbf{E}(x)D$, is the
polarization field. We collect terms order-by-order in the perturbations
$\delta$. The zeroth-order equations are simply the SALT equations
and are already satisfied by construction by $\mathbf{E}$, $\mathbf{P}$,
and $D(\mathbf{x})$. The first-order equations are 
\begin{align}
0 & =-\nabla\times\nabla\times\delta\mathbf{E}+\left(\frac{d}{dt}-i\omega\right)^{2}(\varepsilon\delta\mathbf{E}+\delta\mathbf{P})\nonumber \\
i\delta\dot{\mathbf{P}} & =(\omega_{a}-\omega-i\gamma_{\perp})\delta\mathbf{P}+\gamma_{\perp}(D\delta\mathbf{E}+\mathbf{E}\delta D)\label{eq:delta}\\
\delta\dot{D} & =-\gamma_{\parallel}\delta D+\mathrm{Im}(\mathbf{P}\cdot\delta\mathbf{E}^{\star}+\mathbf{E}^{\star}\cdot\delta\mathbf{P}).\nonumber 
\end{align}
Eq.~(\ref{eq:delta}) can be written as a matrix equation
\begin{equation}
\left(\mathsf{C}\frac{d^{2}}{dt^{2}}+\mathsf{B}\frac{d}{dt}+\mathsf{A}\right)\mathsf{u}(t)=0\label{eq:time-dependent quadratic}
\end{equation}
or alternatively as quadratic eigenvalue problem \cite{ref-tisseur_quadratic_review}
\begin{equation}
\left(\mathsf{C}\sigma^{2}+\mathsf{B}\sigma+\mathsf{A}\right)\mathsf{x}=0\label{eq:quadratic eigenproblem}
\end{equation}
where the unknown vector is
\begin{equation}
\mathsf{u}(t)=\left(\begin{array}{c}
\Re\,\delta\vec{E}\\
\Im\,\delta\vec{E}\\
\Re\,\delta\vec{P}\\
\Im\,\delta\vec{P}\\
\delta\vec{D}
\end{array}\right)=\Re\left(\mathsf{x}e^{\sigma t}\right).
\end{equation}
The goal of the stability analysis is to find the eigenvalues $\sigma$
for a given lasing mode $\mathbf{E}$. If the real part of any of
the eigenvalues $\sigma$ is positive, then the lasing mode is unstable,
while if all the eigenvalues have non-positive real parts, then the
lasing mode is stable (with some technical care required for zero
eigenvalues and structural stability, described in Appendix~\ref{sec: appendix stability calculations}).
Ref. \cite{ref-rotter_degeneracy} discretized Eq.~(\ref{eq:delta})
to obtain Eq.~(\ref{eq:quadratic eigenproblem}) and then solved
the resulting matrix equation numerically to find the eigenvalues
$\sigma$ and hence evaluated the SALT stability for any pump strength
above threshold. Here, we focus on the regime slightly above threshold,
and show that the equations can be solved analytically to lowest order
in $d$, and this is enough to evaluate near-threshold stability.
We begin by noting that the matrices can be expanded as
\begin{align}
\mathsf{A} & =\mathsf{A}_{0}+\mathsf{A}_{1/2}\sqrt{d}+\mathsf{A}_{1}d+O(d^{3/2})\nonumber \\
\mathsf{B} & =\mathsf{B}_{0}+\mathsf{B}_{1}d+O(d^{2})\label{eq:matrix expansion}\\
\mathsf{C} & =\mathsf{C}_{0}\nonumber 
\end{align}
(where $d$ is the relative pump increment above threshold, as introduced
in Sec.~\ref{sec:Threshold-perturbation-theory}), since the matrices
come from the coefficients of Eq.~(\ref{eq:delta}) and contain the
lasing solutions $\mathbf{E}$ and other associated fields $\mathbf{P}$
and $D$. As a result, the eigenvalues and eigenvectors can also be
expanded this way:
\begin{align}
\mathsf{x} & =\mathsf{x}_{0}+\mathsf{x}_{1/2}\sqrt{d}+\mathsf{x}_{1}d+O(d^{3/2})\nonumber \\
\sigma & =\sigma_{0}+\sigma_{1/2}\sqrt{d}+\sigma_{1}d+O(d^{3/2}).\label{eq:eigenpair expansion}
\end{align}
We insert Eqs.~(\ref{eq:matrix expansion}) and (\ref{eq:eigenpair expansion})
into Eq.~(\ref{eq:quadratic eigenproblem}) and solve order-by-order
in $\sqrt{d}$ until a non-zero $\sigma$ is found, as explained in
detail in Appendix~\ref{sec: appendix stability calculations}. At
zeroth order, Eq.~(\ref{eq:quadratic eigenproblem}) is equivalent
to the SALT equation at threshold (Eq.~(\ref{eq:threshold salt})),
and has two kinds of solutions. First, there are the below threshold
``passive'' modes \cite{tureci_self-consistent_2006,ge_steady-state_2010},
which have $\Re(\sigma_{0})<0$ (because for these modes, $\sigma_{0}$
is simply the difference $\tilde{\omega}-\omega_{\mathrm{t}}$ between
the complex pole $\tilde{\omega}$ of the passive mode and the real
threshold frequency $\omega_{\mathrm{t}}$) and hence are stable.
Therefore, when $d$ is small enough, we can say for certain that
having one of these $\sigma_{0}$ above the real axis would make the
system unstable, so having all passive poles of the SALT equation
be below the real axis is a \emph{necessary }condition for lasing
near threshold, for small $d$. (Far above threshold, however, this
is no longer true, as shown in Ref.~\cite{ref-rotter_degeneracy}.)
Second, from the degenerate \emph{threshold} modes, we obtain
\begin{align}
\sigma_{0} & =0\nonumber \\
\mathsf{x}_{0} & =\sum_{k=1}^{4}b_{k}\mathsf{v}_{k},\label{eq:zeroth order}
\end{align}
where $b_{k}$ are arbitrary complex coefficients that will be determined
later at higher order (similarly to linear degenerate perturbation
theory in quantum mechanics \cite{sakurai}), and the vectors $\mathsf{v}_{k}$
are
\begin{equation}
\mathsf{v}_{k}=\left(\begin{array}{c}
\Re\,\vec{e}_{k}\\
\Im\,\vec{e}_{k}\\
D_{\mathrm{t}}\Re\,(\Gamma_{\mathrm{t}}\vec{e}_{k})\\
D_{\mathrm{t}}\Im\,(\Gamma_{\mathrm{t}}\vec{e}_{k})\\
0
\end{array}\right),\label{eq:vk definition}
\end{equation}
where we have defined $\mathbf{e}_{1,2,3,4}=\mathbf{E}_{1},\mathbf{E}_{2},i\mathbf{E}_{1},i\mathbf{E}_{2}$
(again, $\mathbf{E}_{1,2}$ are any two threshold solutions to Eq.~(\ref{eq:threshold salt})).
It is shown in Appendix~\ref{sec: appendix stability calculations}
that the eigenvalue at the next order, $\sigma_{1/2}$ is also zero.
Hence, stability is determined by $\sigma_{1}$. At order $d$, Eq.~(\ref{eq:quadratic eigenproblem})
is 
\begin{equation}
(\mathsf{B}_{0}\sigma_{1}+\mathsf{A}_{1})\mathsf{x}_{0}+\mathsf{A}_{1/2}\mathsf{x}_{1/2}+\mathsf{A}_{0}\mathsf{x}_{1}=0.\label{eq:first order equation}
\end{equation}
Here, all quantities except $\sigma_{1}$ and $\mathsf{x}_{1}$ are
known. Because $\sigma_{0}=0$, we have $\mathsf{A}_{0}\mathsf{v}_{k}=0$.
There are also left eigenvectors \cite{strang_linear_algebra} $\mathsf{w}_{j}$
that satisfy $\mathsf{A}_{0}^{T}\mathsf{w}_{j}=0$. By acting on Eq.~(\ref{eq:first order equation})
with these left eigenvectors, we obtain a $4\times4$ \emph{linear
}eigenvalue problem for the eigenvalue $\sigma_{1}$ and the eigenvectors
(whose elements are the coefficients $b_{k}$ in Eq.~(\ref{eq:zeroth order})).
We can then, in a straightforward fashion, write down the eigenvalues
and eigenvectors in closed form. While the procedure we have just
described can be done with any basis $\mathbf{E}_{1,2}$, again, it
is most convenient to choose the basis $\mathbf{E}_{\pm}$, due to
the symmetry properties which greatly simplify the calculation. Here,
we present the results, leaving the detailed derivation to Appendix~\ref{sec: appendix stability calculations}.

For the circulating lasing modes in Eq.~(\ref{eq:circulating lasing}),
the four eigenvalues are, in no particular order,
\begin{align}
\sigma_{1} & =0\nonumber \\
\sigma_{1} & =2\Im\left(\frac{I}{H}\right)\left|a\right|^{2}\nonumber \\
\sigma_{1} & =\left(\Im\left(\frac{J}{H}\right)+\sqrt{\left|\frac{K}{H}\right|^{2}-\Re\left(\frac{J}{H}\right)^{2}}\right)\left|a\right|^{2}\label{eq:circulating eigenvalues}\\
\sigma_{1} & =\left(\Im\left(\frac{J}{H}\right)-\sqrt{\left|\frac{K}{H}\right|^{2}-\Re\left(\frac{J}{H}\right)^{2}}\right)\left|a\right|^{2},\nonumber 
\end{align}
where $|a|^{2}\equiv\frac{\omega_{1}H+G_{D}}{I}$. The first eigenvalue
comes from the global phase degree of freedom for lasing solutions
\cite{ref-rotter_degeneracy}. For the other three eigenvalues, we
have found empirically that the real part is always negative, indicating
that the circulating modes are stable. Although we have been unable
to prove that $\Re(\sigma_{1})<0$ in general for the last three values
in Eq.~(\ref{eq:circulating eigenvalues}), we have empirically observed
this to be true, and it is easily checked in any specific case by
integrating the threshold modes to compute $H$, $I$, $J$, and $K$.

\begin{figure}
\centerline{\includegraphics[scale=0.42]{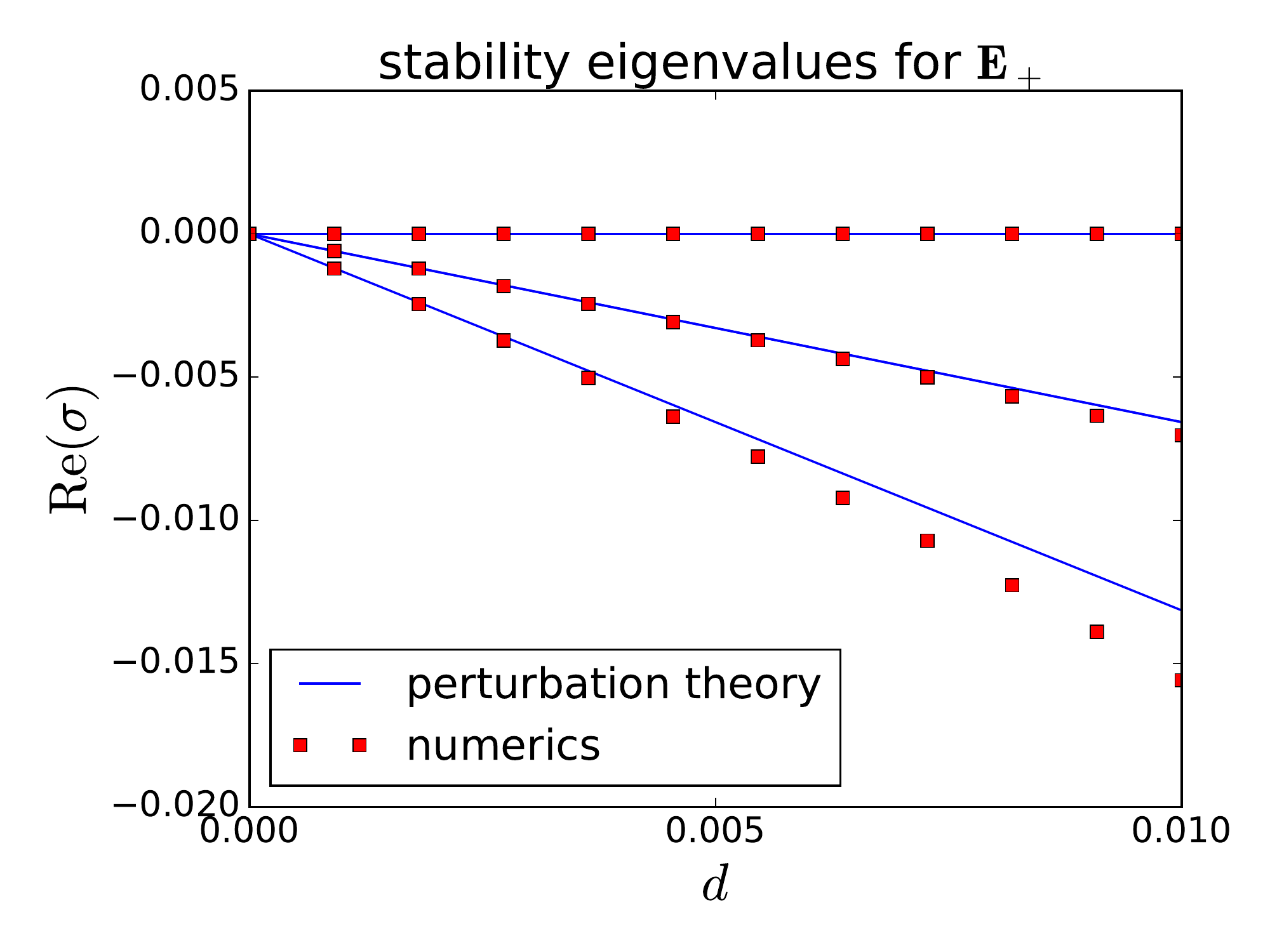}}

\centerline{\includegraphics[scale=0.42]{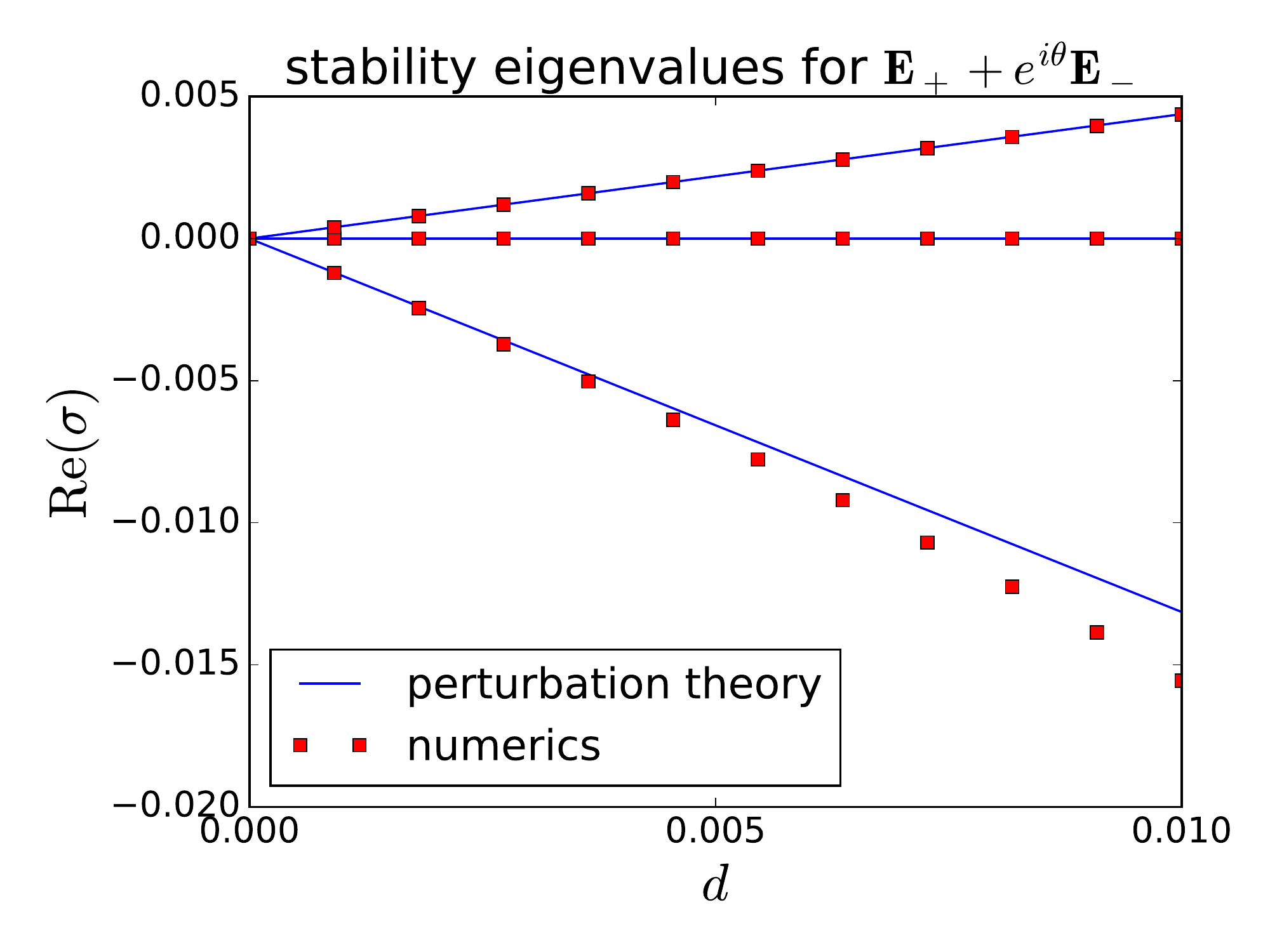}}\caption{(Color online) Stability eigenvalues for the circulating (top) and
standing (bottom) lasing modes from Fig.~\ref{fig:n=00003D5 numerical results}.
For each lasing mode, the four lowest eigenvalues obtained using the
numerical procedure of Ref.~\cite{ref-rotter_degeneracy} are matched
against the four eigenvalues found in perturbation theory. In the
top panel, a zero eigenvalue is clearly seen, coming from the global
phase freedom. The real parts of the third and fourth values of Eq.~(\ref{eq:circulating eigenvalues})
are equal, so two of the curves coincide. None of the other eigenvalues
go above the real axis, indicating that the circulating mode is stable.
For the bottom panel, there are two zero eigenvalues, in agreement
with Eq.~(\ref{eq:n !=00003D 4 stability eigenvalues}). One of the
other two eigenvalues becomes positive, indicating that the standing
lasing mode is not stable. \label{fig:n=00003D5 stability results}}
\end{figure}

For the $n\neq4\ell$ standing lasing modes in Eq.~(\ref{eq:n !=00003D 4 lasing}),
the eigenvalues are given by
\begin{align}
\sigma_{1} & =0\nonumber \\
\sigma_{1} & =0\nonumber \\
\sigma_{1} & =2\Im\left(\frac{2I+J}{H}\right)\left|a\right|^{2}\label{eq:n !=00003D 4 stability eigenvalues}\\
\sigma_{1} & =-2\Im\left(\frac{J}{H}\right)\left|a\right|^{2},\nonumber 
\end{align}
where $|a|^{2}\equiv\frac{\omega_{1}H+G_{D}}{2I+J}$. Here, both zero
eigenvalues come from continuous degrees of freedom: one comes from
the global phase degree of freedom, while the other comes from the
relative phase between $\mathbf{E}_{+}$ and $\mathbf{E}_{-}$ in
Eq.~(\ref{eq:n !=00003D 4 lasing}), which can take any value (as
explained previously, this degree of freedom is likely removed at
higher orders in $d$, so that only certain linear combinations, namely
the $n$-fold rotations of $\mathbf{E}_{\mathrm{even}}$ and $\mathbf{E}_{\mathrm{odd}}$,
are actually lasing solutions). For TM modes in 2d, which have $\mathbf{E}=E\hat{\mathbf{z}}$
\cite{phc_book}, it can be shown that $I=J$, so at least one of
the two non-zero eigenvalues here must have a positive real part (in
practice, it is always the last eigenvalue), indicating that these
standing lasing modes are always unstable. Figure~\ref{fig:n=00003D5 stability results}
shows a comparison between the theoretical first-order approximation
and the exact numerical values of the circulating and standing-wave
stability eigenvalues for a $C_{n\mathrm{v}}$ case with $n=5$.

For the $n=4\ell$ standing-wave modes, we first have the $\mathbf{E}_{+}\pm\mathbf{E}_{-}$
solutions in Eq.~(\ref{eq:E+ + E- lasing}), which have the eigenvalues
\begin{align}
\sigma_{1} & =0\nonumber \\
\sigma_{1} & =2\Im\left(\frac{2I+J+K}{H}\right)|a|^{2}\nonumber \\
\sigma_{1} & =\left[-\Im\left(\frac{J+3K}{H}\right)+\rho\right]|a|^{2}\label{eq:stability eigenvalues for E+ + E-}\\
\sigma_{1} & =\left[-\Im\left(\frac{J+3K}{H}\right)-\rho\right]|a|^{2}\nonumber \\
\rho & \equiv\sqrt{\Im\left(\frac{J-K}{H}\right)^{2}-8\Re\left(\frac{K}{H}\right)\Re\left(\frac{J+K}{H}\right)},\nonumber 
\end{align}
where $|a|^{2}\equiv\frac{\omega_{1}H+G_{D}}{2I+J+K}$. Again, there
is a zero eigenvalue coming from the global phase degree of freedom.
For almost all cases, we have empirically observed that the second
eigenvalue has a negative real part, but also $-\Im\left(\frac{J+3K}{H}\right)>0$,
and hence the third and fourth eigenvalues are unstable. However,
there are pathological cases where the gain profile $D_{0}(\mathbf{x})$
can be chosen (e.g., in terms of $4\ell$ delta functions) so that
$J=-K$, upon which the third eigenvalue is stable, and the last eigenvalue
is zero. This zero eigenvalue turns out to become positive (unstable)
for physical, finite-sized gain regions, as discussed further in Appendix.~\ref{sec: appendix stability calculations}.

Finally, we have the $\mathbf{E}_{+}\pm i\mathbf{E}_{-}$ solutions
for the $n=4\ell$ case in Eq.~(\ref{eq:E+ + i E- lasing}). The
eigenvalues are
\begin{align}
\sigma_{1} & =0\nonumber \\
\sigma_{1} & =2\Im\left(\frac{2I+J-K}{H}\right)|a|^{2}\nonumber \\
\sigma_{1} & =\left[\Im\left(\frac{3K-J}{H}\right)+\eta\right]|a|^{2}\label{eq:stability eigenvalues for E+ + iE-}\\
\sigma_{1} & =\left[\Im\left(\frac{3K-J}{H}\right)-\eta\right]|a|^{2}\nonumber \\
\eta & \equiv\sqrt{\Im\left(\frac{J+K}{H}\right)^{2}+8\Re\left(\frac{K}{H}\right)\Re\left(\frac{J-K}{H}\right)},\nonumber 
\end{align}
where $|a|^{2}\equiv\frac{\omega_{1}H+G_{D}}{2I+J-K}$. Again, in
most cases we have empirically found that the second eigenvalue is
stable while the third and fourth are unstable. However, there are
pathological delta-function cases where the external pump profile
$D_{0}(\mathbf{x})$ can be chosen in a specific way (different from
that for the $\mathbf{E}_{+}\pm\mathbf{E}_{-}$ solutions) such that
$J=K$, and the third eigenvalue is stable while the fourth is zero.
Again, this zero eigenvalue becomes positive (unstable) for finite-size
gain regions, and so this marginal case is unlikely to be of practical
importance.

We note that the absence of a positive value for $\sigma_{1}$ for
the circulating lasing mode does not guarantee that the true eigenvalue
$\sigma$ (Eq.~(\ref{eq:eigenpair expansion})) will remain below
the real axis for all $d$. Indeed, Ref.~\cite{ref-rotter_degeneracy}
(Fig.~1 in the reference) found an instance of a 1d uniform ring
laser where, for certain regimes, the circulating mode actually becomes
unstable above a certain $d_{\mathrm{cutoff}}$. The instability comes
from one of the four Maxwell--Bloch stability eigenvalues associated
with the degenerate threshold pair (whose first-order coefficients
are given in Eq.~(\ref{eq:circulating eigenvalues})) going above
the real axis. However, the onset of the instability depends on the
value of $\gamma_{\parallel}$, the relaxation rate of the inversion,
and the eigenvalues in Eq.~(\ref{eq:circulating eigenvalues}) are
independent of that parameter, so the effect must come from higher
orders. Closer inspection of the data in the ring laser of Ref.~\cite{ref-rotter_degeneracy}
shows that for very small $\gamma_{\parallel}$, the cutoff pump strength
$d_{\mathrm{cutoff}}$ is linear with $\gamma_{\parallel}$; that
is, the circulating mode is stable for $d<z_{0}\gamma_{\parallel}$,
where $z_{0}$ is a constant independent of $d$ and $\gamma_{\parallel}$.
In Appendix~\ref{sec: appendix stability calculations}, we rigorously
explain this criterion by extending the perturbation theory used to
obtain $\sigma_{1}$ and finding the $\gamma_{\parallel}$ dependence
to all orders in $d$ (Eq.~(\ref{eq:s expansion})).

\subsection{Threshold perturbation examples\label{sec:threshold_perturbation_Examples}}

\begin{figure}
\centerline{\includegraphics[scale=0.25]{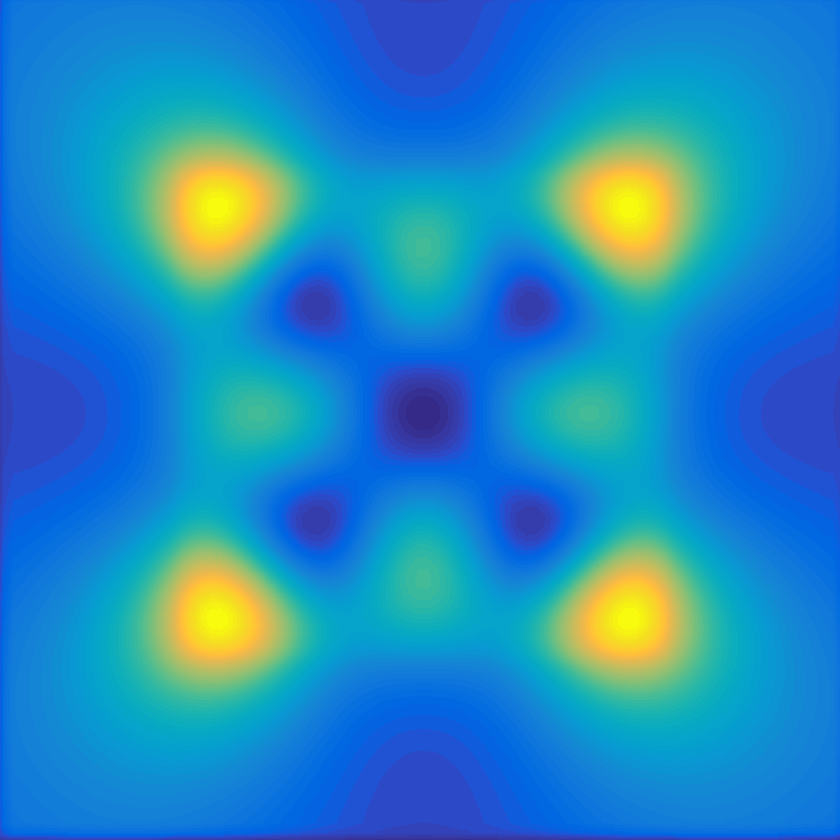}$\;$\includegraphics[scale=0.25]{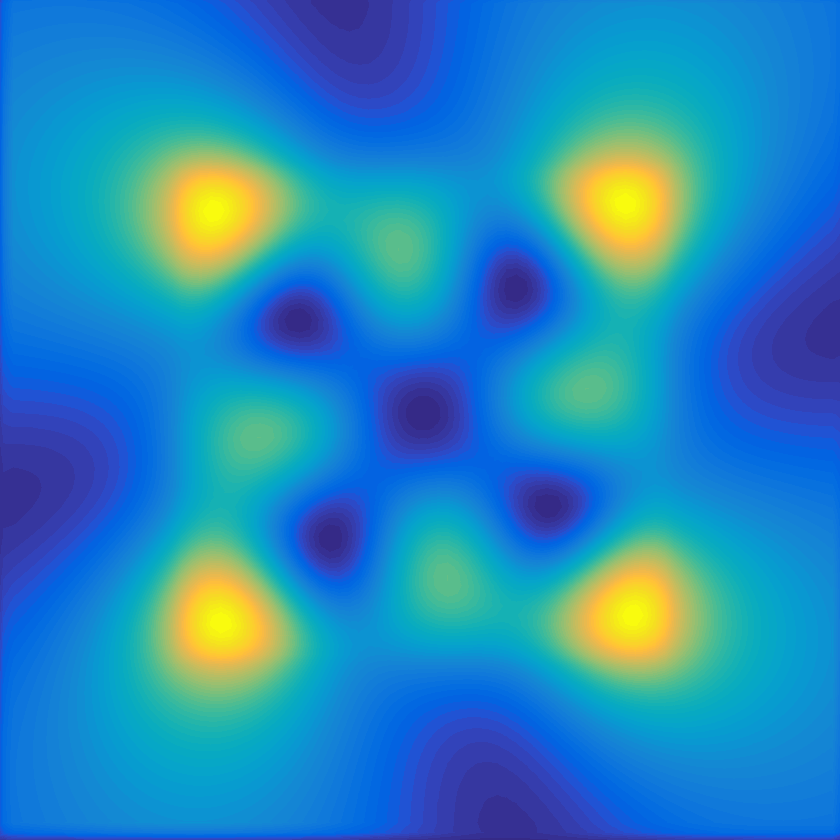}$\;$\includegraphics[scale=0.25]{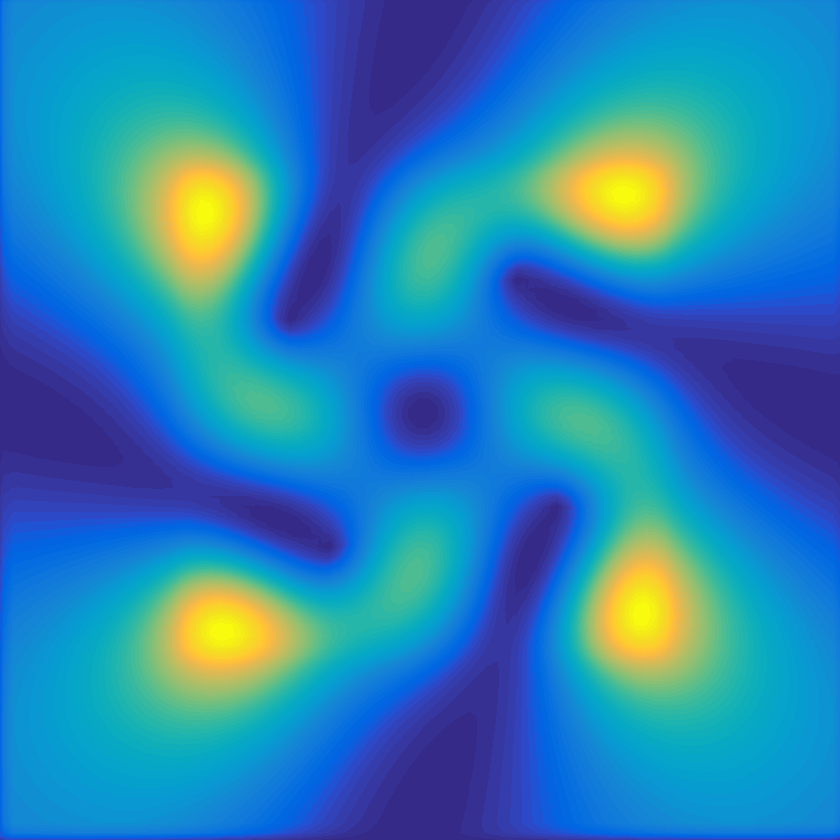}}\caption{(Color online) The incorrect (left) intensity pattern, correct (center)
intensity pattern for the pair of low-$Q$ dielectric square modes,
slightly above threshold (the pattern remains essentially the same
even for much higher pump strengths), and degenerate passive mode
of opposite chirality (right) of low-$Q$ dielectric square for lasing
very high above threshold ($D_{0}=100\Dt)$. The left intensity pattern
was obtained by solving two-mode SALT without any interference effects,
while the center pattern was obtained by constructing the stable linear
combination predicted by symmetry and perturbation theory and then
solving single-mode SALT. The correct pattern clearly has a chirality,
which the incorrect pattern lacks. The profile on the right is not
simply a mirror flip of the lasing mode, but is in fact degenerate
with it, as explained in Appendix~\ref{sec: appendix allowed lasing modes}.
\label{fig:lowQ_chirality}}
\end{figure}

Now, we illustrate the ideas of threshold perturbation theory with
an example of a symmetric geometry with degeneracies: a dielectric
square. Unlike in a metal square (with Dirichlet boundary conditions),
the equation for the electric field is not separable in the $x$ and
$y$ directions. The modes $\mathbf{E}_{1}$ and $\mathbf{E}_{2}$
are shown in the middle panel of Fig.~\ref{fig:overview}. The stable
linear combination is the circulating mode predicted by Sec.~\ref{sub:Stability-analysis summary}.
As a consequence of including interference between the two standing-wave
modes $\mathbf{E}_{\mathrm{even}}$ and $\mathbf{E}_{\mathrm{odd}}$,
this intensity pattern $\left|\mathbf{E}_{\mathrm{even}}\pm i\mathbf{E}_{\mathrm{odd}}\right|$
is chiral (with $C_{4}$ symmetry) while a naive summation of the
individual intensities $|\mathbf{E}_{\mathrm{even}}|^{2}+|\mathbf{E}_{\mathrm{odd}}|^{2}$
would still yield a $C_{4\mathrm{v}}$ pattern, as shown in Fig.~\ref{fig:lowQ_chirality}.
Because $C_{n}$ symmetry groups have no 2d irreps, one would normally
not expect there to be a degeneracy. However, a key point is that
the degeneracy indeed persists even when $C_{n\mathrm{v}}$ symmetry
becomes $C_{n}$, as a consequence of electromagnetic reciprocity~\cite{dichroism},
as explained in Appendix~\ref{sec: appendix allowed lasing modes}.
Since this degeneracy does not come from geometric symmetry alone,
there is no simple symmetry operation that takes the lasing mode (center
panel of Fig.~\ref{fig:lowQ_chirality}) to its degenerate partner
(right panel), e.g., they are not mirror flips. However, there is
still an exact degeneracy.

\section{Effects of chirality \label{sec:Effects-of-chirality}}

In this section, we discuss the effects of chirality. So far, we have
worked with $C_{n\mathrm{v}}$ geometries, which are symmetric under
$n$-fold rotations and flip operations across the mirror planes \cite{inui_group,tinkham}.
However, if the mirror symmetry is broken, then we no longer have
$C_{n\mathrm{v}}$ symmetry: while there is still symmetry under $n$-fold
rotations, the geometry aquires a certain ``handedness'', as in
the last two panels of Fig.~\ref{fig:lowQ_chirality}. These symmetry
groups are known as $C_{n}$, and have different consquences for lasing
modes arising from these geometries, as we discuss below. 

There are two ways for mirror symmetry to be broken: first, for a
$C_{n\mathrm{v}}$-symmetric geometry {[}dielectric function $\varepsilon(\mathbf{x})$
and gain profile $D_{0}(\mathbf{x})${]}, the intensity patterns of
the circulating modes (given in Eq.~(\ref{eq:chiral definition}),
and which we have observed are always the only stable modes) turn
out to have $C_{n}$ symmetry as opposed to $C_{n\mathrm{v}}$ symmetry,
leading to spontaneous breaking of mirror symmetry as soon as the
mode starts lasing (the only exception to this rule is for a $C_{\infty\mathrm{v}}$
system, in which the circulating lasing modes have an intensity pattern
$|\mathbf{E}|^{2}\propto|e^{i\ell\phi}|^{2}=1$, and hence still have
mirror symmetry). We have also observed that circulating lasing modes
in \emph{lossy} $C_{n\mathrm{v}}$-symmetric cavities (such as that
in Fig.~\ref{fig:lowQ_chirality}), those with low quality factor
\cite{phc_book} $Q\equiv-\Re\,\omega^{\prime}/\Im\,\omega^{\prime}$,
where $\omega^{\prime}$ is the passive {[}zero pump{]} pole in the
Green's function, tend to have greater chirality than circulating
modes in cavities with \emph{high }$Q$ (such as that in Fig.~\ref{fig:phc_stable}).
Second, the geometry itself can already have $C_{n}$ symmetry, e.g.,
the dielectric and gain functions themselves have chirality. Whether
the chirality is due to the intensity pattern of a lasing mode or
due to the geometry itself, the effects are similar.

First, the presence of chirality affects the nature of the degeneracy
between the lasing mode and its passive pole, e.g., the solution to
Eq.~(\ref{eq:passive pole salt}). For a laser with $C_{n\mathrm{v}}$
symmetry at threshold, the two chiral circulating modes, Eq.~(\ref{eq:chiral definition})
are exactly related to each other by a mirror-flip operation. As soon
as $\mathbf{E}_{+}$ starts lasing, the mirror symmetry is broken
(for $n\neq\infty$) and the passive pole, which we denote as $\tilde{\mathbf{E}}_{-}$,
will move further and further away from being the mirror flip of $\mathbf{E}_{+}$
(as seen in the right panel of Fig.~\ref{fig:lowQ_chirality}). It
is important to note that if $\mathbf{E}_{-}$ were to lase instead
of $\mathbf{E}_{+}$, then the \emph{lasing }mode $\mathbf{E}_{-}$
will be an exact mirror flip of $\mathbf{E_{+}}$, while the eigenfunction
of its passive pole $\tilde{\mathbf{E}}_{+}$ will be an exact mirror
flip of $\tilde{\mathbf{E}}_{-}$, due to the $C_{n\mathrm{v}}$ symmetry
at threshold. The lasing frequency $\omega$ will also be independent
of whether $\mathbf{E}_{+}$ or $\mathbf{E}_{-}$ lases, as confirmed
by Eq.~(\ref{eq:circulating lasing}). On the other hand, if the
laser already had $C_{n}$ symmetry at threshold (either due to chirality
in a previously lasing mode or in the dielectric or gain functions),
then the threshold eigenfunctions $\mathbf{E}_{\pm}$ are no longer
mirror flips of each other, even though their threshold frequencies
are both the same $\omega_{\mathrm{t}}$ (as an interesting consequence
of Lorentz reciprocity of Maxwell's equations, as reviewed in Appendix.~\ref{sec: degeneracy in Cn}).
The fact that $\mathbf{E}_{+}$ and $\mathbf{E}_{-}$ are not mirror
flips of each other causes a splitting between the overlap integrals
$I_{+}$ and $I_{-}$, as well as in $J_{\pm}$ and $H_{\pm}$. As
a consequence, the expressions for the amplitude $|a|$ and the frequency
shift $\omega_{1}$ in Eq.~(\ref{eq:circulating lasing}) will have
$I_{+}$ if $\mathbf{E}_{+}$ lases, and $I_{-}$ if $\mathbf{E}_{-}$
lases. Hence, if the symmetry is only $C_{n}$ at threshold, the clockwise
and counterclockwise lasing modes would also have different amplitudes
and frequencies, in addition to not being related to each other by
a mirror flip operation. These facts allow us to imagine a situation
in which there is a ``binary tree'' of allowed possibilities, e.g.,
the first degenerate pair lases in clockwise mode, the second in counterclockwise,
and so on, and each branch of the tree has distinct lasing amplitudes
and frequencies. 

Second, the presence of chirality affects the perturbation theory
results for standing-mode lasing solutions in Sec.~\ref{sec:Threshold-perturbation-theory}.
When the threshold symmetry is $C_{n\mathrm{v}}$, there exist standing-mode
solutions of the form $\mathbf{E}\propto\mathbf{E}_{+}+e^{i\theta}\mathbf{E}_{-}$
(Eq.~(\ref{eq:n !=00003D 4 lasing})) when $n\neq4\ell$, and standing-wave
modes of the form $\mathbf{E}\propto\mathbf{E}_{+}\pm\mathbf{E}_{-}$
and $\mathbf{E}_{+}\pm i\mathbf{E}_{-}$ when $n=4\ell$. However,
when the threshold symmetry is $C_{n}$, there are no longer any standing-mode
lasing solutions for the $n\neq4\ell$ case. In the $n=4\ell$ case,
however, we have found empirically that the standing-mode solutions
$\mathbf{E}_{+}\pm\mathbf{E}_{-}$ and $\mathbf{E}_{+}\pm i\mathbf{E}_{-}$
still exist (provided that the correct normalization and overall phase
of $\mathbf{E}_{-}$ is chosen appropriately). However, because of
the splitting in the values of the overlap integrals $I_{\pm}$, $J_{\pm}$,
and $K_{\pm}$ for the $C_{n}$ case, the stability eigenvalues (Sec.~\ref{sub:Stability-analysis summary})
for these standing-mode solutions will no longer be given by the simple
expressions in Eq.~(\ref{eq:stability eigenvalues for E+ + E-})
and Eq.~(\ref{eq:stability eigenvalues for E+ + iE-}), and will
have to be numerically computed (nevertheless, we have empirically
found that these standing-mode solutions are still unstable).

\subsection{Multimode lasing}

So far, the discussion and examples in this paper have dealt with
the case in which only one pair of degenerate modes are lasing. The
generalization to the case of multimode lasing (i.e. multiple nondegenerate
and degenerate lasing modes all lasing simultaneously) is straightforward.
Since our method combines degenerate pairs into a single mode that
is the stable linear combination (as given by the perturbation theory
in Sec.~\ref{sec:Threshold-perturbation-theory}, the multimode treatment
is exactly the same as for SALT without degeneracies: the degenerate
pairs are always treated as a single mode. As in previous work on
SALT \cite{ge_steady-state_2010,tureci_self-consistent_2006,direct_salt},
all lasing modes are solved simultaneously at first, and there the
collective effect of their spatial hole-burning is used to track the
passive modes and add any mode (degenerate or nondegenerate) that
crosses threshold to the list of lasing modes. As in the case of non-degenerate
SALT, the spacing between modes with different frequencies must remain
much larger than $\gamma_{\parallel}$ in order for the stationary
inversion approximation to remain valid (usually, lasing frequencies
do not appreciably deviate from their threshold values, so this condition
is often safely satisfied). The only aspects of our method requiring
generalization are the threshold perturbation theory of Sec.~\ref{sec:Threshold-perturbation-theory}
and the \emph{quadratic program} \cite{boyd} (QP) method of Sec.~\ref{sub:Forcing-the-degeneracy}.
For both aspects, we describe small tweaks to the methods presented
in those sections that make them valid for the case of multimode lasing.

A general situation in which multimode-lasing is occuring can be described
by Eq.~(\ref{eq:salt general}) \cite{ge_steady-state_2010,tureci_self-consistent_2006},
where there are $M$ lasing modes ($\mu=1,2,\ldots,M$). If we start
with $C_{n\mathrm{v}}$ symmetry and have lasing modes that are either
circulating modes (as in Sec.~\ref{sub:Stability-analysis summary})
or non-degenerate modes (partners of \emph{real} 1d irreps without
a corresponding complex-conjugate irrep of the opposite chirality),
then each of the $\left|\mathbf{E}_{\nu}\right|^{2}$ terms has at
least $C_{n}$ symmetry, so the full stationary inversion $D(\mathbf{x})$,
which includes the effects of spatial hole-burning, has $C_{n}$ symmetry.
Suppose that the pump strength is at the threshold of mode $M$ so
that this mode has just started lasing, and that only modes $1$ through
$M-1$ contribute to the spatial hole-burning. Then all results in
Sec.~\ref{sec:Threshold-perturbation-theory} still hold, except
with gain profile $D_{0}(\mathbf{x})$ replaced by $D(\mathbf{x})$.
While the gain profile is now $C_{n}$ symmetric rather than $C_{n\mathrm{v}}$,
the presence of chiral degenerate pairs (which requires only $C_{n}$
symmetry and Lorentz reciprocity, as explained in Appendix~\ref{sec: appendix allowed lasing modes})
still remains, as explained in Sec.~\ref{sub:Stability-analysis summary}
(the arguments in that section do not assume $C_{n\mathrm{v}}$ symmetry,
so they still hold even if $D_{0}(\mathbf{x})$ is replaced by a function
with only $C_{n}$ symmetry.

\section{$C_{n\mathrm{v}}$ Symmetry broken by discretization\label{sec:discretization_breaking}}

In many cases, the $C_{n\mathrm{v}}$-symmetric geometry we are trying
to solve has a degeneracy that is broken when the geometry is approximated
by a discretized grid for numerical solution on a computer \cite{champagne_fdfd,taflove,direct_salt},
since the grid may no longer have the original $C_{n\mathrm{v}}$
symmetry. For linear equations, this unphysical splitting is not an
issue because it is usually straightforward to tell whether a pair
of modes is ``really'' degenerate by how it corresponds to the eigenfunctions
of the ``real'' symmetry group, and since all linear superpositions
solve the equation in the infinite-resolution limit, we can construct
arbitrary superpositions as needed \emph{after }solving for both of
the modes. However, for SALT (which is nonlinear), the coefficients
of the superposition are physical quantities that must be found by
our solution method, as explained in Sec.~\ref{sub:Stability-analysis summary}.
As explained in Ref. \cite{direct_salt}, the process for solving
for lasing modes begins with the linear problem for the passive poles.
Because both the real and imaginary parts of the passive poles are
split by the discretization error, the modes will lase at different
pump strengths, and even after both modes lase we cannot construct
a linear combination of them because the two modes satisfy equations
with different real eigenfrequencies. 

When the pump strength is sufficiently high above threshold, however,
the two near-degenerate modes can interact with each other via the
nonlinear spatial hole burning interaction to form a single stable
laser mode. This effect is commonly known as cooperative frequency
locking \cite{ref-lugiato-josa,ref-tamm}. For example, Ref.~\cite{ref-rotter_degeneracy}
found instances where intentionally breaking a degeneracy, such as
by introducing a wedge at a single location on the rim of a ring laser,
can result in the circulating mode not existing near threshold as
expected, but coming back \emph{into} existence (once the pump strength
is high enough above threshold and the nonlinearity is strong enough)
as a modified version that is nearly the degenerate circulating mode.
We have found similar results in hexagonal ($C_{6\mathrm{v}})$ structures
\emph{without }artifically-introduced defects, in which the degeneracy
is broken by discretization alone. There, a circulating single-mode
lasing solution starts existing above a certain pump strength (somewhat
higher than threshold), even though there is no degeneracy at threshold.
The reason the circulating lasing mode requires a minimum pump strength
is that the nonlinearity must be strong enough to counteract the broken
degeneracy and to lock the two modes to a single frequency. In many
cases such as these, numerically solving the single mode problem,
using an artificially-constructed circulating solution as an initial
guess, results in the solver correctly converging to a circulating
lasing mode. However, this effect is not yet completely understood,
and it is not entirely predictable under what circumstances such a
circulating mode exists. Furthermore, the pump window between the
original lasing threshold and the threshold at which the stable circular
mode emerges can not be described with SALT as the electric field
no longer shows a multi-periodic time dependency \cite{ref-lugiato-josa,ref-tamm}.
Moreover, a discretization-induced error and a physical perturbation
breaking the degeneracy are two distinct effects (even if their consequences
are mathematically similar), and it is useful to be able to study
them independently. When one is studying a physical symmetry-breaking
defect, one does not want to accidentally observe an artificial effect
of discretization instead. To eliminate numerical symmetry breaking
at arbitrary pump strengths, we therefore devised a solution: we construct
a minimal perturbation to the dielectric function that restores the
degeneracy in both the pump strengths and the frequencies at threshold.
We discuss this method below. 
\begin{figure}
\centerline{\includegraphics[scale=0.31]{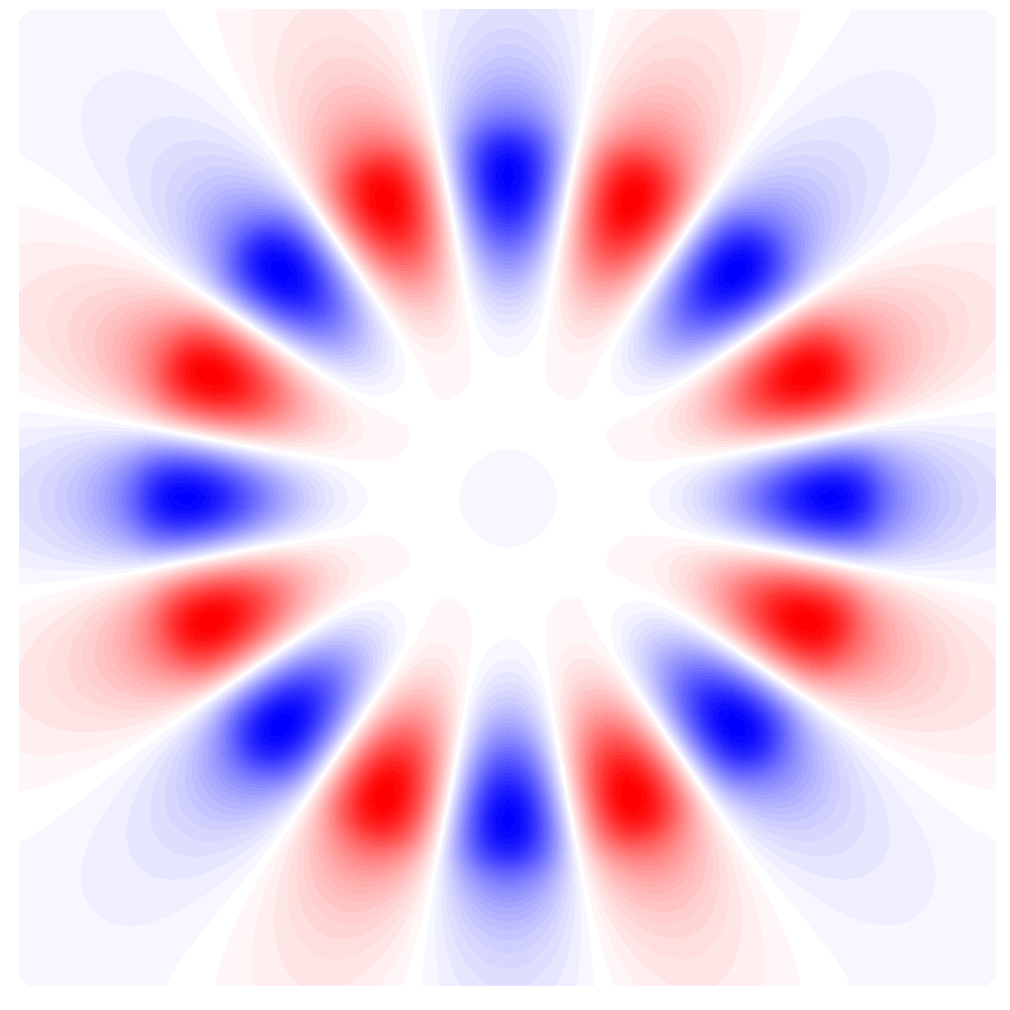}$\quad$\includegraphics[scale=0.31]{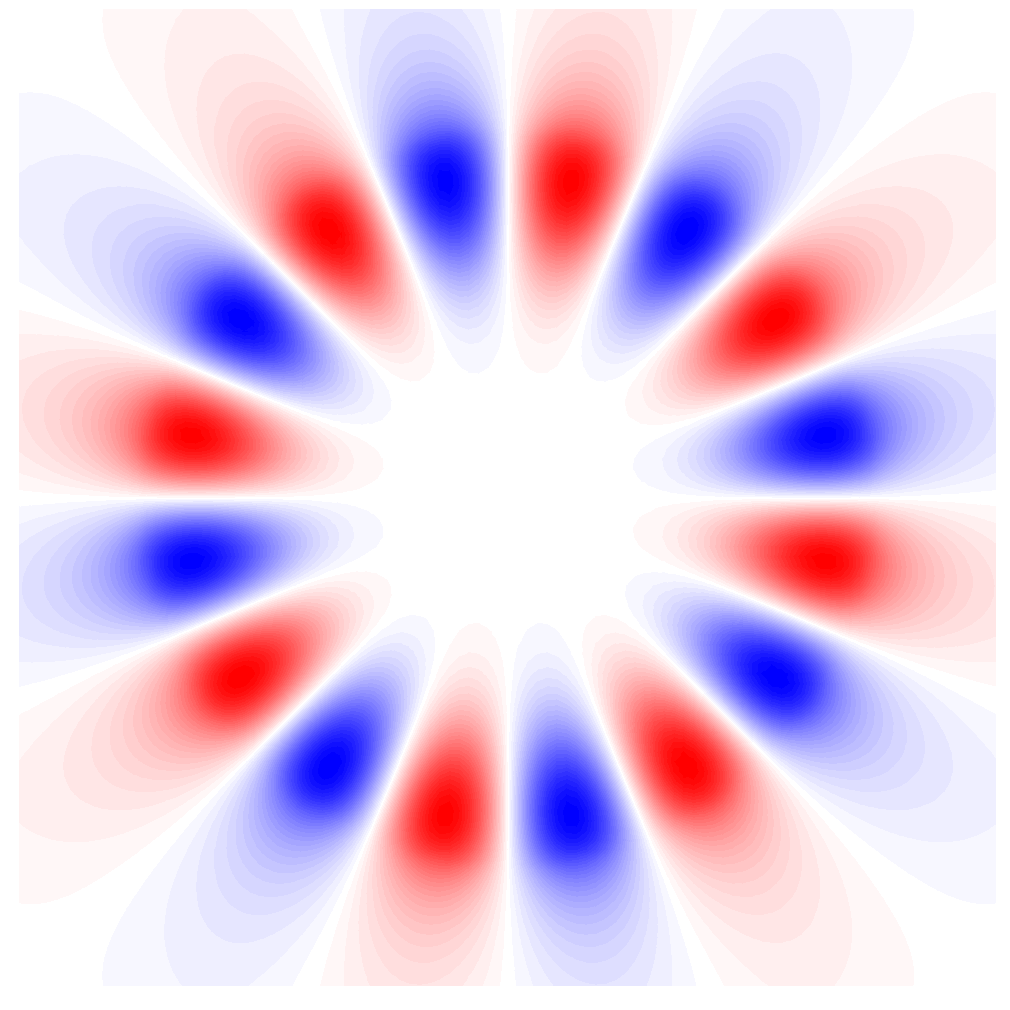}}\caption{(Color online) $\ell=8$ threshold modes {[}$\Re(\mathbf{E}_{\mathrm{e,o}})${]}
for a cylinder with uniform dielectric $\varepsilon=5$ and radius
$r=1$. Unlike in the odd-$\ell$ case (Fig.~\ref{fig:overview}),
however, the discretized modes are \emph{not} $\frac{\pi}{2}$ rotations
from each other. Consequently, there is an unphysical splitting, due
to discretization, of $0.11\%$ in $\mathrm{Re}(\omega_{1}-\omega_{2})$
(for a resolution of 14 pixcels per wavelength) and $11.5\%$ in $\mathrm{Im}(\omega_{1}-\omega_{2})$
at zero pump strength (the latter being larger only because these
are high-$Q$ modes and $\mathrm{Im}(\omega_{\mu})$ is already very
small at zero pump strength). A difference in imaginary parts also
means a splitting in the threshold pump strength $D_{\mathrm{t}}$.
\label{fig:even_cylinder_modes}}
\end{figure}

\subsection{Restoring degeneracy by minimal perturbations \label{sub:Forcing-the-degeneracy}}

\begin{figure}
\centerline{\includegraphics[bb=0bp 160bp 612bp 630bp,clip,scale=0.4]{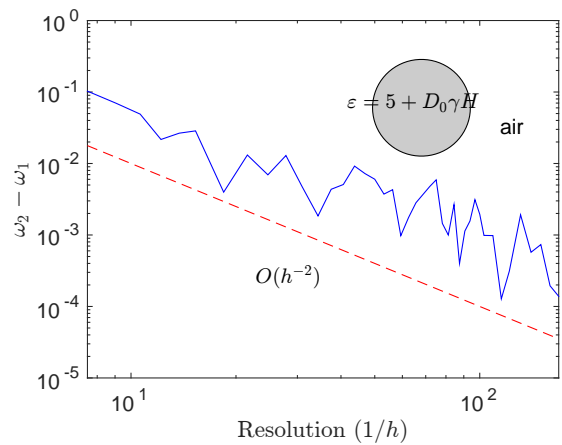}}

\caption{(Color online) Splitting in degeneracy due to discretization error
for even-$\ell$ modes of dielectric cylinder versus the resolution
$1/h$ of the discretization, where $h$ is the distance between adjacent
gridpoints. The oscillations, which are due to the discontinuous interfaces
between dielectric and air that ``jump'' when the resolution is
changed, could in principle be smoothed by using subpixel averaging
techniques for the discretization \cite{ref-subpixel}. \label{fig:cylinder_resolution}}
\end{figure}

The basic idea is that we construct an artificial perturbation $\delta\varepsilon(\mathbf{x})$
to the dielectric permittivity that forces the degeneracy in both
the frequency and threshold, and then we solve the perturbed single-mode
SALT equation. There are infinitely many possible functions that can
achieve this goal, so we look for the one with the smallest $L_{2}$
norm $\left\Vert \delta\varepsilon(\mathbf{x})\right\Vert _{2}^{2}=\int\left|\delta\varepsilon(\mathbf{x})\right|^{2}$.
This is a good choice because in the limit of infinite resolution,
the perturbation $\delta\varepsilon(\mathbf{x})$ approaches zero.
We construct $\delta\varepsilon(\mathbf{x})$ by solving a \emph{quadratic
program} \cite{boyd} (QP) with linear constraints that we obtain
using perturbation theory. Not only does this uniquely (and cheaply)
determine $\delta\varepsilon$, as described below, but it also guarantees
convergence to the solution of the unperturbed (physical) single-mode
SALT equation in the limit of infinite resolution. The reason it guarantees
convergence is that the frequency splitting vanishes in the limit
of infinite resolution, as shown in Fig.~\ref{fig:cylinder_resolution},
and so the minimum-norm $\delta\varepsilon$ to force a degeneracy
also vanishes in the limit of infinite resolution, recovering the
unperturbed SALT.

It turns out that determining the minimum-norm $\delta\varepsilon$
requires only that we solve a sequence of QP problems: minimizing
a convex quadratic function ($\left\Vert \delta\varepsilon\right\Vert _{2}^{2}$)
of $\delta\varepsilon$ subject to a linear constraint on $\delta\varepsilon$.
QPs are convex optimization problems with a unique global minimum
that can be efficiently found simply by solving a system of linear
equations \cite{boyd}. In particular, the linear constraint (Eq.~(\ref{eq:degeneracy_forcing})),
which coalesces the eigenvalues, can be derived from perturbation
theory. Because the perturbation theory is only first-order, however,
the $\delta\varepsilon$ that we find by solving the QP only approximately
eliminates the splitting, but we can simply re-solve SALT and solve
a new QP, iterating the process a few times (twice is typically enough)
to force a degeneracy to machine precision. The full details of the
procedure are given in Appendix~\ref{sec:Determining-the-dielectric-perturbation}.

\begin{figure}
\centerline{\includegraphics[scale=0.1]{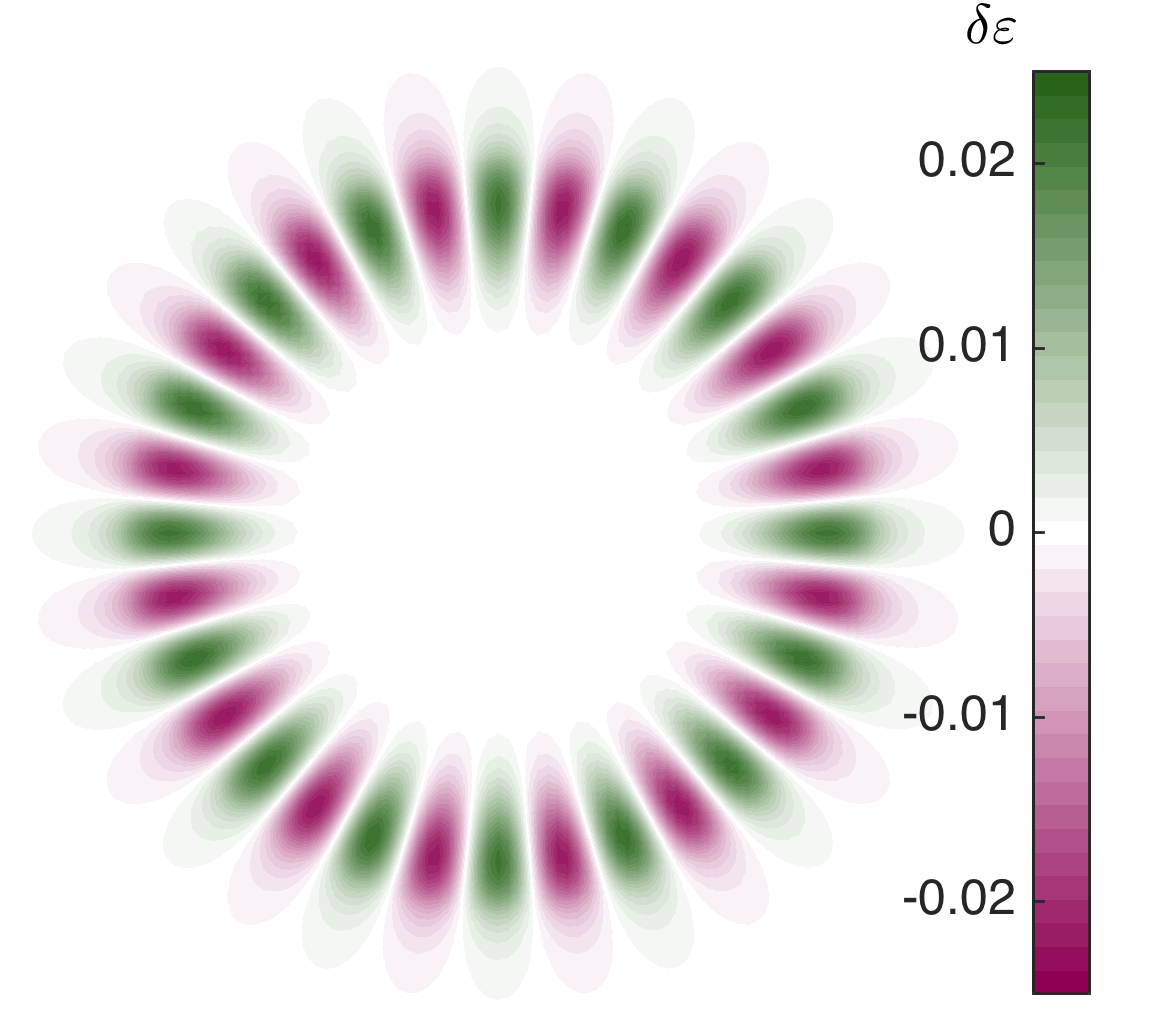}$\quad$\includegraphics[scale=0.1]{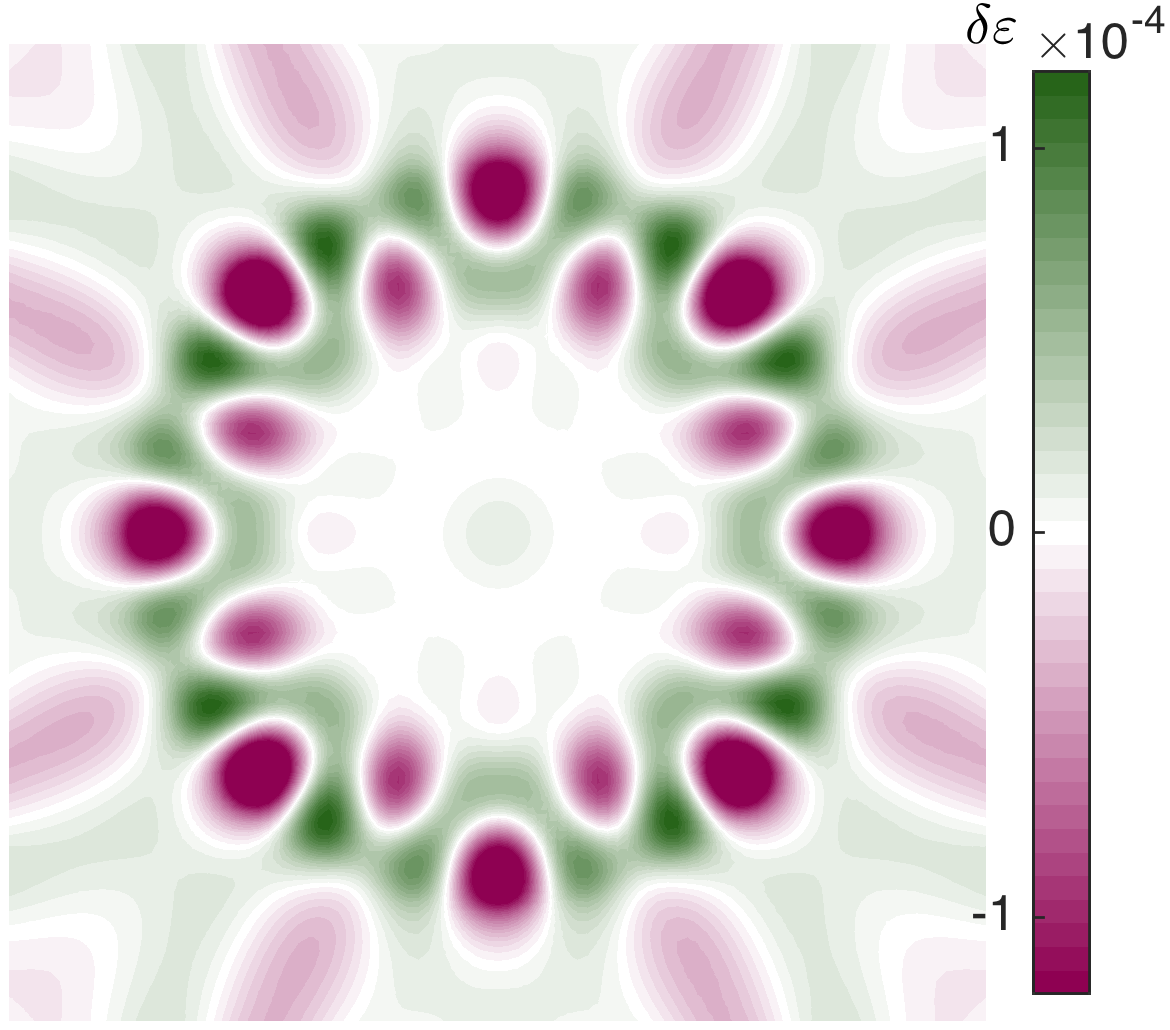}}\caption{(Color online) Dielectric perturbation $\delta\varepsilon$ obtained
by solving QP for threshold modes with $\ell=8$. The real part (left)
has a dependence $\cos(2\ell\phi)$, while the imaginary part (right)
is a more complicated function. \label{fig:even_cylinder_deps}}
\end{figure}

\begin{figure}
\centerline{\includegraphics[bb=0bp 160bp 612bp 630bp,clip,scale=0.4]{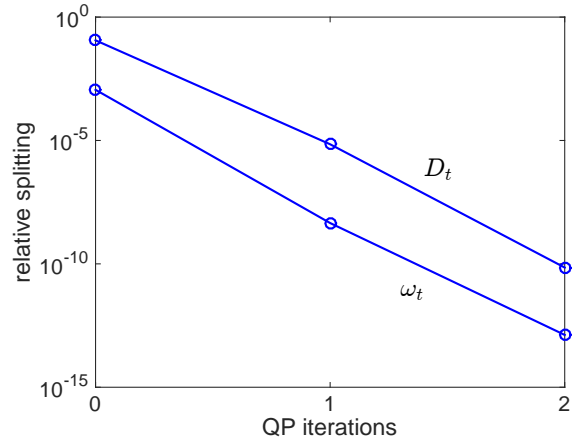}}

\caption{(Color online) Relative splitting in threshold pump strength $D_{t}$
and frequency $\omega_{t}$ for even-$\ell$ cylinder modes after
QP iterations. The relative splitting in frequency is defined in the
usual way as $2\left|\frac{\omega_{1}-\omega_{2}}{\omega_{1}+\omega_{2}}\right|$,
and similarly for the pump strength. Only two iterations of QP were
required reduce the splitting in both the threshold frequency $\omega_{t}$
and the threshold pump strength $D_{\mathrm{t}}$ to $10^{-10}$ or
smaller. \label{fig:even_cylinder_QP}}
\end{figure}

\begin{figure}
\centerline{\includegraphics[bb=0bp 160bp 612bp 630bp,clip,scale=0.4]{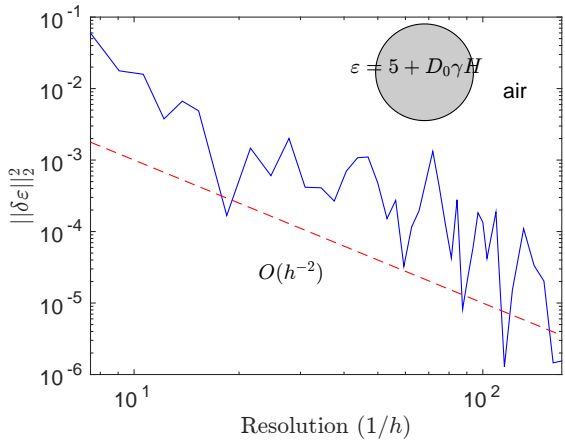}}

\caption{(Color online) $L_{2}$ norm of resulting $\delta\varepsilon(\mathbf{x})$
function obtained from QP procedure versus discretization resolution
$1/h$ for nearly degenerate even-$\ell$ modes of the cylinder, where
$h$ is the spacing between adjacent gridpoints. The same resolutions
as in Fig.~\ref{fig:cylinder_resolution} were used, and the oscillations
resemble the curve for splitting very closely. This is because the
larger the splitting $\omega_{2}-\omega_{1}$, the larger the $\delta\varepsilon(\mathbf{x})$
function needed to enforce the degeneracy. The fact that $\left\Vert \delta\varepsilon\right\Vert _{2}^{2}$
appears to be going to zero as the resolution increases indicates
that our QP procedure is convergent.\label{fig:deps_scale}}
\end{figure}

The resulting $\delta\varepsilon$ of this procedure applied to the
even-$\ell$ threshold modes in Fig.~\ref{fig:even_cylinder_modes}
is shown in Fig.~\ref{fig:even_cylinder_deps}, and the convergence
of the splitting to zero is shown in Fig.~\ref{fig:even_cylinder_QP}.
As verified in Fig.~\ref{fig:deps_scale}, the $L_{2}$ norm of $\delta\varepsilon(\mathbf{x})$
decreases with resolution, satisfying our requirement that the dielectric
perturbation should go to zero in the continuum limit. In principle,
one must resolve for $\delta\varepsilon$ at each pump strength, since
the hole-burning term changes the problem. However, in practice we
have found changes in $\delta\varepsilon$ with pump strength to be
negligible, as in Fig.~\ref{fig:QP_stability}, and one can typically
use the same $\delta\varepsilon$ for all pump strengths. In Appendix~\ref{sec:Determining-the-dielectric-perturbation},
we give a method that re-forces the degeneracy for pump strengths
above threshold, if a machine precision degeneracy is desired.


\begin{figure}
\centerline{\includegraphics[bb=0bp 160bp 612bp 630bp,clip,scale=0.4]{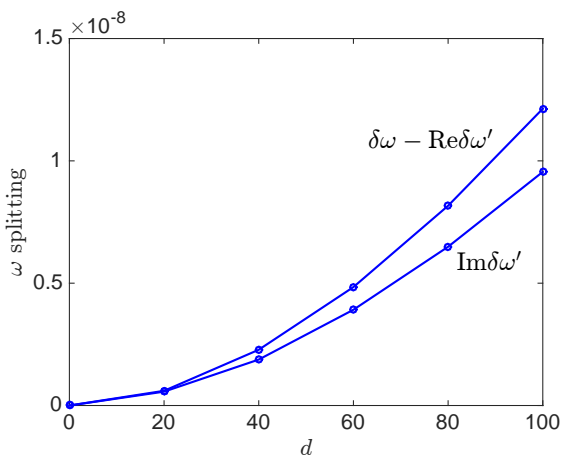}}

\caption{(Color online) Above-threshold splitting in real and imaginary parts
of $\delta\omega^{\prime}$ after performing QP procedure for even-$\ell$
modes. The magnitude is very small because the intensity profile is
very close to\emph{ }rotationally symmetric. \label{fig:QP_stability}}
\end{figure}

\subsection{Example with $C_{6\mathrm{v}}$ symmetry\label{sub:Examples}}

\begin{figure}
\centerline{\includegraphics[scale=0.1]{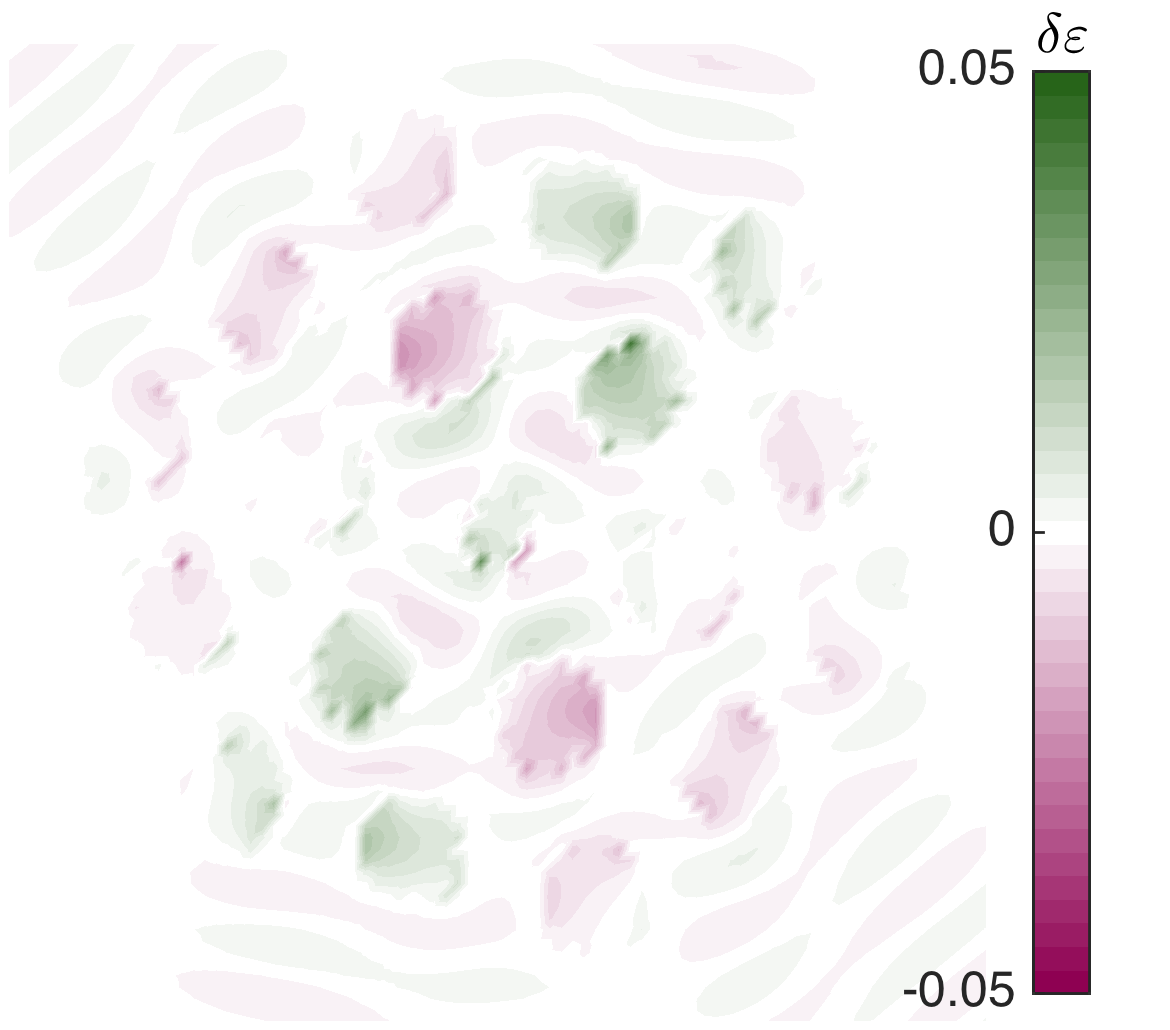}$\qquad$\includegraphics[scale=0.1]{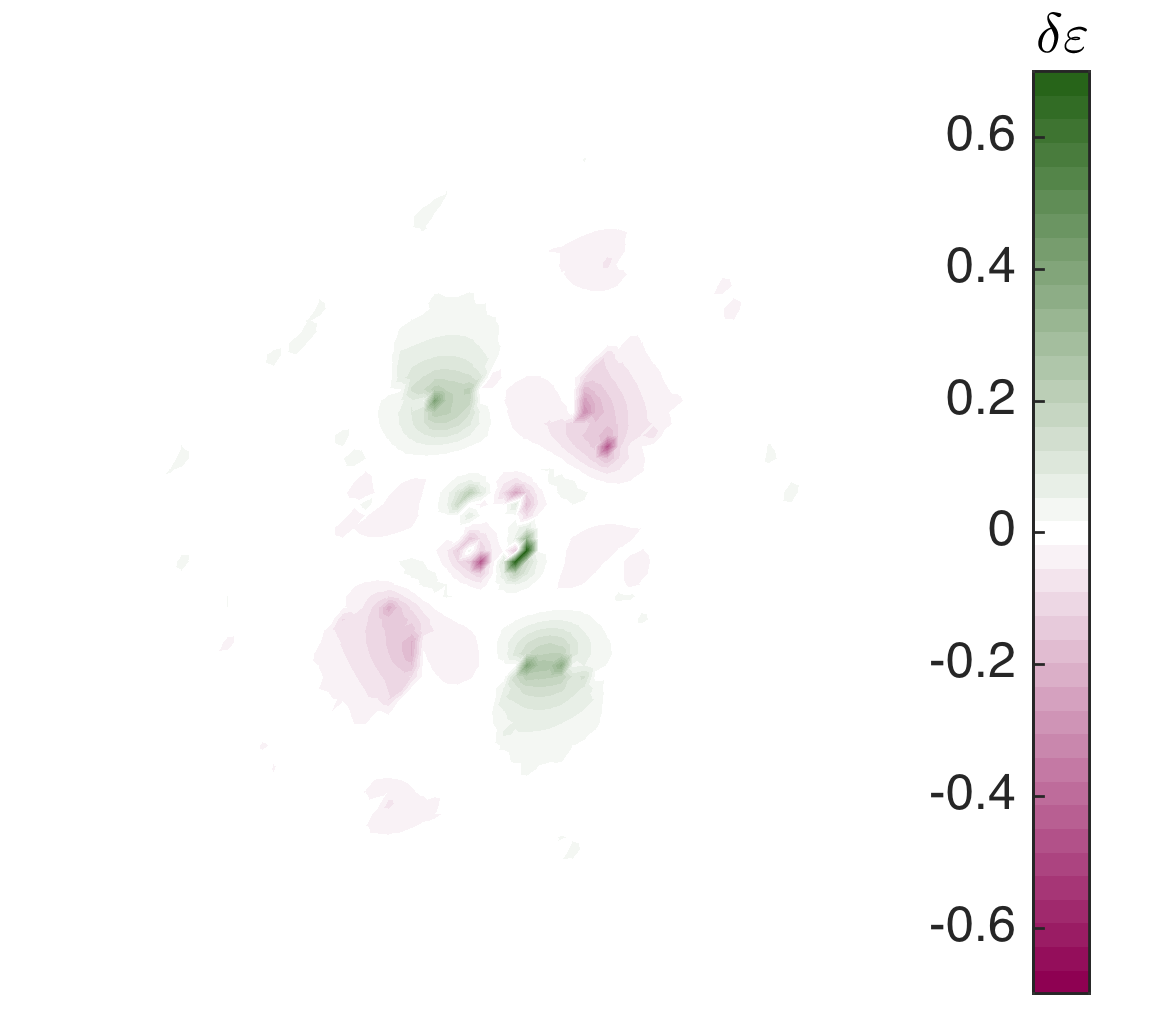}}\caption{(Color online) Dielectric perturbation obtained from QP procedure
for hexagonal cavity. Since the mode is TE ($\mathbf{E}=E_{\mathrm{even}}\hat{\mathbf{x}}+E_{\mathrm{odd}}\hat{\mathbf{y}}$),
we have allowed the perturbation to be a diagonally-anisotropic tensor,
as in Eq.~(\ref{eq:tensor_delta_epsilon}). Shown here are the real
(left) and imaginary (right) parts of $\delta\varepsilon_{xx}$. The
$\delta\varepsilon_{yy}$ looks similar except rotated by 60 degrees.
\label{fig:phc_deps}}
\end{figure}

\begin{figure}
\centerline{\includegraphics[scale=0.31]{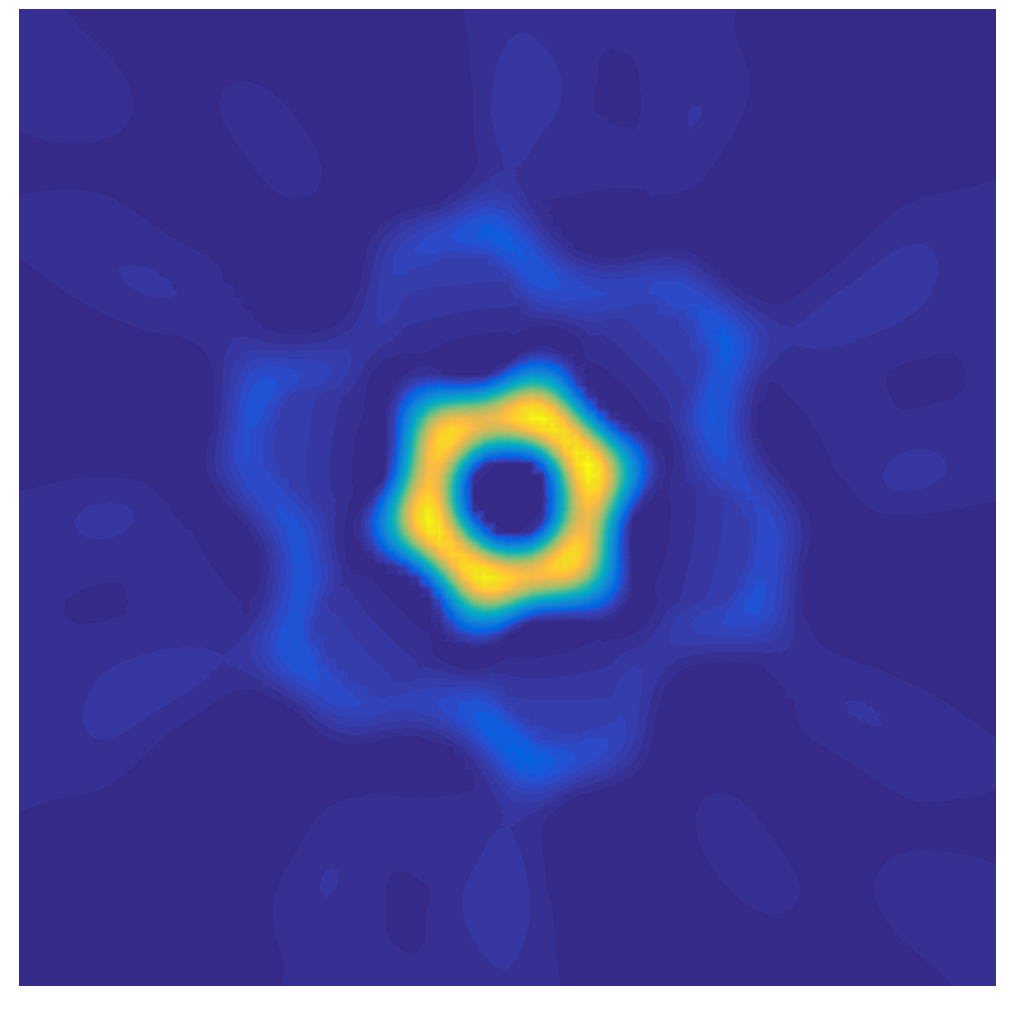}}\caption{(Color online) Intensity pattern for stable lasing mode for hexagonal
cavity. The pattern appears to be six-fold symmetric, which is expected.
Unlike in the right panel of Fig.~\ref{fig:lowQ_chirality} however,
the chirality is not significant enough to be visible, since the hexagonal
cavity is not as lossy as the square cavity in Fig.~\ref{fig:lowQ_chirality}.
In the ideal system, the second pole $\delta\omega^{\prime}$ stays
degenerate with the lasing eigenvalue $\delta\omega$, and this linear
combination stays stable for all pump strengths above threshold. In
the discretized system, there is not a true $C_{6\mathrm{v}}$ symmetry,
so there is a small splitting similar to that of the even-$\ell$
cylinder modes. Again, this splitting is too small to affect physically
meaningful results of the simulation, but can be removed using the
QP procedure if desired. \label{fig:phc_stable}}
\end{figure}

An example of a hexagonal cavity is shown in Fig.~\ref{fig:overview}.
This geometry was adapted from an infinite lattice of period $a$
with air holes of radius $0.3a$. A single hole in the middle has
a reduced radius $0.2a$ to create a defect in the band gap. The dielectric
is $\varepsilon_{c}=11.56$ everywhere except in the holes, where
there is air. A perfectly matched layer (PML) is added to the boundaries
to simulate the radiation loss, and the axes of the hexagon have been
aligned with the diagonals rather than the $x$ and $y$ axes because
the finite-difference Yee discretization \cite{taflove} happens to
only have mirror symmetry along the diagonals. Here, the lasing modes
are TE (electric field in-plane and magnetic field out of plane),
and there is a pair of degenerate threshold modes from the hexagon's
$C_{6\mathrm{v}}$ symmetry, as shown in Fig.~\ref{fig:overview}.
For a $100\times100$ finite-difference discretization, there is about
a 1.5\% splitting between the threshold eigenvalues, so again we must
use the QP procedure to force the threshold degeneracy. Since these
are TE modes, we now have \emph{two }components of the electric field,
and consequently we may treat $\delta\varepsilon$ as a tensor, as
in Eq.~(\ref{eq:tensor_delta_epsilon}). We only consider the diagonal
components $\delta\varepsilon_{xx}$ and $\delta\varepsilon_{yy}$
here for simplicity. Only two iterations of QP are necessary to force
the degeneracy down to machine precision, and the perturbation used
to force the degeneracy is shown in Fig.~\ref{fig:phc_deps}. We
then use Eq.~(\ref{eq:chiral definition}) as an initial guess for
our numerical solver, and the intensity pattern of the resulting circulating
solution is shown in Fig.~\ref{fig:phc_stable}. 

\subsection{Multimode case \label{sub:multimode discretization method}}

Now we consider how to treat the problem of discretization-broken
symmetry (Sec.~\ref{sec:discretization_breaking}) in the case of
multimode lasing. The method of Sec.~\ref{sub:Forcing-the-degeneracy}
gives a $\delta\varepsilon(\mathbf{x})$ that forces the threshold
degeneracy for a \emph{single }mode pair. When there are multiple
pairs of nearly degenerate modes the generalization is straightforward:
we allow each pair to have its \emph{own} $\delta\varepsilon_{\mu}$,
so that their degeneracies can be forced \emph{independently}. As
a result, the dielectric for each pair in Eq.~(\ref{eq:salt general})
will become
\begin{equation}
\varepsilon_{\mu}(\mathbf{x})=\varepsilon(\mathbf{x})+\delta\varepsilon_{\mu}(\mathbf{x})+\frac{D_{0}(\mathbf{x})\Gamma(\omega_{\mu})}{1+\gamma_{\parallel}^{-1}\sum\left|\Gamma_{\nu}\mathbf{E}_{\nu}(\mathbf{x})\right|^{2}}.
\end{equation}
Since our QP method finds the $\delta\varepsilon_{\mu}$ with the
lowest $L_{2}$ norm (as described in Sec.~\ref{sub:Forcing-the-degeneracy})
and the splitting decreases with resolution as seen in Fig.~\ref{fig:cylinder_resolution},
each $\delta\varepsilon_{\mu}$ will independently go to zero as we
increase the resolution, so this generalized method is also \emph{convergent}:
the unphysical frequency-dependent $\delta\varepsilon_{\mu}$ vanishes
with increasing resolution.

\section{Concluding Remarks}

In this paper, we have reduced the problem of identifying the stable
lasing modes of a degenerate laser from the full nonlinear Maxwell--Bloch
equations to a small \emph{semi-analytical} solution evaluated in
terms of integrals of  the threshold modes (solutions to the \emph{linear}
Maxwell partial-differential equation). Our perturbative solution
near threshold confirms an ansatz in the earlier degenerate SALT work
\cite{ref-rotter_degeneracy}, in which the circulating and standing
wave solutions were guessed as starting points for a SALT solver and
it was conjectured that the resulting four solutions were the only
possibilities. Furthermore, we have presented an efficient numerical
scheme to track these solutions far above threshold via numerical
SALT solvers \cite{direct_salt} combined with a simple technique
to correct for numerical symmetry breaking. And finally, we have shown
that the degeneracy of the $C_{n}$ group Sec.~\ref{sec:Effects-of-chirality}
means that circulating lasing modes will retain a degenerate passive
pole even far above threshold, where the hole-burning term breaks
mirror symmetry. In addition, our work poses some intriguing open
questions for future research.

First, although we have reduced the question of stability of circulating
modes near threshold to a simple semi-analytical criterion (checking
whether a certain integral expression is positive), one would like
to additionally have a fully analytical proof that the circulating
modes are stable, or alternatively a counter-example of a $C_{n\mathrm{v}}$-symmetric
problem with a degenerate lasing threshold in which the circulating
mode is unstable. Since Ref.~\cite{ref-rotter_degeneracy} found
that circulating modes can become unstable at higher pump strengths,
one might naively hope that the technique of Ref.~\cite{ref-lige_condensation_arxiv}
(in which a lasing SALT solution at some pump strength above threshold
is transformed into a threshold SALT solution by designing the pump
profile at threshold to match the hole-burning term at the higher
pump strength) could be used to translate these into an unstable threshold
circulating mode. However, the technique of Ref.~\cite{ref-lige_condensation_arxiv}
only translates one SALT solution into another, and does not translate
the full Maxwell--Bloch solutions since it does not keep track of
$\gamma_{\parallel}$. Because the exact Maxwell--Bloch stability
eigenvalue $\sigma$ depends on $\gamma_{\parallel}$, as explained
in Appendix~\ref{sub:Region-of-validity} (only the first-order term
$\sigma_{1}$ is independent), and SALT solutions do not, this method
of translating a hole-burning term to an artificially-designed pump
profile at a higher threshold does not account for the effects of
$\gamma_{\parallel}$. 

Second, it would be interesting to extend this sort of perturbative
SALT/stability analysis to other lasing systems besides $C_{n}$ and
$C_{n\mathrm{v}}$ symmetries. For example, in a 3d photonic-crystal
cavity \cite{phc_book} one could have cubic symmetry and threefold
degeneracies, or one could have even greater degeneracies in spherical
resonators. Alternatively, in a surface-emitting distributed feedback
\cite{ref-surface_emitting,ref-surface_emitting2,ref-turnbull,ref-distributed_feedback}
or photonic-crystal laser \cite{ref-imada_phc_distributed,ref-continuous_phc_bandedge,ref-ryu_bandedge,ref-coherent_surface_emitting_phc},
one might have lasing occur at a ``band edge'' \cite{yariv} in
the dispersion relation. While a band edge may or may not be degenerate
\emph{per se}, it coincides with a singularity in the density of states
\cite{ashcroft-mermin} where a continuum of resonances occurs in
a small neighborhood of the lasing resonance, and perturbative analysis
might be very helpful in understanding its stability. Finally, it
would be interesting to apply semi-analytical perturbative stability
analysis to cases where a small imperfection slightly splits the degeneracy,
which was studied numerically in Ref.~\cite{ref-rotter_degeneracy}.
\begin{acknowledgments}
This work was supported in part by the Army Research Office through
the Institute for Soldier Nanotechnologies (ISN), Grant No. W911NF-07-D-0004,
the Air Force Research Laboratory under agreement number FA8650-15-
2-5220, and by the Austrian Science Fund (FWF) through Project No.
SFB NextLite F49-P10. L. G. acknowledges partial support by NSF under
Grant No. DMR-1506987. We are also grateful to A. D. Stone for helpful
discussions. The U.S. Government is authorized to reproduce and distribute
reprints for governmental purposes notwithstanding any copyright notation
thereon. The views and conclusions contained herein are those of the
authors and should not be interpreted as necessarily representing
the official policies or endorsements, either expressed or implied,
of Air Force Research Laboratory or the U.S. Government.
\end{acknowledgments}

\appendix

\section{Degeneracy in $C_{n}$\label{sec: degeneracy in Cn}}

We review the result, given in Ref. \cite{dichroism}, of the fact
that there are two-fold degeneracies (due to Lorentz reciprocity)
in geometries with $C_{n}$ but not $C_{n\mathrm{v}}$ symmetry, even
though there are only one-dimensional irreps. We give a slightly simpler
and more general proof by exploiting the differential form of Maxwell's
equations, as opposed to the integral form in Ref. \cite{dichroism}. 

Consider a field $\mathbf{E}^{+}$ that satisfies the equation $\hat{L}(\omega^{+})\mathbf{E}^{+}=0$,
where we define the linear operator (as in Ref. \cite{ref-adi})
\begin{equation}
\hat{L}(\omega)\equiv-\nabla\times\frac{1}{\mu(\mathbf{x},\omega)}\nabla\times+\,\omega^{2}\varepsilon(\mathbf{x},\omega)
\end{equation}
where $\omega$ is the eigenfrequency and $\varepsilon$ and $\mu$
have $C_{n}$ symmetry: that is, $R_{n}\varepsilon R_{n}^{-1}=\varepsilon$,
where $R_{n}$ is an $n$-fold rotation and $R_{n}^{n}=1$, the identity
operator. Suppose that the field transforms like one of the chiral
irreps of $C_{n}$: that is, $R_{n}\mathbf{E}^{+}=\exp\left(-\frac{2\pi im}{n}\right)\mathbf{E}^{+}$,
with $0<\left|m\right|\leq\mathrm{floor}(\frac{n-1}{2})$. We want
to show that there exists some other function $\mathbf{E}^{-}$ that
transforms according to the irrep of the opposite chirality \emph{and
}has the same eigenfrequency: that is, $R_{n}\mathbf{E}^{-}=\exp\left(\frac{2\pi im}{n}\right)\mathbf{E}^{-}$
and $\hat{L}(\omega^{+})\mathbf{E}^{-}=0$. 

The key step is to use the right basis: we \emph{could }find the Maxwell
eigenfrequencies (Green's-function poles) by solving the nonlinear
(in $\omega$) eigenvalue problem $\hat{L}(\omega)\mathbf{E}=0$.
However, these make a poor basis because they diagonalize \emph{different
}operators $\hat{L}(\omega)$ with $\omega\neq\omega^{+}$. Instead,
we fix $\omega=\omega^{+}$ and examine the set of eigenfunctions
$\mathbf{E}_{j}^{-}$ that satisfy $\hat{L}(\omega^{+})\mathbf{E}_{j}^{-}=\lambda_{j}\mathbf{E}_{j}^{-}$
and that transform as the $\exp\left(\frac{2\pi im}{n}\right)$ irrep.
A similar strategy was employed in Ref.~\cite{ge_steady-state_2010}
to introduce the threshold constant-flux (TCF) states basis. (The
TCF approach is slightly different, for it assumes that the eigenvalues
$\lambda_{j}$ are followed by a spatial function that specifies the
pump profile.) Note that $\lambda_{j}$ are \emph{not }squared eigenfrequencies
and $\mathbf{E}_{j}^{-}$ are \emph{not }Maxwell solutions, except
for $\lambda_{j}=0$. Because this set is a complete basis for functions
of this chirality, the function $(\mathbf{E}^{+})^{\star}$ (which
transforms in the same way as $\mathbf{E}_{j}^{-}$ because the rotation
operator $R_{n}$ is real) can be expanded in this basis:
\begin{equation}
\left(\mathbf{E}^{+}\right)^{\star}=\sum b_{j}\mathbf{E}_{j}^{-},
\end{equation}
assuming that $\hat{L}(\omega^{+})$ is diagonalizable (which is generically
true for matrices except at exceptional points; the situation for
infinite-dimensional operators is more complicated, but diagonalizability
is typically assumed there too in physics). We will now show that
at least \emph{one }of these $\mathbf{E}_{j}^{-}$ is exactly the
$\mathbf{E}^{-}$ satisfying $\hat{L}(\omega^{+})\mathbf{E}^{-}=0$
that we are looking for.

First, we define the \emph{unconjugated }inner product $\left(\mathbf{f},\mathbf{g}\right)\equiv\int d^{3}x\,\mathbf{f}\cdot\mathbf{g}$.
Then, for appropriate boundary conditions, $\hat{L}(\omega^{+})$
is complex symmetric, that is: $\left(\mathbf{f},\hat{L}\mathbf{g}\right)=\left(\hat{L}\mathbf{f},\mathbf{g}\right)$
for reciprocal materials $\varepsilon=\varepsilon^{T}$, $\mu=\mu^{T}$,
and this is known as Lorentz reciprocity \cite{phc_book}. Because
$\hat{L}(\omega^{+})$ is complex symmetric, its eigenfunctions with
distinct eigenvalues are orthogonal; that is: $\left(\mathbf{E}_{i}^{-},\mathbf{E}_{j}^{-}\right)=0$
for $\lambda_{i}\neq\lambda_{j}$. Now write
\begin{align}
\int d^{3}x\,\left|\mathbf{E}^{+}\right|^{2} & =\sum b_{j}\left(\mathbf{E}^{+},\mathbf{E}_{j}^{-}\right).
\end{align}
If all $\mathbf{E}_{j}^{-}$ had $\lambda_{j}\neq0$, then $\left(\mathbf{E}^{+}\cdot\mathbf{E}_{j}^{-}\right)=0$
for all $j$. However, the left-hand side is obviously positive, so
at least one term in the sum on the right-hand side must be non-vanishing.
Hence, this term has the eigenvalue $\lambda_{j}=0$, and it is precisely
the $\mathbf{E}^{-}$ that is degenerate to $\mathbf{E}^{+}$.

\section{Allowed lasing modes\label{sec: appendix allowed lasing modes}}

In this appendix, we show that the only allowed lasing modes for $C_{n\mathrm{v}}$
geometries above threshold are the circulating modes $\mathbf{E}_{\pm}$
and standing-wave modes $\mathbf{E}_{+}+e^{i\theta}\mathbf{E}_{-}$,
with $\theta$ an arbitrary angle for $n\neq4\ell$ (where $\ell$
is the order of the 2d irrep that the degenerate modes transform as),
and $\theta$ an integer multiple of $\frac{\pi}{2}$ for $n=4\ell$.
We begin by writing $D_{0}=D_{\mathrm{t}}(1+d)$ and inserting into
Eq.~(\ref{eq:single-mode-salt}). We expand to lowest order in $d$
and the mode intensity, and we have
\begin{equation}
\nabla\times\nabla\times\mathbf{E}=\omega^{2}\left[\varepsilon+D_{\mathrm{t}}\Gamma\left(1+d-\gamma_{\parallel}^{-1}\left|\Gamma\mathbf{E}\right|^{2}\right)\right]\mathbf{E}.\label{eq:expanded salt}
\end{equation}
Comparing with Eq.~(\ref{eq:threshold salt}) at threshold, we conclude
that $\mathbf{E}=O(\sqrt{d})$, and that the profile should be some
linear combination of the threshold modes $\mathbf{E}_{1,2}$, as
in Eq.~(\ref{eq:lasing_ansatz}). Inserting Eq.~(\ref{eq:lasing_ansatz})
into Eq.~(\ref{eq:expanded salt}), noting that the zeroth-order
terms vanish due to Eq.~(\ref{eq:threshold salt}), we obtain
\begin{multline}
\left[\nabla\times\nabla\times-\:\omega_{\mathrm{t}}^{2}\left(\varepsilon+D_{\mathrm{t}}\Gamma_{\mathrm{t}}\right)\right]\delta\mathbf{E}\\
=\left\{ \omega_{1}\frac{\partial}{\partial\omega_{\mathrm{t}}}\left[\omega_{\mathrm{t}}^{2}\left(\varepsilon+D_{\mathrm{t}}\Gamma_{\mathrm{t}}\right)\right]+\omega_{\mathrm{t}}^{2}D_{\mathrm{t}}\Gamma_{\mathrm{t}}\left(1-\left|\mathbf{f}\right|^{2}\right)\right\} \mathbf{f}
\end{multline}
where $\mathbf{f}\equiv a_{1}\mathbf{E}_{1}+a_{2}\mathbf{E}_{2}$.
Now multiply by $\mathbf{E}_{1,2}$ and integrate over all space,
and we obtain Eq.~(\ref{eq:a eigenproblem}), where we have used
the fact that 
\begin{equation}
\int d^{3}x\,\mathbf{E}_{1,2}\cdot\left[\nabla\times\nabla\times-\:\omega_{\mathrm{t}}^{2}\left(\varepsilon+D_{\mathrm{t}}\Gamma_{\mathrm{t}}\right)\right]\delta\mathbf{E}=0
\end{equation}
because the threshold Maxwell operator is complex-symmetric \cite{moiseyev,siegman,phc_book}
so that it acts to the left and annihilates $\mathbf{E}_{1,2}$. (The
fact that the $\nabla\times\nabla\times$ operator acts to the left
can be understood using integration by parts, with the boundary terms
vanishing due to the limiting-absorption principle \cite{ref-limiting-absorption}.)
Now, choose $\mathbf{E}_{1}=\mathbf{E}_{+}$ and $\mathbf{E}_{2}=\mathbf{E}_{-}$,
where $\mathbf{E}_{\pm}$ is defined in Eq.~(\ref{eq:chiral definition}).
For generality, we assume that the geometry {[}i.e. the functions
$\varepsilon(\mathbf{x})$ and $D_{\mathrm{t}}(\mathbf{x})${]} has
at least $C_{n}$ symmetry. In Eq.~(\ref{eq:a eigenproblem}), we
see that $\int d^{3}x\varepsilon(\mathbf{x})\mathbf{E}_{+}\cdot\mathbf{E}_{+}$
vanishes, because it is the conjugated inner product of $\mathbf{E}_{+}$
and $\mathbf{E}_{+}^{\star}$, which transform as the clockwise and
counterclockwise 1d irreps in the $C_{n}$ group, and from the great
orthogonality theorem \cite{inui_group,tinkham}, conjugated inner
products between functions belonging to different irreps always vanish
{[}e.g., for circulating modes on a uniform ring, the integral $\int d\phi\,e^{i\ell\phi}(e^{-i\ell\phi})^{\star}$
vanishes{]}. Also, the integral over $D_{\mathrm{t}}(\mathbf{x})$
is the same, because it also has at least $C_{n}$ symmetry. Excluding
the intensity term in Eq.~(\ref{eq:a eigenproblem}), the rest of
the terms become
\begin{equation}
a_{\mp}\left(\omega_{1}H+G_{D}\right),
\end{equation}
where we have defined
\begin{align}
G_{\varepsilon} & \equiv\int d^{3}x\,\varepsilon(\mathbf{x})\mathbf{E}_{+}\cdot\mathbf{E}_{-}\nonumber \\
G_{D} & \equiv\int d^{3}x\,D_{\mathrm{t}}(\mathbf{x})\mathbf{E}_{+}\cdot\mathbf{E}_{-}.\label{eq:G and H definitions}\\
H & \equiv(\omega_{\mathrm{t}}^{2}\Gamma_{\mathrm{t}})^{-1}\frac{\partial}{\partial\omega_{\mathrm{t}}}\left\{ \omega_{\mathrm{t}}^{2}\left(G_{\varepsilon}+G_{D}\Gamma_{\mathrm{t}}\right)\right\} .\nonumber 
\end{align}
For the intensity term, we have $\int d^{3}x\,D_{\mathrm{t}}\mathbf{E}_{\pm}\cdot\mathbf{f}\left|\mathbf{f}\right|^{2}$.
We note that the quantities $\left|\mathbf{E}_{\pm}\right|^{2}$ and
$\mathbf{E}_{+}\cdot\mathbf{E}_{-}$ have $C_{n}$ symmetry, so by
symmetry arguments, the only surviving integrals are
\begin{align}
I_{\pm} & =\int d^{3}x\,D_{\mathrm{t}}\left|\vec{E}_{\pm}\right|^{2}\vec{E}_{+}\cdot\vec{E}_{-}\nonumber \\
J_{\pm} & =\int d^{3}x\,D_{\mathrm{t}}(\vec{E}_{\pm}^{\star}\cdot\vec{E}_{\mp})\vec{E}_{\pm}\cdot\vec{E}_{\pm}\label{eq:I J K definitions}\\
K_{\pm} & =\int d^{3}x\,D_{\mathrm{t}}(\vec{E}_{\mp}^{\star}\cdot\vec{E}_{\pm})\vec{E}_{\pm}\cdot\vec{E}_{\pm}.\nonumber 
\end{align}
Additionally, note that for $C_{n\mathrm{v}}$, we have $I_{+}=I_{-}$,
and the same for $J$ and $K$. Further, for TM modes ($\mathbf{E}_{\pm}=E_{\pm}\hat{\mathbf{z}}$),
we have $I_{\pm}=J_{\pm}$. Finally, $K_{\pm}$ is only non-vanishing
for $n=4\ell$, since in that special case, $\mathbf{E}_{+}$ picks
up a factor of $i$ under four-fold rotation, as seen in Eq.~(\ref{eq:chiral definition}).
With these definitions, Eq.~(\ref{eq:a eigenproblem}) straightforwardly
reduces to Eq.~(\ref{eq:reduced a plus minus}).

To solve for the coefficients $a_{\pm}$ and the frequency shift $\omega_{1}$,
first consider the case $a_{-}=0$. Dividing Eq.~(\ref{eq:reduced a plus minus})
(with the bottom sign) by $a_{+}$ then yields
\begin{equation}
\left|a_{+}\right|^{2}=\frac{\omega_{1}H+G_{D}}{I_{+}}.
\end{equation}
Taking the imaginary part of both sides, and noting that $\omega_{1}$
is real, we obtain
\begin{equation}
0=\omega_{1}\Im\left(\frac{H}{I_{+}}\right)+\Im\left(\frac{G_{D}}{I_{+}}\right),\label{eq:imaginary part of |a|^2}
\end{equation}
which leads to the circulating solution. (Eq.~(\ref{eq:circulating lasing})
holds for $C_{n\mathrm{v}}$; the same expression with $I$ replaced
by $I_{\pm}$ also holds for $C_{n}$. In this case $I_{+}\neq I_{-}$,
so the two circulating lasing modes will actually have slightly different
amplitudes and frequencies.) Note that the fact that $\left|a_{+}\right|^{2}$
must be a positive number also gives a cutoff condition
\begin{equation}
\Re(G_{D}/I_{\pm})>\frac{\Im\left(G_{D}/I_{\pm}\right)}{\Im\left(H/I_{\pm}\right)}\Re(H/I_{\pm}).
\end{equation}

Next, we consider the case that both $a_{\pm}$ are nonzero. Write
$a_{\pm}=\left|a_{\pm}\right|e^{i\theta_{\pm}}$ and define the relative
phase $z=e^{i(\theta_{-}-\theta_{+})}$. Divide Eq.~(\ref{eq:reduced a plus minus})
by $a_{\mp}$, and we obtain
\begin{equation}
\left|a_{\mp}\right|^{2}I_{\mp}+\left|a_{\pm}\right|^{2}\left(I_{\pm}+J_{\pm}+z^{\mp2}K_{\pm}\right)=\omega_{1}H+G_{D}.
\end{equation}
Solving this linear equation for the unknowns $\left|a_{\pm}\right|^{2}$,
we obtain
\begin{equation}
\left|a_{\pm}\right|^{2}=(\omega_{1}H+G_{D})T_{\pm},
\end{equation}
where
\begin{equation}
T_{\pm}=\frac{J_{\mp}+z^{\pm2}K_{\mp}}{\left(I_{+}+J_{+}+z^{-2}K_{+}\right)\left(I_{-}+J_{-}+z^{2}K_{-}\right)-I_{+}I_{-}}.
\end{equation}
Again, since $\left|a_{\pm}\right|^{2}$ and $\omega_{1}$ are real,
we have
\begin{equation}
\omega_{1}=-\frac{\Im\left(G_{D}T_{+}\right)}{\Im\left(HT_{+}\right)}=-\frac{\Im\left(G_{D}T_{-}\right)}{\Im\left(HT_{-}\right)}.\label{eq:omega1 constraint}
\end{equation}
The second equality here is a constraint that must be satisfied. For
$C_{n\mathrm{v}}$ with $n\neq4\ell$, we have $I_{+},J_{+}=I_{-},J_{-}$
and $K_{\pm}=0$ (as explained previously), so $T_{+}=T_{-}$ and
the constraint is automatically satisfied, indicating that $z$ is
free to have any phase, and yielding the solution in Eq.~(\ref{eq:n !=00003D 4 lasing}).
For $C_{n}$ with $n\neq4\ell$, we again have $K_{\pm}=0$, but there
is no mirror symmetry so $I_{+},J_{+}\neq I_{-},J_{-}$, and no choice
of $z$ will allow Eq.~(\ref{eq:omega1 constraint}) to be satisfied.
Hence, there are no standing lasing modes for this case.

For $C_{n\mathrm{v}}$ with $n=4\ell$, we again have $I_{+},J_{+}=I_{-},J_{-}$,
but we also have $K_{+}=K_{-}\neq0$. Hence, for $T_{+}=T_{-}$ to
be true, we must have $z^{2}=z^{-2}=\pm1$. Hence, there are two cases,
$z=\pm1$, for which the solution is given in Eq.~(\ref{eq:E+ + E- lasing}),
and $z=\pm i$, for which the solution is given in Eq.~(\ref{eq:E+ + i E- lasing}).
For $C_{n}$ with $n=4\ell$, we now have $I_{+},J_{+}\neq I_{-},J_{-}$,
and $K_{+}\neq K_{-}$, with both $K_{\pm}$ nonzero. Empirically,
we have found that Eq.~(\ref{eq:omega1 constraint}) still has solutions
(which must obtained by solving the equation numerically) at four
allowed phases $z$, with angles separated by $\frac{\pi}{2}$, just
as in the $C_{4\ell\mathrm{v}}$ case. Of course, it is straightforward
to choose the overall normalization and phase of the threshold basis
$\mathbf{E}_{\pm}$ such that the standing modes are still $\mathbf{E}_{+}\pm\mathbf{E}_{-}$
and $\mathbf{E}_{+}\pm i\mathbf{E}_{-}$, just as in the $C_{4\ell\mathrm{v}}$
case.

\section{Stability calculations\label{sec: appendix stability calculations}}

In this appendix, we provide details for the derivation of the stability
eigenvalues given in Sec.~\ref{sub:Stability-analysis summary}.
Comparing Eq.~(\ref{eq:delta}) and Eq.~(\ref{eq:time-dependent quadratic}),
we see that the matrices are

\begin{equation}
\mathsf{A}=\left(\begin{array}{ccccc}
\Delta & -\varepsilon_{\mathrm{I}}\omega^{2} & \omega^{2} & 0 & 0\\
\varepsilon_{\mathrm{I}}\omega^{2} & \Delta & 0 & \omega^{2} & 0\\
\gamma_{\perp}D & 0 & \omega_{a}-\omega & \gamma_{\perp} & \gamma_{\perp}\mathbf{E}_{\mathrm{R}}\\
0 & \gamma_{\perp}D & -\gamma_{\perp} & \omega_{a}-\omega & \gamma_{\perp}\mathbf{E}_{\mathrm{I}}\\
-\mathbf{P}_{\mathrm{I}} & \mathbf{P}_{\mathrm{R}} & \mathbf{E}_{\mathrm{I}} & -\mathbf{E}_{\mathrm{R}} & \gamma_{\parallel}
\end{array}\right),\label{eq:A matrix}
\end{equation}
where $\Delta=\varepsilon_{\mathrm{R}}\omega^{2}-\nabla\times\nabla\times$,
$\mathbf{E}_{\mathrm{R}}\equiv\Re(\mathbf{E})$, and $\mathbf{E}_{\mathrm{I}}\equiv\Im(\mathbf{E})$,
\begin{equation}
\mathsf{B}=\left(\begin{array}{ccccc}
-2\varepsilon_{\mathrm{I}}\omega & -2\varepsilon_{\mathrm{R}}\omega & 0 & -2\omega & 0\\
2\varepsilon_{\mathrm{R}}\omega & -2\varepsilon_{\mathrm{I}}\omega & 2\omega & 0 & 0\\
0 & 0 & 0 & 1 & 0\\
0 & 0 & -1 & 0 & 0\\
0 & 0 & 0 & 0 & 1
\end{array}\right),
\end{equation}
and
\begin{equation}
\mathsf{C}=\left(\begin{array}{ccccc}
-\varepsilon_{\mathrm{R}} & \varepsilon_{\mathrm{I}} & -1 & 0 & 0\\
-\varepsilon_{\mathrm{I}} & -\varepsilon_{\mathrm{R}} & 0 & -1 & 0\\
0 & 0 & 0 & 0 & 0\\
0 & 0 & 0 & 0 & 0\\
0 & 0 & 0 & 0 & 0
\end{array}\right).
\end{equation}
We expand the matrices in powers of $\sqrt{d}$, as in Eq.~(\ref{eq:matrix expansion}),
by noting that $\omega=\omega_{\mathrm{t}}+\omega_{1}d+O(d^{2})$,
$D_{0}=D_{\mathrm{t}}(1+d)$, and $\mathbf{E},\mathbf{P}=O(\sqrt{d})$,
according to Eq.~(\ref{eq:lasing_ansatz}). Now expand Eq.~(\ref{eq:quadratic eigenproblem})
and keep track of terms order-by-order using Eq.~(\ref{eq:eigenpair expansion}).

\subsection{Zeroth and lowest ($d^{1/2}$) orders}

At zeroth order, we have
\begin{equation}
\left(\mathsf{C}\sigma_{0}^{2}+\mathsf{B}_{0}\sigma_{0}+\mathsf{A}_{0}\right)\mathsf{x}_{0}=0,\label{eq:zeroth order quadratic}
\end{equation}
which turns out to be equivalent to the SALT equation at threshold,
Eq.~(\ref{eq:threshold salt}). The solution is easily seen to be
any linear combination of the threshold modes (with the associated
polarizations), given in Eq.~(\ref{eq:vk definition}). Because these
modes already solve the SALT equation, they necessarily have $\sigma_{0}=0$
(there are other solutions to Eq.~(\ref{eq:zeroth order quadratic}),
corresponding to the below-threshold modes. However, by definition,
they are stable, so we are not concerned with them). Note that there
are now four linearly-independent eigenvectors (from Eq.~(\ref{eq:vk definition})),
even though we only have a double degeneracy in the threshold modes.
This is because we have separated the problem into real and imaginary
parts, and this separation will become significant at higher orders
in $\sqrt{d}$, when the non-analyticity appears in the equations.

Inserting these results into Eq.~(\ref{eq:quadratic eigenproblem})
at order $\sqrt{d}$, we have
\begin{equation}
(\mathsf{B}_{0}\sigma_{1/2}+\mathsf{A}_{1/2})\sum b_{k}\mathsf{v}_{k}+\mathsf{A}_{0}\mathsf{x}_{1/2}=0.\label{eq:half order quadratic}
\end{equation}
We now define the vectors
\begin{equation}
\mathsf{w}_{j}=\left(\begin{array}{c}
\Re\vec{e}_{j}\\
-\Im\vec{e}_{j}\\
\omega_{\mathrm{t}}^{2}\Re\left(\vec{e}_{j}\Gamma_{\mathrm{t}}\right)/\gamma_{\perp}\\
-\omega_{\mathrm{t}}^{2}\Im\left(\vec{e}_{j}\Gamma_{\mathrm{t}}\right)/\gamma_{\perp}\\
0
\end{array}\right).
\end{equation}
where $\mathbf{e}_{k}$ forms the four-component complex basis defined
in Eq.~(\ref{eq:vk definition}). It is straightforward to show that
$\mathsf{A}_{0}^{T}\mathsf{w}_{j}=0$. Additionally, due to the nonzero
pattern of $\mathsf{A}_{1/2}$, it is easy to see that $\mathsf{A}_{1/2}\mathsf{v}_{k}$
has all zero elements except the last, and hence $\mathsf{w}_{j}^{T}\mathsf{A}_{1/2}\mathsf{v}_{k}=0$.
Acting on Eq.~(\ref{eq:half order quadratic}) with $\mathsf{w}_{j}^{T}$,
we then obtain $\sigma_{1/2}\sum\left(\mathsf{w}_{j}^{T}\mathsf{B}_{0}\mathsf{v}_{k}\right)b_{k}=0$.
The matrix $\mathsf{w}_{j}^{T}\mathsf{B}_{0}\mathsf{v}_{k}$ is nonsingular,
which we will see later after we explicitly compute it in Eq.~(\ref{eq:B0 element final}).
Hence, we conclude that $\sigma_{1/2}=0$. 

Next, we compute $\mathsf{A}_{1/2}\mathsf{v}_{k}$. Since the only
nonzero element of $\mathsf{A}_{1/2}\mathsf{v}_{k}$ is the last,
we define this element, after straightforward evaluation, as 
\begin{equation}
g_{k}\equiv2D_{\mathrm{t}}\left|\Gamma_{\mathrm{t}}\right|\sqrt{\gamma_{\parallel}}\Re\left[\vec{e}_{k}^{\star}\cdot(a_{+}\vec{E}_{+}+a_{-}\vec{E}_{-})\right].
\end{equation}
Inserting the result $\sigma_{1/2}=0$ into Eq.~(\ref{eq:half order quadratic}),
we obtain
\begin{equation}
\mathsf{A}_{0}\mathsf{x}_{1/2}=-\mathsf{A}_{1/2}\sum b_{k}\mathsf{v}_{k}=\left(\begin{array}{c}
0\\
0\\
0\\
0\\
-\sum b_{k}g_{k}
\end{array}\right).
\end{equation}
The vector $\mathsf{x}_{1/2}$ must also have this same nonzero pattern,
because if it had any nonzero elements in the first four, they must
be annihilated by $\mathsf{A}_{0}$ and hence must be some linear
combination of $\mathsf{v}_{k}$ (which is already accounted for in
$\mathsf{x}_{0}$ and would be redundant). Hence, we immediately conclude
by inspection of $\mathsf{A}_{0}$ that
\begin{equation}
\mathsf{x}_{1/2}=\left(\begin{array}{c}
0\\
0\\
0\\
0\\
-\gamma_{\parallel}^{-1}\sum b_{k}g_{k}
\end{array}\right).\label{eq:x1/2 exact}
\end{equation}

\subsection{First order}

At the next order, $O(d)$, Eq.~(\ref{eq:quadratic eigenproblem})
is
\begin{equation}
(\mathsf{B}_{0}\sigma_{1}+\mathsf{A}_{1})\sum b_{k}\mathsf{v}_{k}+\mathsf{A}_{1/2}\mathsf{x}_{1/2}+\mathsf{A}_{0}\mathsf{x}_{1}=0.\label{eq:first order quadratic}
\end{equation}
(Note that without the $\mathsf{A}_{1/2}\mathsf{x}_{1/2}$ term, Eq.~(\ref{eq:first order quadratic})
would yield identical results to finding the two passive poles of
the SALT equation that come from the threshold degenerate lasing modes.)
Again, act on this equation with $\mathsf{w}_{j}^{T}$. By direct
evaluation, we have
\begin{multline}
\mathsf{w}_{j}^{T}\mathsf{A}_{1}\mathsf{v}_{k}=\Re\left[\omega_{1}\frac{\partial}{\partial\omega_{\mathrm{t}}}\omega_{\mathrm{t}}^{2}\int d^{3}x\,\vec{e}_{j}\cdot(\varepsilon+\Gamma_{\mathrm{t}}D_{\mathrm{t}})\vec{e}_{k}\right]+\\
\Re\left[\omega_{\mathrm{t}}^{2}\Gamma_{\mathrm{t}}\int d^{3}x\,D_{\mathrm{t}}\vec{e}_{j}\cdot\left(1-\left|\vec{f}\right|^{2}\right)\vec{e}_{k}\right]
\end{multline}
and
\begin{equation}
\mathsf{w}_{j}^{T}\mathsf{B}_{0}\mathsf{v}_{k}=-\Im\left[\frac{\partial}{\partial\omega_{\mathrm{t}}}\omega_{\mathrm{t}}^{2}\int d^{3}x\,\vec{e}_{j}\cdot(\varepsilon+\Gamma_{\mathrm{t}}D_{\mathrm{t}})\vec{e}_{k}\right].
\end{equation}
 By the same symmetry arguments used to evaluate the integrals in
Appendix~\ref{sec: appendix allowed lasing modes}, it is straightforward
to show that 
\begin{equation}
\frac{\partial}{\partial\omega_{\mathrm{t}}}\omega_{\mathrm{t}}^{2}\int d^{3}x\,\vec{e}_{j}\cdot(\varepsilon+\Gamma_{\mathrm{t}}D_{\mathrm{t}})\vec{e}_{k}=\omega_{\mathrm{t}}^{2}\Gamma_{\mathrm{t}}H\left(\begin{array}{cc}
\mathsf{X} & i\mathsf{X}\\
i\mathsf{X} & -\mathsf{X}
\end{array}\right)_{jk},
\end{equation}
where $H$ is given in Eq.~(\ref{eq:G and H definitions}) and
\begin{equation}
\mathsf{X}\equiv\left(\begin{array}{cc}
0 & 1\\
1 & 0
\end{array}\right).
\end{equation}
Next, we have
\begin{multline}
\int d^{3}x\,D_{\mathrm{t}}\vec{e}_{j}\cdot\left|\vec{f}\right|^{2}\vec{e}_{k}=\left(\begin{array}{cc}
\mathsf{M} & i\mathsf{M}\\
i\mathsf{M} & -\mathsf{M}
\end{array}\right)_{jk}+\\
\left(\left|a_{+}\right|^{2}I_{+}+\left|a_{-}\right|^{2}I_{-}\right)\left(\begin{array}{cc}
\mathsf{X} & i\mathsf{X}\\
i\mathsf{X} & -\mathsf{X}
\end{array}\right)_{jk}
\end{multline}
where
\begin{equation}
\mathsf{M}\equiv\left(\begin{array}{cc}
a_{+}^{\star}a_{-}J_{+}+a_{+}a_{-}^{\star}K_{+} & 0\\
0 & a_{+}^{\star}a_{-}K_{-}+a_{+}a_{-}^{\star}J_{-}
\end{array}\right).
\end{equation}
Putting these results together, we have
\begin{multline}
\mathsf{w}_{j}^{T}\mathsf{A}_{1}\mathsf{v}_{k}=\\
\omega_{\mathrm{t}}^{2}\Re\left[\Gamma_{\mathrm{t}}W\left(\begin{array}{cc}
\mathsf{X} & i\mathsf{X}\\
i\mathsf{X} & -\mathsf{X}
\end{array}\right)_{jk}-\Gamma_{\mathrm{t}}\left(\begin{array}{cc}
\mathsf{M} & i\mathsf{M}\\
i\mathsf{M} & -\mathsf{M}
\end{array}\right)_{jk}\right]\label{eq:A1 element final}
\end{multline}
where 
\begin{equation}
W\equiv\omega_{\mathrm{1}}H+G_{D}-\left|a_{+}\right|^{2}I_{+}-\left|a_{-}\right|^{2}I_{-},
\end{equation}
and
\begin{equation}
\mathsf{w}_{j}^{T}\mathsf{B}_{0}\mathsf{v}_{k}=-\omega_{\mathrm{t}}^{2}\Im\left[\Gamma_{\mathrm{t}}H\left(\begin{array}{cc}
\mathsf{X} & i\mathsf{X}\\
i\mathsf{X} & -\mathsf{X}
\end{array}\right)_{jk}\right].\label{eq:B0 element final}
\end{equation}

Next, by straightforward computation, we obtain
\begin{multline}
\mathsf{w}_{j}^{T}\mathsf{A}_{1/2}\mathsf{x}_{1/2}=\\
-\sum_{k}b_{k}\int d^{3}x\,\omega_{\mathrm{t}}^{2}D_{\mathrm{t}}\Re\left[\Gamma_{\mathrm{t}}\left(\vec{f}\cdot\vec{e}_{j}\right)\left(\vec{f}\cdot\vec{e}_{k}^{\star}+\vec{f}^{\star}\cdot\vec{e}_{k}\right)\right].\label{eq:A1/2 matrix element}
\end{multline}
Again, by straightforward computation, we see that
\begin{align}
\int d^{3}x\,D_{\mathrm{t}}\left(\vec{f}\cdot\vec{e}_{j}\right)\left(\vec{f}\cdot\vec{e}_{k}^{\star}\right) & =\left(\begin{array}{cc}
\mathsf{Q} & -i\mathsf{Q}\\
i\mathsf{Q} & \mathsf{Q}
\end{array}\right)_{jk}\nonumber \\
\int d^{3}x\,D_{\mathrm{t}}\left(\vec{f}\cdot\vec{e}_{j}\right)\left(\vec{f}^{\star}\cdot\vec{e}_{k}\right) & =\left(\begin{array}{cc}
\mathsf{P} & i\mathsf{P}\\
i\mathsf{P} & -\mathsf{P}
\end{array}\right)_{jk}
\end{align}
where
\begin{align}
\mathsf{Q} & \equiv\left(\begin{array}{cc}
a_{+}a_{-}(I_{+}+J_{+}) & a_{-}^{2}I_{-}+a_{+}^{2}K_{+}\\
a_{+}^{2}I_{+}+a_{-}^{2}K_{-} & a_{+}a_{-}(I_{-}+J_{-})
\end{array}\right)\nonumber \\
\mathsf{P} & \equiv\left(\begin{array}{cc}
a_{-}a_{+}^{\star}I_{+}+a_{+}a_{-}^{\star}K_{+} & \left|a_{-}\right|^{2}I_{-}+\left|a_{+}\right|^{2}J_{+}\\
\left|a_{+}\right|^{2}I_{+}+\left|a_{-}\right|^{2}J_{-} & a_{+}a_{-}^{\star}I_{-}+a_{-}a_{+}^{\star}K_{-}
\end{array}\right).
\end{align}
Putting this together, Eq.~(\ref{eq:A1/2 matrix element}) then becomes
\begin{multline}
\mathsf{w}_{j}^{T}\mathsf{A}_{1/2}\mathsf{x}_{1/2}=\\
-\sum_{k}b_{k}\omega_{\mathrm{t}}^{2}\Re\left[\Gamma_{\mathrm{t}}\left(\begin{array}{cc}
\mathsf{Q}+\mathsf{P} & i(\mathsf{P}-\mathsf{Q})\\
i(\mathsf{P}+\mathsf{Q}) & \mathsf{Q}-\mathsf{P}
\end{array}\right)_{jk}\right].
\end{multline}
Combining this with Eq.~(\ref{eq:A1 element final}) and Eq.~(\ref{eq:B0 element final}),
defining $\tilde{\mathsf{P}}\equiv\mathsf{P}+\mathsf{M}-W\mathsf{X}$,
and evaluating the real and imaginary components, Eq.~(\ref{eq:first order quadratic})
becomes
\begin{multline}
\left(\begin{array}{cc}
\Re\left[\Gamma_{\mathrm{t}}(\mathsf{Q}+\tilde{\mathsf{P}})\right] & \Im\left[\Gamma_{\mathrm{t}}(\mathsf{Q}-\tilde{\mathsf{P}})\right]\\
-\Im\left[\Gamma_{\mathrm{t}}(\mathsf{Q}+\tilde{\mathsf{P}})\right] & \Re\left[\Gamma_{\mathrm{t}}(\mathsf{Q}-\tilde{\mathsf{P}})\right]
\end{array}\right)\mathsf{b}\\
=-\sigma_{1}\left(\begin{array}{cc}
\Im(\Gamma_{\mathrm{t}}H)\mathsf{X} & \Re(\Gamma_{\mathrm{t}}H)\mathsf{X}\\
\Re(\Gamma_{\mathrm{t}}H)\mathsf{X} & -\Im(\Gamma_{\mathrm{t}}H)\mathsf{X}
\end{array}\right)\mathsf{b}.
\end{multline}
Using the fact that $\mathsf{X}$ is its own inverse, we multiply
this equation by the matrix on the right-hand side, and obtain
\begin{equation}
\left(\begin{array}{cc}
\Im\left[\mathsf{X}(\mathsf{Q}+\tilde{\mathsf{P}})/H\right] & -\Re\left[\mathsf{X}(\mathsf{Q}-\tilde{\mathsf{P}})/H\right]\\
-\Re\left[\mathsf{X}(\mathsf{Q}+\tilde{\mathsf{P}})/H\right] & -\Im\left[\mathsf{X}(\mathsf{Q}-\tilde{\mathsf{P}})/H\right]
\end{array}\right)\mathsf{b}=\sigma_{1}\mathsf{b},\label{eq:final linear eigenvalue problem}
\end{equation}
which is a $4\times4$ linear eigenvalue problem for $\sigma_{1}$.

\subsection{Closed-form stability eigenvalues}

We now diagonalize Eq.~(\ref{eq:final linear eigenvalue problem})
for each of the lasing mode solutions in Sec.~\ref{sec:Threshold-perturbation-theory}.

\subsubsection*{Circulating lasing mode}

For the circulating solution in Eq.~(\ref{eq:circulating lasing}),
we have $a_{-}=0$, leading to $W=0$ and $\tilde{\mathsf{P}}=\mathsf{P}$.
The matrix in Eq.~(\ref{eq:final linear eigenvalue problem}) then
becomes
\begin{equation}
\left(\begin{array}{cccc}
2\Im\left(\frac{I_{+}}{H}\right) & 0 & 0 & 0\\
0 & \Im\left(\frac{K_{+}+J_{+}}{H}\right) & 0 & -\Re\left(\frac{K_{+}-J_{+}}{H}\right)\\
-2\Re\left(\frac{I_{+}}{H}\right) & 0 & 0 & 0\\
0 & -\Re\left(\frac{K_{+}+J_{+}}{H}\right) & 0 & -\Im\left(\frac{K_{+}-J_{+}}{H}\right)
\end{array}\right)|a_{+}|^{2},
\end{equation}
where $|a_{+}|^{2}\equiv\frac{\omega_{1}H+G_{D}}{I_{+}}$. By inspection,
there is an eigenpair with
\begin{equation}
\sigma_{1}=0,\;\mathsf{b}=\left(\begin{array}{c}
0\\
0\\
1\\
0
\end{array}\right).
\end{equation}
Since the third component of the basis is $\mathbf{e}_{3}=i\mathbf{E}_{+}$,
this eigenvector corresponds to a global phase rotation $\mathbf{E}_{+}\rightarrow(1+i\delta)\mathbf{E}_{+}$,
which is a continuous symmetry of the original Maxwell--Bloch equations.
A second eigenpair is
\begin{equation}
\sigma_{1}=2\left|a_{+}\right|^{2}\Im\left(\frac{I_{+}}{H}\right),\;\mathsf{b}=\left(\begin{array}{c}
\Im\left(\frac{I_{+}}{H}\right)\\
0\\
-\Re\left(\frac{I_{+}}{H}\right)\\
0
\end{array}\right).
\end{equation}
The remaining two eigenvalues are
\begin{equation}
\sigma_{1}=\left[\Im\left(\frac{J_{+}}{H}\right)\pm\sqrt{\left|\frac{K_{+}}{H}\right|^{2}-\Re\left(\frac{J_{+}}{H}\right)^{2}}\right]\left|a_{+}\right|^{2},
\end{equation}
with
\begin{equation}
\mathbf{b}=\left(\begin{array}{c}
0\\
\Re\left(\frac{K_{+}-J_{+}}{H}\right)\\
0\\
\Im\left(\frac{K_{+}}{H}\right)\mp\sqrt{\left|\frac{K_{+}}{H}\right|^{2}-\Re\left(\frac{J_{+}}{H}\right)^{2}}
\end{array}\right).
\end{equation}

\subsubsection*{Standing-wave modes, $n\protect\neq4\ell$}

We now diagonalize Eq.~(\ref{eq:final linear eigenvalue problem})
for the standing-wave modes. First, for $n\neq4\ell$, standing-wave
modes only occur in $C_{n\mathrm{v}}$, as discussed in Appendix~\ref{sec: appendix allowed lasing modes}.
In this case, all $K_{\pm}=0$, and $I_{+},J_{+}=I_{-},J_{-}$. The
matrix in Eq.~(\ref{eq:final linear eigenvalue problem}) then becomes
$2|a|^{2}\times$
\begin{equation}
\left(\begin{array}{cccc}
\Im\left(\frac{I}{H}\right) & \Re(z)\Im\left(\frac{I+J}{H}\right) & 0 & \Im(z)\Im\left(\frac{I+J}{H}\right)\\
\Im\left[z\left(\frac{I+J}{H}\right)\right] & \Re(z)\Im\left(\frac{Iz}{H}\right) & 0 & \Im(z)\Im\left(\frac{Iz}{H}\right)\\
-\Re\left(\frac{I}{H}\right) & -\Re(z)\Re\left(\frac{I+J}{H}\right) & 0 & -\Im(z)\Re\left(\frac{I+J}{H}\right)\\
-\Re\left[z\left(\frac{I+J}{H}\right)\right] & -\Re(z)\Re\left(\frac{Iz}{H}\right) & 0 & -\Im(z)\Re\left(\frac{Iz}{H}\right)
\end{array}\right),
\end{equation}
where $|a|^{2}=\frac{\omega_{1}H+G_{D}}{2I+J}$. This matrix has \emph{two}
zero eigenvectors that have $\sigma_{1}=0$:
\begin{equation}
\mathsf{b}=\left(\begin{array}{c}
0\\
0\\
1\\
0
\end{array}\right),\;\mathsf{b}=\left(\begin{array}{c}
0\\
\Im(z)\\
0\\
\Re(z)
\end{array}\right),
\end{equation}
which comes from the two continuous degrees of freedom in the solution:
the overall global phase freedom, as well as the relative phase $z$
in Eq.~(\ref{eq:n !=00003D 4 lasing}). It is straightforward to
find the other two eigenpairs, which are
\begin{equation}
\sigma_{1}=2\Im\left(\frac{2I+J}{H}\right)|a|^{2},\;\mathsf{b}=\left(\begin{array}{c}
1\\
\Re(z)\\
0\\
\Im(z)
\end{array}\right),
\end{equation}
and
\begin{equation}
\sigma_{1}=-2\Im\left(\frac{J}{H}\right)|a|^{2},\;\mathsf{b}=\left(\begin{array}{c}
-1\\
\Re(z)\\
0\\
\Im(z)
\end{array}\right).
\end{equation}

\subsubsection*{Standing-wave modes, $n=4\ell$ }

Next, we examine the case $C_{n\mathrm{v}}$ for $n=4\ell$ (standing-wave
modes also exist here in the $C_{n}$ case, but Eq.~(\ref{eq:final linear eigenvalue problem})
must be diagonalized numerically). For the case of a $\mathbf{E}_{+}\pm\mathbf{E}_{-}$
lasing mode (Eq.~(\ref{eq:E+ + E- lasing})), the matrix in Eq.~(\ref{eq:final linear eigenvalue problem})
is $2|a|^{2}\times$ 
\begin{equation}
\left(\begin{array}{cccc}
\Im\left(\frac{I}{H}\right) & \pm\Im\left(\frac{I+J+K}{H}\right) & -\Re\left(\frac{K}{H}\right) & \pm\Re\left(\frac{K}{H}\right)\\
\pm\Im\left(\frac{I+J+K}{H}\right) & \Im\left(\frac{I}{H}\right) & \pm\Re\left(\frac{K}{H}\right) & -\Re\left(\frac{K}{H}\right)\\
-\Re\left(\frac{I}{H}\right) & \mp\Re\left(\frac{I+J+K}{H}\right) & -\Im\left(\frac{K}{H}\right) & \pm\Im\left(\frac{K}{H}\right)\\
\mp\Re\left(\frac{I+J+K}{H}\right) & -\Re\left(\frac{I}{H}\right) & \pm\Im\left(\frac{K}{H}\right) & -\Im\left(\frac{K}{H}\right)
\end{array}\right),
\end{equation}
where $|a|^{2}=\frac{\omega_{1}H+G_{D}}{2I+J+K}$, and the $\pm$
symbols correspond to $z=\pm1$. There is a single zero eigenvalue:
\begin{equation}
\sigma_{1}=0,\;\mathsf{b}=\left(\begin{array}{c}
0\\
0\\
1\\
\pm1
\end{array}\right).
\end{equation}
A second eigenpair is
\begin{equation}
\sigma_{1}=2|a|^{2}\Im\left(\frac{2I+J+K}{H}\right),\;\mathsf{b}=\left(\begin{array}{c}
1\\
\pm1\\
0\\
0
\end{array}\right).
\end{equation}
Empirically, we have found that this eigenvalue is always stable.
The final two eigenvalues are
\begin{multline}
\sigma_{1}=-|a|^{2}\Im\left(\frac{J+3K}{H}\right)\pm\\
|a|^{2}\sqrt{\Im\left(\frac{J-K}{H}\right)^{2}-8\Re\left(\frac{K}{H}\right)\Re\left(\frac{J+K}{H}\right)}\label{eq:z =00003D 1 double eigenvalues}
\end{multline}
(the eigenvectors can be written down in closed form, but are tedious
and not illuminating). Empirically, we have found that at least one
of these two eigenvalues are unstable (except for an isolated case,
that we explain below). Next, for the case of a $\mathbf{E}_{+}\pm i\mathbf{E}_{-}$
lasing mode (Eq.~(\ref{eq:E+ + i E- lasing}), the matrix in Eq.~(\ref{eq:final linear eigenvalue problem})
is $2|a|^{2}\times$
\begin{equation}
\left(\begin{array}{cccc}
\Im\left(\frac{I}{H}\right) & \pm\Re\left(\frac{K}{H}\right) & \Re\left(\frac{K}{H}\right) & \pm\Im\left(\frac{I+J-K}{H}\right)\\
\pm\Re\left(\frac{I+J-K}{H}\right) & \Im\left(\frac{K}{H}\right) & \pm\Im\left(\frac{K}{H}\right) & \Re\left(\frac{I}{H}\right)\\
-\Re\left(\frac{I}{H}\right) & \pm\Im\left(\frac{K}{H}\right) & \Im\left(\frac{K}{H}\right) & \mp\Re\left(\frac{I+J-K}{H}\right)\\
\pm\Im\left(\frac{I+J-K}{H}\right) & -\Re\left(\frac{K}{H}\right) & \mp\Re\left(\frac{K}{H}\right) & \Im\left(\frac{I}{H}\right)
\end{array}\right),
\end{equation}
where $|a|^{2}=\frac{\omega_{1}H+G_{D}}{2I+J-K}$, and the $\pm$
signs correspond to $z=\pm i$. There is an eigenpair with zero eigenvalue:
\begin{equation}
\sigma_{1}=0,\;\mathsf{b}=\left(\begin{array}{c}
0\\
1\\
\mp1\\
0
\end{array}\right)
\end{equation}
and another eigenpair
\begin{equation}
\sigma_{1}=2|a|^{2}\Im\left(\frac{2I+J-K}{H}\right),\;\mathsf{b}=\left(\begin{array}{c}
\pm1\\
0\\
0\\
1
\end{array}\right).
\end{equation}
Empirically, we have found that this eigenvalue is always stable.
Finally, the remaining two eigenvalues are
\begin{multline}
\sigma_{1}=|a|^{2}\Im\left(\frac{3K-J}{H}\right)\pm\\
|a|^{2}\sqrt{\Im\left(\frac{J+K}{H}\right)^{2}+8\Re\left(\frac{K}{H}\right)\Re\left(\frac{J-K}{H}\right)}\label{eq:z =00003D i double eigenvalues}
\end{multline}
(the eigenvectors can be written down in closed form, but are tedious
and not illuminating). Empirically, we have found that at least one
of these two eigenvalues are unstable, except for an isolated case
that we will now explain.

We note that for the previous two cases, where $n=4\ell$, it is possible
to choose the shape of the gain profile $D_{\mathrm{t}}(\mathbf{x})$
such that $J=\pm K$, in which case the \emph{one }of the two pairs
$\mathbf{E}_{+}\pm\mathbf{E}_{-}$ and $\mathbf{E}_{+}\pm i\mathbf{E}_{-}$
actually becomes \emph{stable}. To see this, we consider a $C_{n\mathrm{v}}$
geometry with $n=4\ell$. Equation.~(\ref{eq:chiral definition})
then becomes
\begin{equation}
\mathbf{E}_{\pm}=\sum_{b=1}^{n}(\pm i)^{b}R_{b/n}\mathbf{E}_{\mathrm{even}}.
\end{equation}
The specific choice of geometry requires that we place radially-symmetric
lines of gain on the faces or diagonals of the $C_{n\mathrm{v}}$
geometry, which preserves the $C_{n\mathrm{v}}$ symmetry. We can
write this as
\begin{equation}
D_{\mathrm{t}}(\mathbf{x})=\sum_{a=1}^{n}G(r)\delta\left(\theta-\theta_{a}\right),
\end{equation}
where $\theta_{a}=\frac{2\pi a}{n}$ for the faces, and $\theta_{a}=(2a+1)\frac{\pi}{n}$
for the diagonals. For a TM geometry, we then have the overlap integrals
(Eq.~(\ref{eq:I J K definitions}))
\begin{align}
J & =\sum_{a=1}^{n}\int rdr\,G(r)\left|E_{+}\left(r,\theta_{a}\right)\right|^{2}E_{-}(r,\theta_{a})E_{+}(r,\theta_{a})\nonumber \\
K & =\sum_{a=1}^{n}\int rdr\,G(r)E_{+}(r,\theta_{a})^{\star}E_{-}(r,\theta_{a})^{3}.
\end{align}
It can be shown that depending on the choice of $\theta_{a}$ being
the faces or diagonals, we will have either $K=J$ or $K=-J$. For
$K=J$, the eigenvalues in Eq.~(\ref{eq:z =00003D i double eigenvalues});
of the $\mathbf{E}_{+}\pm i\mathbf{E}_{-}$ standing-wave modes, become
0 and a stable eigenvalue. For $K=-J$, the same happens for those
in Eq.~(\ref{eq:z =00003D 1 double eigenvalues}); of the $\mathbf{E}_{+}\pm\mathbf{E}_{-}$
standing-wave modes.

\subsection{Region of validity in small-$\gamma_{\parallel}$ limit\label{sub:Region-of-validity}}

In this section, we work out the $\gamma_{\parallel}$ dependence
of higher-order terms in the perturbation theory, and demonstrate
that the regime of validity of the perturbation theory depends on
$d$ being small compared to a constant multiple of $\gamma_{\parallel}$.
We show that the exact expansion of $\sigma/d$ to all orders in $d$
only contains terms of the form $d^{\ell}/\gamma_{\parallel}^{j}$
with $\ell\geq j$, and that in the limit where $\gamma_{\parallel},d\to0$,
with $d$ vanishing at least as rapidly as $\gamma_{\parallel}$,
the terms with $\ell>j$ vanish and the stability eigenvalue takes
the asymptotic functional form $\sigma/d=f(d/\gamma_{\parallel})$.
Here, $f(0)$ is exactly the first-order stability eigenvalue $\sigma_{1}$.
For circulating modes that have $\mathrm{Re}(\sigma_{1})<0$, the
the smallest positive solution $z_{0}$ of the equation $\Re[f(z)]=0$
gives a boundary of stability, for which a circulating lasing mode
becomes \emph{unstable }for $d>\gamma_{\parallel}z_{0}$, as seen
in the ring-laser example in Ref.~\cite{ref-rotter_degeneracy}.

First, since we have already obtained closed-form expressions for
$\sigma_{1}$ and the coefficients $b_{k}$ of $\mathsf{x}_{0}$,
Eq.~(\ref{eq:first order quadratic}) can be solved for $\mathsf{x}_{1}$:
$\mathsf{x}_{1}=\mathsf{A}_{0}^{-1}\left[\left(\mathsf{B}_{0}\sigma_{1}+\mathsf{A}_{1}\right)\mathsf{x}_{0}+\mathsf{A}_{1/2}\mathsf{x}_{1/2}\right]$.
Next, at order $d^{3/2}$, we have
\begin{multline}
\mathsf{A}{}_{3/2}\mathsf{x}{}_{0}+\mathsf{A}{}_{1}\mathsf{x}{}_{1/2}+\mathsf{A}{}_{1/2}\mathsf{x}{}_{1}+\mathsf{A}{}_{0}\mathsf{x}{}_{3/2}\\
=-\mathsf{B}_{0}\sigma_{3/2}\mathsf{x}{}_{0}-\mathsf{B}_{0}\sigma_{1}\mathsf{x}{}_{1/2}.
\end{multline}
The first three terms on the left-hand side as well as the very last
term on the right all have the same nonzero pattern as $\mathsf{x}_{1/2}$.
Hence, multiplying both sides by $\mathsf{x}_{0}^{T}$ (which is now
known after having solved the degenerate problem at order $d$), we
obtain $0=\sigma_{3/2}\mathsf{x}_{0}^{T}\mathsf{B}_{0}\mathsf{x}{}_{0}$,
which leads to $\sigma_{3/2}=0$. The only remaining unknown is then
$\mathsf{x}_{3/2}$. By the same arguments leading to Eq.~(\ref{eq:x1/2 exact}),
we have $\mathsf{A}{}_{0}\mathsf{x}{}_{3/2}=\gamma_{\parallel}\mathsf{x}{}_{3/2}$,
which yields $\mathsf{x}_{3/2}=-\gamma_{\parallel}^{-1}\mathsf{f}_{3/2}$,
where
\begin{equation}
\mathsf{f}_{3/2}\equiv\mathsf{A}{}_{3/2}\mathsf{x}{}_{0}+\mathsf{A}{}_{1}\mathsf{x}{}_{1/2}+\mathsf{A}{}_{1/2}\mathsf{x}{}_{1}+\mathsf{B}_{0}\sigma_{1}\mathsf{x}{}_{1/2},
\end{equation}
and $\mathsf{f}_{3/2}$ is $O(1)$ with respect to $\gamma_{\parallel}$,
i.e. it goes to a constant as $\gamma_{\parallel}\rightarrow0$. Moving
onto order $d^{2}$, we have
\begin{multline}
\mathsf{C}\sigma_{1}^{2}\mathsf{x}_{0}+\mathsf{B}_{0}\left(\sigma_{2}\mathsf{x}{}_{0}+\sigma_{1}\mathsf{x}{}_{1}\right)+\mathsf{B}_{1}\sigma_{1}\mathsf{x}_{0}\\
=-\sum_{k=0}^{4}\mathsf{A}_{k/2}\mathsf{x}_{2-\frac{k}{2}}.
\end{multline}
Multiplying both sides by $\mathsf{x}_{0}^{T}$ annihilates the $k=0$
term, leaving only a single unknown $\sigma_{2}$ and a single term
of $O(\gamma_{\parallel}^{-1})$: $-\mathsf{A}_{1/2}\mathsf{x}_{3/2}=\gamma_{\parallel}^{-1}\mathsf{A}_{1/2}\mathsf{f}_{3/2}$.
We then have
\begin{equation}
\gamma_{\parallel}\sigma_{2}=\frac{\mathsf{x}_{0}^{T}\mathsf{A}_{1/2}\mathsf{f}_{3/2}}{\mathsf{x}_{0}^{T}\mathsf{B}_{0}\mathsf{x}_{0}}+O(\gamma_{\parallel}).
\end{equation}
Carrying on to the next order results in $\sigma_{5/2}=0$ and $\mathsf{x}_{5/2}=-\gamma_{\parallel}^{-1}\mathsf{f}_{5/2}$,
where now $\mathsf{f}_{5/2}=O(\gamma_{\parallel}^{-1})$ due to it
including terms with $\sigma_{2}$ and $\mathsf{x}_{2}$. By continuing
this process, we find that $\sigma_{m+\frac{1}{2}}=0$, and $\mathsf{x}_{m+\frac{1}{2}}$,
$\sigma_{m+1}$, and $\mathsf{x}_{m+1}$ are all $O(\gamma_{\parallel}^{-m})$.
This is made possible by the fact that the only place $\gamma_{\parallel}$
occurs in the entire problem is the very last matrix element of $\mathsf{A}_{0}$
in Eq.~(\ref{eq:A matrix}), as well as the fact that the $\mathsf{x}_{m+\frac{1}{2}}$
and $\mathsf{x}_{m}$ have predictable nonzero patterns, due to the
nonzero patterns of $\mathsf{A}_{m+\frac{1}{2}}$ being different
from those of all the other matrices. Defining $s_{m+1}\equiv\gamma_{\parallel}^{m}\sigma_{m+1}=O(1)$,
we obtain a full expansion for the exact Maxwell--Bloch eigenvalue
(Eq.~(\ref{eq:eigenpair expansion})):
\begin{equation}
\sigma\approx d\sum_{k=1}^{\infty}\left(\frac{d}{\gamma_{\parallel}}\right)^{k-1}s_{k},\label{eq:s expansion}
\end{equation}
where the $\approx$ comes from the fact that we have thrown away
terms of the form $d^{\ell}/\gamma_{\parallel}^{j}$ with $\ell>j$,
which are negligible compared to $(d/\gamma_{\parallel})^{k}$ and
vanish in the limit $d,\gamma_{\parallel}\rightarrow0$ (provided
that $\gamma_{\parallel}$ does not go zero more rapidly than $d$).
In this limit, we can then infer a generic functional form of $\sigma$:
\begin{equation}
\lim_{\gamma_{\parallel},d\rightarrow0}\frac{\sigma}{d}=f\left(\frac{d}{\gamma_{\parallel}}\right),\label{eq:limit gamma parallel}
\end{equation}
where $f$ is a complex-valued function (with a real argument) whose
Taylor expansion is the sum in Eq.~(\ref{eq:s expansion}). For an
eigenvalue $\sigma$ of the Maxwell--Bloch equation linearized about
a circulating mode, we have $f(0)=\sigma_{1}$ having a negative real
part. The smallest positive zero $z_{0}$ of $\Re f$ then gives the
equation for a boundary of stability $d=\gamma_{\parallel}z_{0}$.
Hence, in the limit that both $\gamma_{\parallel}$ and $d$ go to
zero, the region of stability for a circulating mode is given by $d<\gamma_{\parallel}z_{0}$,
where $z_{0}$ is a constant independent of $\gamma_{\parallel}$
and $d$.

\section{Determining the dielectric perturbation $\delta\varepsilon$\label{sec:Determining-the-dielectric-perturbation}}

In this Appendix, we describe the processe used to force a degeneracy
in a geometry whose symmetry has been broken by the discretization
scheme (e.g., a $C_{6\mathrm{v}}$ geometry discretized into a rectangular
grid). We first analyze the effect of a small $\delta\varepsilon$
on the eigenfrequencies (of the lasing mode and the passive pole)
by well-known first-order perturbation theory for Maxwell's equations
\cite{phc_book} (some modification is required to handle the nonlinearity
of the hole-burning term above threshold). However, we first force
the degeneracy below threshold (repeating as needed as the pump strength
is increased), so that both passive poles reach threshold simultaneously.
(In practice, we achieved the fastest convergence by allowing passive
poles to have positive imaginary parts, and then setting the pump
strength so that the two poles ``straddle'' the real axis; this
way, when they meet in the middle they are both exactly at threshold.)
Below threshold, the eigenproblem is linear in the eigenvector $\mathbf{E}$
(the nonlinearity in $\omega$ is still present but easy to deal with
using standard methods), and we can apply standard perturbation theory
(albeit for a complex-symmetric operator, not a Hermitian operator)
as follows:

Consider two nonlasing modes that satisfy
\begin{align}
0 & =-\nabla\times\nabla\times\mathbf{E}_{\mu}+\omega_{\mu}^{2}\varepsilon_{\mu}\mathbf{E}_{\mu}\nonumber \\
\varepsilon_{\mu} & \equiv\varepsilon_{c}+\Dt\Gamma(\omega_{\mu})
\end{align}
Adding a perturbation to the dielectric $\delta\varepsilon$ will
result in corresponding responses $\delta\mathbf{E}_{\mu}$ and $\delta\omega_{\mu}$.
As in the threshold perturbation theory, we multiply both sides by
$\mathbf{E}_{\mu}$ and keep only first-order terms. Terms involving
$\delta\mathbf{E}_{\mu}$ again vanish because the operators act to
the left, and we are left with \cite{ref-raman,ref-microcavity,ref-adi}
\begin{equation}
\delta\omega_{\mu}=-\frac{\int d^{3}x\,\mathbf{E}_{\mu}\cdot\delta\varepsilon\mathbf{E}_{\mu}}{\int d^{3}x\,\mathbf{E}_{\mu}\cdot\left(\frac{2\varepsilon_{\mu}}{\omega_{\mu}}+\frac{\partial\varepsilon_{\mu}}{\partial\omega_{\mu}}\right)\mathbf{E}_{\mu}}.\label{eq:delta_omega_QP}
\end{equation}
We write this frequency shift as an inner product 
\begin{equation}
\delta\omega_{\mu}=-p_{\mu}^{T}\delta\varepsilon.\label{eq:delta_omega_mu}
\end{equation}
As an aside, while it is fine to use a scalar $\delta\varepsilon$
function for this procedure, in the case when the $\mathbf{E}_{\mu}$
are TE modes or fully-vectorial fields, then it is also possible to
allow $\delta\varepsilon(\mathbf{x})$ to be a diagonally anisotropic
tensor
\begin{equation}
\overleftrightarrow{\delta\varepsilon}(\mathbf{x})=\left(\begin{array}{ccc}
\delta\varepsilon_{xx}(\mathbf{x}) & 0 & 0\\
0 & \delta\varepsilon_{yy}(\mathbf{x}) & 0\\
0 & 0 & \delta\varepsilon_{zz}(\mathbf{x})
\end{array}\right).\label{eq:tensor_delta_epsilon}
\end{equation}
The column-vector form of $\delta\varepsilon$ in Eq.~(\ref{eq:delta_omega_mu})
would then have as its elements all the real and imaginary components
of $\overleftrightarrow{\delta\varepsilon}(\mathbf{x})$ at each Yee
point \cite{taflove} {[}$\delta\varepsilon_{xx}(\mathbf{x})$, $\delta\varepsilon_{yy}(\mathbf{x})$,
and $\delta\varepsilon_{zz}(\mathbf{x})$ for all the grid points
$\mathbf{x}${]}, while the row-vector $p_{\mu}^{T}$ would have as
its elements the real and imaginary parts of $E_{x}(\mathbf{x})^{2}$,
$E_{y}(\mathbf{x})^{2}$, and $E_{z}(\mathbf{x})^{2}$ at all the
grid points. If we take this option, then the norm we minimize would
be
\begin{equation}
\left\Vert \delta\varepsilon\right\Vert _{2}^{2}=\int d^{3}x\,\left\Vert \overleftrightarrow{\delta\varepsilon}(\mathbf{x})\right\Vert _{F}^{2},
\end{equation}
where the Frobenius norm \cite{trefethen} at each point $\mathbf{x}$
is defined as
\begin{equation}
\left\Vert \overleftrightarrow{\delta\varepsilon}(\mathbf{x})\right\Vert _{F}^{2}\equiv\left|\delta\varepsilon_{xx}(\mathbf{x})\right|^{2}+\left|\delta\varepsilon_{yy}(\mathbf{x})\right|^{2}+\left|\delta\varepsilon_{zz}(\mathbf{x})\right|^{2}.
\end{equation}
Whether we take $\delta\varepsilon$ to be a scalar or a tensor, the
degeneracy-forcing condition $\omega_{1}+\delta\omega_{1}=\omega_{2}+\delta\omega_{2}$
then becomes
\begin{equation}
(p_{2}-p_{1})^{T}\delta\varepsilon=\omega_{2}-\omega_{1}.\label{eq:degeneracy_forcing}
\end{equation}
It turns out that the solution of a quadratic program with equality
constraints can be obtained directly by solving a linear \emph{dual}
problem \cite{boyd}, which in this case is
\begin{equation}
\left(\begin{array}{cccc}
1 & 0 & q^{R} & q^{I}\\
0 & 1 & -q^{I} & q^{R}\\
\left(q^{R}\right)^{T} & -\left(q^{I}\right)^{T} & 0 & 0\\
\left(q^{I}\right)^{T} & \left(q^{R}\right)^{T} & 0 & 0
\end{array}\right)\left(\begin{array}{c}
\delta\varepsilon^{R}\\
\delta\varepsilon^{I}\\
\lambda_{1}\\
\lambda_{2}
\end{array}\right)=\left(\begin{array}{c}
0\\
0\\
\omega_{2}^{R}-\omega_{1}^{R}\\
\omega_{2}^{I}-\omega_{1}^{I}
\end{array}\right).
\end{equation}
Here, the superscripts $^{R}$ and $^{I}$ denote real and imaginary
parts, and we have defined $q\equiv p_{2}-p_{1}$, and the $\lambda_{1,2}$
are Lagrange multipliers that are not needed. When $\omega_{2}$ is
very close to $\omega_{1}$, we can improve the condition number of
the matrix by freely multiplying the second-to-last row and column
of the matrix by a constant factor, provided that the second-to-last
element of the right-hand side is \emph{divided }by the same factor.
The same can be done for the last row and column, with the last element
of the right-hand side. 

Note that even after the thresholds and threshold frequencies have
been made exactly degenerate using the QP procedure illustrated above,
we are still in principle forcing the degeneracy. Above threshold,
the delicate balance created by $\delta\varepsilon$ to force the
frequencies together is slightly broken. This results in an \emph{approximate
}degeneracy that is maintained very far above threshold, as shown
in Fig.~\ref{fig:QP_stability}, with only a $10^{-8}$ splitting
for pump strengths up to 100 times threshold. In practice, these results
are already accurate enough to give all the desired physical information
about the degenerate pair. If we wanted to be absolutely correct and
force the degeneracy to machine precision (as it was in the exactly
symmetric case for odd-$\ell$ modes), we could simply perform QP
again at some given $d>0$ to force $\delta\omega$ and $\delta\omega^{\prime}$
back together. One extra caveat in this case is that $\delta\omega$
is now a lasing pole, so the spatial hole-burning term needs to be
accounted for in the perturbation theory ($\delta\omega^{\prime}$
is still a passive pole, so the previous perturbation theory still
applies), and instead of Eq.~(\ref{eq:delta_omega_QP}) we now have
\begin{align}
\delta\omega & =-\frac{\int d^{3}x\,\mathbf{E}\cdot\left(\delta\varepsilon+\Dt\Gamma(\omega_{\mu})\delta H\right)\mathbf{E}}{\int d^{3}x\,\mathbf{E}\cdot\left(\frac{2\varepsilon}{\omega}+\frac{\partial\varepsilon}{\partial\omega}\right)\mathbf{E}}\nonumber \\
\varepsilon & \equiv\varepsilon_{c}+\Dt\Gamma(\omega)H\\
\delta H & \equiv\frac{1}{1+\left|\mathbf{E}+\delta\mathbf{E}\right|^{2}}-\frac{1}{1+\left|\mathbf{E}\right|^{2}}.\nonumber 
\end{align}
Here, $\delta H$ is the change in the spatial-hole burning term arising
from the dielectric perturbation $\delta\varepsilon$. However, since
there is no easy way to determine $\delta\mathbf{E}$ without numerically
solving the full problem, $\delta H$ is hard to determine semi-analytically.
A simple work-around is to set $\delta H=0$ above, which makes this
procedure no longer a true first-order perturbation theory. However,
since the splitting is already so small as shown in Fig.~\ref{fig:QP_stability},
the $\delta\varepsilon$ needed is also extremely small, so $\delta H$
is also negligible. Although $\delta\omega$ is not zero to first
order, the $\delta H=0$ approximation is enough to find a $\delta\varepsilon$
that greatly decreases $\delta\omega$. We find empirically that it
usually takes one iteration of this above-threshold QP procedure to
restore the degeneracy of the lasing pole $\omega$ and its passive
mode $\omega^{\prime}$ to machine precision, since $\delta\omega$
is already very small. Practically speaking, this entire extra step
is rarely needed since the solutions obtained from $\delta\varepsilon$
for the linear problem below threshold are already close enough for
most pump strengths of physical interest.

\bibliographystyle{apsrev4-1}

\end{document}